\documentclass[aps,twocolumn,prd,reprint,superscriptaddress,amssymb]{revtex4-1}

\usepackage{amsmath}

\usepackage{graphicx}
\usepackage{floatrow}
\usepackage[leftcaption]{sidecap}
\sidecaptionvpos{figure}{right}
\usepackage{amsmath,amsfonts,amsbsy,amssymb,array,accents,dsfont,empheq}
\usepackage{enumerate,array,latexsym,graphicx,mathrsfs,verbatim,psfrag}
\usepackage{bm} 
\usepackage[normalem]{ulem}
\usepackage{multirow}
\usepackage{cancel}
\usepackage{chngpage} 

\usepackage{booktabs} 
\usepackage[usenames]{color}
\usepackage[utf8]{inputenc}
\usepackage[english]{babel}
\usepackage{fancybox}
\usepackage[dvipsnames]{xcolor}

\usepackage{tikz}
\usepackage{pgfplots}
\usetikzlibrary{decorations.text,intersections, pgfplots.fillbetween}
\usetikzlibrary{math}
 \usetikzlibrary{snakes,3d,shapes.geometric,shadows.blur}
\usepackage{tikz-3dplot}

\usetikzlibrary{positioning}
\usetikzlibrary{automata}
\usetikzlibrary{arrows}
\usetikzlibrary{calc}
\usetikzlibrary{decorations.markings}
\usetikzlibrary{decorations.pathreplacing}
\usetikzlibrary{intersections}
\usetikzlibrary{positioning}
\usetikzlibrary{topaths}
\usetikzlibrary{shapes.geometric}
\usetikzlibrary{shapes.misc}
\tikzset{cf-group/.style = {
    shape = rounded rectangle, minimum size=1.0cm,
    rotate=90,
    rounded rectangle right arc = none,
    draw}}
\tikzset{cross/.style={path picture={ 
  \draw[black]
(path picture bounding box.south east) -- (path picture bounding box.north west) (path picture bounding box.south west) -- (path picture bounding box.north east);
}}}

\tikzset{unode/.style=
{black, circle,draw,thick,fill=black!100 ,minimum size=1mm}}

\tikzset{sunode/.style=
{black, circle,draw,thick,fill=yellow!100 ,minimum size=1mm}}

\tikzset{fnode/.style=
{black, rectangle,draw,thick,minimum size=1mm}}

\tikzset{afnode/.style=
{blue,rectangle,draw,thick,minimum size=1mm}}

\tikzset{rnode/.style={black, circle,draw,thick,fill=gray!100 ,minimum size=4mm}}

\tikzset{hnode/.style={black, hexagon,draw,thick,fill=gray!100 ,minimum size=8mm}}

\usepackage{sidecap}

\sidecaptionvpos{figure}{c}

\usepackage{mciteplus}

\usepackage[colorlinks=true,      linkcolor=blue,      urlcolor=blue,      
            filecolor=blue,      citecolor=blue,       pdfstartview=FitH,     
						pdfpagemode=UseNone,      bookmarksopen=true]{hyperref}  
\usepackage[all]{hypcap}     

\newcommand{\be}{\begin{equation}}
\newcommand{\ee}{\end{equation}}
\newcommand{\ba}{\begin{array}}
\newcommand{\ea}{\end{array}} 
\newcommand{\bi}{\begin{itemize}}
\newcommand{\ei}{\end{itemize}}
\def\vec#1{\bm{#1}}
\def\bea#1\eea{\allowdisplaybreaMs \begin{align}#1\end{align}}
 \newcommand{\ben}{\begin{enumerate}}
\newcommand{\een}{\end{enumerate}}
\newcommand{\bean}{\begin{eqnarray*}}
\newcommand{\eean}{\end{eqnarray*}}
\newcommand{\eref}[1]{(\ref{#1})}

\newcommand{\BC}{\mathbb{C}}
\newcommand{\BR}{\mathbb{R}}

\newcommand{\BZ}{\mathbb{Z}}

\newcommand{\CT}{{\cal T}}

\newcommand{\CO}{{\cal O}}

\newcommand{\CN}{{\cal N}}

\newcommand{\CK}{{\cal K}}

\newcommand{\fru}{\mathfrak{u}}

\newcommand{\frg}{\mathfrak{g}}

\newcommand{\hk}{hyperk\"ahler }
\newcommand{\wt}{\widetilde}

\newcommand{\Secref}[1]{Section~\ref{#1}}

\newcommand{\Appref}[1]{Appendix~\ref{#1}}

\newcommand{\Figref}[1]{Figure~\ref{#1}}

\renewcommand{\eqref}[1]{(\ref{#1})}

\usepackage{titlesec}
\titlespacing*{\section}
{0pt}{0.5cm}{0.2cm}
\titlespacing*{\subsection}
{0pt}{0.25cm}{0.1cm}

\begin{document}

\title{Argyres-Douglas Theories, IR N-ality and Complete Graphs}


\author{Anindya Dey}
\email{anindya.hepth@gmail.com}


\begin{abstract}

We show that for a large subclass of Argyres-Douglas-type theories, the Higgs branch admits  
multiple \hk quotient realizations as Higgs branches of three dimensional $\CN=4$ quiver gauge theories, 
which are related by a sequence of Seiberg-like IR dualities. We refer to this phenomenon as the 
\textit{Hyperk\"ahler Quotient N-ality} of the four dimensional Higgs branch. The associated set of 
3d theories contains a special subset of \textit{maximal unitary quivers}: quiver gauge theories for which the resolution/deformation 
parameters of the Higgs branch are manifest in the Lagrangian as Fayet-Iliopoulas parameters. Starting from 
the Type IIB description for a given SCFT, we present an explicit construction to determine the aforementioned set of 3d quivers, 
including the subset of maximal unitary quivers. As a byproduct, we find a simple method for constructing the three 
dimensional mirror associated with the SCFT. We demonstrate the construction for the $(A_k, A_k)$ theories of Cecotti, 
Neitzke and Vafa, focusing on the cases $k=3$ and $k=4$. The associated maximal unitary quiver is unique up to field 
redefinitions and turns out to be an Abelian quiver gauge theory. The three dimensional mirror obtained in this fashion 
reproduces the well-known complete graph. In the appendices to the main paper, we study the quotient N-ality in the 
closely related family of $D^b_p (SU(N))$ SCFTs, for which both the maximal unitary quiver as well as the 3d mirror turn out to be  
non-Abelian gauge theories generically.

%

\end{abstract}

\maketitle

\section{Introduction}

Four dimensional $\CN=2$ Superconformal Field Theories (SCFTs) for which the Coulomb branch (CB) operators 
have fractional scaling dimensions are collectively referred to as Argyres-Douglas (AD) \cite{Argyres:1995xn, Argyres:1995jj} theories. 
While the earliest examples of such SCFTs were obtained as low energy effective field theories at special points 
on the moduli space of asymptotically free quiver gauge theories,  the program of Geometric 
Engineering -- pioneered by the papers \cite{Klemm:1996bj, Katz:1996fh} -- gives an efficient algorithm to construct a large class of 
these SCFTs. A particularly interesting subclass of such theories can be 
obtained by compactifying Type IIB String Theory on an isolated hypersurface singularity $X$ in four complex dimensions \cite{Shapere:1999xr} :
\be  \label{IHS-def}
X = \{\vec x = (x_1, x_2, x_3, x_4) \in \BC^4 | F(\vec x) =0\},
\ee
where $F(\vec x)$ is a quasi-homogenous polynomial with an isolated singularity at the origin, and has the following 
$\BC^\star$-action with weights $q_i > 0\, \forall i$:
\be
F(\lambda^{q_i}\, x_i) = \lambda\, F(\vec x) \quad {\rm with} \quad \sum_i q_i > 1.
\ee
The Seiberg-Witten (SW) curve of the 4d SCFT arises from the deformed singularity:
\be
\widehat{F}(\vec x) = F(\vec x) + \sum_{i,j,k,l}\, u_{ijkl}\, x^i_1\, x^j_2\, x^k_3 \, x^l_4 =0,
\ee
where the coefficients $\{ u_{ijkl} \}$ include the Coulomb branch operators, marginal and relevant couplings, and mass parameters associated 
with the flavor symmetry of the SCFT.  The scaling dimensions of the Coulomb branch operators and the various couplings 
can also be read off from the above equation.

In the Geometric Engineering picture, the CB physics of the 4d SCFT is encoded in the SW curve which in turn is 
completely determined by the classical geometry of the deformed singularity. On the other hand, the 4d 
Higgs branch (HB) turns out to be a more subtle object to study, given that the metric receives non-trivial 
quantum corrections from the D-brane instantons. There exists a large literature on the various 
approaches for studying the 4d HB of an AD theory -- we refer the reader to 
\cite{Argyres:2012fu, Gaiotto:2009jjh, Cecotti:2010fi, Nanopoulos:2010bv, Closset:2020afy, Giacomelli:2020ryy} for a partial 
list. One such approach, possibly the most direct, is to consider an appropriate circle reduction of the 4d SCFT 
and flow to a 3d IR SCFT. Since the 4d HB is not affected by this reduction, one may identify it with the 3d HB 
of this IR SCFT. In addition, there may exist a 3d $\CN=4$ quiver gauge theory which flows to the aforementioned 
SCFT in the IR, with or without an exchange of the CB and the HB. In the former case, the 4d HB can be realized as 
the 3d CB of the quiver gauge theory -- a \hk manifold but not necessarily a \hk quotient. For the latter case, we 
have a realization of the 4d HB as a \hk quotient in the standard fashion from the quiver representation.

In this paper, we show that for a large class of AD-type SCFTs, the 4d HB generically admits $N \geq 2$ realizations as 
\hk quotients, where each quotient realization is associated with the HB of a 3d $\CN=4$ quiver gauge theory. 
We refer to this phenomenon as the \textit{\hk quotient N-ality} for a given 4d HB. We also show that this quotient N-ality directly follows from three dimensional 
IR N-ality \cite{Dey:2023xhq, Dey:2022abc} : a set of 3d $\CN=4$ quiver gauge theories flowing to the same IR SCFT without an exchange of the CBs 
and the HBs. As discussed in \cite{Dey:2023xhq, Dey:2022abc}, the theories in the N-al set can be related by a sequence of quiver mutations. 
The principal objective of this paper is to present an explicit construction of this N-al set of quivers for a given 4d SCFT, starting from its Seiberg-Witten geometry. 

The number of Fayet-Iliopoulas (FI) parameters that can be turned on for a 3d quiver is given by the 
UV-manifest rank of the CB global symmetry.  A generic quiver in the N-al set, however, has emergent CB symmetry : 
the rank of the IR symmetry is greater than the rank of the UV-manifest symmetry, which implies that the IR SCFT has 
mass deformations that are not visible in the Lagrangian. Since the FI parameters resolve/deform the HB geometry, a generic quiver 
therefore only makes some of the resolution/deformation parameters manifest. An important result of our paper is to show that there always exists a 
subset of quivers in the N-al set for which all the resolution/deformation parameters of the HB become manifest as FI parameters.
We will refer to these quivers as \textit{maximal unitary quivers}, and construct them explicitly for certain families of 4d SCFTs. 
Finally, we will show that our construction leads to a simple method for computing the 3d mirror of the SCFT. 

We will first demonstrate our construction using the $(A_k, A_k)$ SCFTs \cite{Cecotti:2010fi} which arise from the following hypersurface singularity:
\be
X = \{\vec x \in \BC^4 | x^2_1 +  x^2_2 + x^{k+1}_3 + x^{k+1}_4 =0\}. 
\ee
We will also study the $D^b_p(SU(N))$ theories \cite{Cecotti:2012jx, Cecotti:2013lda} which are engineered by compactifying Type IIB String Theory on a hypersurface singularity in $\BC^3 \times \BC^\star$ :
\be 
X=  \begin{cases} \{(\vec x,z)  | x^2_1 + x^2_2 + x^N_3 + z^p=0 \} \quad b=N,\\
\{(\vec x,,z) | x^2_1 + x^2_2 + x^N_3 + x_3 z^p=0 \} \quad b=N-1. \end{cases} \label{DpSUN-undef}
\ee
where $(\vec x,z) \in \BC^3 \times \BC^\star$ denote the local coordinates. Our construction, however, applies to a much larger class of SCFTs, as we will specify momentarily. A larger set of examples as well as additional details of the construction will be discussed in a longer version of this 
paper \cite{Dey:2024gen}. \\

The rest of the paper is organized as follows. In \Secref{sec: circlered}, we will give a brief review of the circle reduction of a 4d 
SCFT, and discuss certain S-duality frames of the $(A_{k}, A_{k})$ SCFTs which will be relevant for our construction. 
In \Secref{sec: maxquiv}, we will discuss the basic recipe for constructing the N-al set of 3d quivers and in particular, the maximal 
unitary quivers. We will determine the N-al set for the $(A_{k}, A_{k})$ SCFTs in \Secref{sec: IREx}, explicitly working out the $k=3$ and $k=4$ 
cases.  The general procedure of constructing the 3d mirror from a maximal unitary quiver will be discussed in \Secref{sec: 3dmirr}, 
which will be used to determine the 3d mirrors of the $(A_{k}, A_{k})$ SCFTs in \Secref{sec: mainexmirr}. The details of quotient N-ality for the 
$D^b_p(SU(N))$ theories can be found in \Appref{app: 3dLagDp} -- \Appref{app: DpEx}. We summarize our results and comment on future directions of work in \Secref{sec: concl}.

\begin{figure*}[htbp]
\begin{tabular}{c}
\scalebox{0.6}{\begin{tikzpicture}
\node[rnode] (1) at (2,0){$D_2(SU(3))$};
\node[sunode] (2) at (4,0){};
\node[rnode] (3) at (6,0){$D_2(SU(5))$};
\node[sunode] (4) at (8,0){};
\node[] (5) at (9,0){};
\node[] (6) at (10,0){};
\node[rnode] (7) at (12,0){$D_2(SU(k))$};
\node[sunode] (8) at (14,0){};
\node[fnode] (9) at (14, 2){};
\node[rnode] (10) at (16,0){$D_2(SU(k))$};
\node[] (11) at (18,0){};
\node[] (12) at (19,0){};
\node[sunode] (13) at (20,0){};
\node[rnode] (14) at (22,0){$D_2(SU(5))$};
\node[sunode] (15) at (24,0){};
\node[rnode] (16) at (26,0){$D_2(SU(3))$};
\draw[-] (1) -- (2);
\draw[-] (2)-- (3);
\draw[-] (3) -- (4);
\draw[-] (4) -- (5);
\draw[-, dotted] (5) -- (6);
\draw[-] (6) -- (7);
\draw[-] (7) -- (8);
\draw[-] (8) -- (9);
\draw[-] (8) -- (10);
\draw[-] (10) -- (11);
\draw[-, dotted] (11) -- (12);
\draw[-] (12) -- (13);
\draw[-] (13) -- (14);
\draw[-] (14) -- (15);
\draw[-] (15) -- (16);
\node[text width= 0.1 cm](41) [below=0.1cm of 2]{2};
\node[text width= 0.1 cm](42) [below=0.1cm of 4]{3};
\node[text width= 0.5 cm](43) [below=0.1cm of 8]{$\frac{k+1}{2}$};
\node[text width= 0.1 cm](44) [right=0.1cm of 9]{1};
\node[text width= 0.1 cm](45) [below=0.1cm of 13]{3};
\node[text width= 0.1 cm](46) [below=0.1cm of 15]{2};
\end{tikzpicture}}\\
\qquad \qquad  \\
\qquad \qquad  \\
\scalebox{0.6}{\begin{tikzpicture}
\node[rnode] (1) at (2,0){$D_2(SU(3))$};
\node[sunode] (2) at (4,0){};
\node[rnode] (3) at (6,0){$D_2(SU(5))$};
\node[sunode] (4) at (8,0){};
\node[] (5) at (9,0){};
\node[] (6) at (10,0){};
\node[rnode] (7) at (12,0){$D_2(SU(k-1))$};
\node[sunode] (8) at (14,0){};
\node[rnode] (9) at (16,0){$D_2(SU(k+1))$};
\node[sunode] (10) at (18,0){};
\node[rnode] (11) at (20,0){$D_2(SU(k-1))$};
\node[] (12) at (22,0){};
\node[] (13) at (23,0){};
\node[sunode] (14) at (24,0){};
\node[rnode] (15) at (26,0){$D_2(SU(5))$};
\node[sunode] (16) at (28,0){};
\node[rnode] (17) at (30,0){$D_2(SU(3))$};
\draw[-] (1) -- (2);
\draw[-] (2)-- (3);
\draw[-] (3) -- (4);
\draw[-] (4) -- (5);
\draw[-, dotted] (5) -- (6);
\draw[-] (6) -- (7);
\draw[-] (7) -- (8);
\draw[-] (8) -- (9);
\draw[-] (9) -- (10);
\draw[-] (10) -- (11);
\draw[-] (11) -- (12);
\draw[-, dotted] (12) -- (13);
\draw[-] (13) -- (14);
\draw[-] (14) -- (15);
\draw[-] (15) -- (16);
\draw[-] (16) -- (17);
\node[text width= 0.1 cm](41) [below=0.1cm of 2]{2};
\node[text width= 0.1 cm](42) [below=0.1cm of 4]{3};
\node[text width= 0.1 cm](43) [below=0.1cm of 8]{$\frac{k}{2}$};
\node[text width= 0.1 cm](44) [below=0.1cm of 10]{$\frac{k}{2}$};
\node[text width= 0.1 cm](45) [below=0.1cm of 14]{3};
\node[text width= 0.1 cm](46) [below=0.1cm of 16]{2};
\end{tikzpicture}}
\caption{\footnotesize{An S-duality frame for the $(A_k, A_k)$ SCFT for $k$ odd (top) and $k$ even (bottom) with $k > 2$.}}
\label{fig: Sdualityframes}
\end{tabular}
\end{figure*}
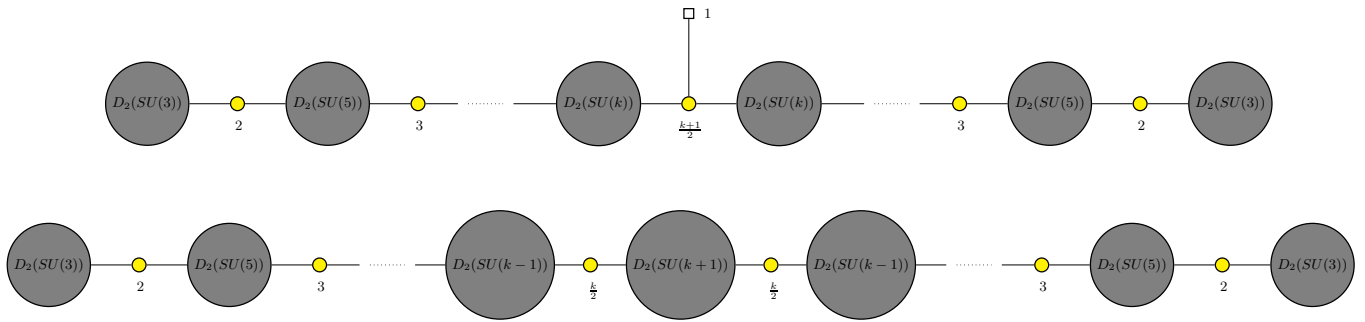

\section{Circle Reduction and S-duality frames}\label{sec: circlered}

Let us begin by discussing the circle reduction of an isolated 4d SCFT i.e. an SCFT for which the associated conformal 
manifold is zero-dimensional, following the treatment in \cite{Closset:2020afy}. The first step is to realize the SCFT as the low energy effective theory 
at the origin of the CB of a 4d Lagrangian theory $\CK$ with appropriately scaled mass deformations. In the next step, 
one compactifies the theory on $\BR^{1,2} \times S^1_R$ and considers the limit $R \to 0$ for the energy scales $E \ll \Lambda$,
holding the 4d RG scale $\Lambda = \exp{(-\frac{8\pi^2}{g^2_4})}$ fixed. 
This limit corresponds to the strong-coupling limit of the 3d gauge coupling : $g^2_3 = \frac{g^2_4}{2\pi R} \to \infty$.  
By studying the low energy effective theory in the neighborhood of the origin of the CB, one can often realize the resultant 3d SCFT as the 
IR SCFT of a 3d Lagrangian field theory $\CT$ with no mass deformations at its CB origin. Since the 4d HB remains 
unaffected by the circle reduction, it can be realized as the HB of the theory $\CT$ as a \hk quotient.

For a generic non-isolated 4d SCFT, one can implement a similar program if the SCFT admits an S-duality frame where the 
SCFT can be described as a network of isolated SCFTs connected by conformally gauged vector multiplets in the following 
fashion :

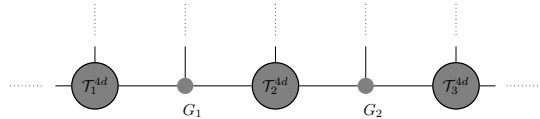
\begin{center}
\scalebox{0.6}{\begin{tikzpicture}
\node[] (0) at (0,0){};
\node[] (1) at (1,0){};
\node[rnode] (2) at (2,0){$\CT^{4d}_1$};
\node[unode, gray] (3) at (4,0){};
\node[rnode] (4) at (6,0){$\CT^{4d}_2$};
\node[unode, gray] (5) at (8,0){};
\node[rnode] (6) at (10,0){$\CT^{4d}_3$};
\node[] (7) at (11,0){};
\node[] (8) at (12,0){};
\node[] (9) at (2,1){};
\node[] (10) at (2,2){};
\node[] (11) at (4,1){};
\node[] (12) at (4,2){};
\node[] (13) at (6,1){};
\node[] (14) at (6,2){};
\node[] (15) at (8,1){};
\node[] (16) at (8,2){};
\node[] (17) at (10,1){};
\node[] (18) at (10,2){};
\draw[-, dotted] (0) -- (1);
\draw[-] (1) -- (2);
\draw[-] (2)-- (3);
\draw[-] (3) -- (4);
\draw[-] (4) -- (5);
\draw[-] (5) -- (6);
\draw[-] (6) -- (7);
\draw[-, dotted] (7) -- (8);
\draw[-] (2)-- (9);
\draw[-, dotted] (9)-- (10);
\draw[-] (3) -- (11);
\draw[-, dotted] (11)-- (12);
\draw[-] (4) -- (13);
\draw[-, dotted] (13)-- (14);
\draw[-] (5) -- (15);
\draw[-, dotted] (15)-- (16);
\draw[-] (6) -- (17);
\draw[-, dotted] (17)-- (18);
\node[text width= 0.1 cm](11) [below=0.1cm of 3]{$G_1$};
\node[text width= 0.1 cm](12) [below=0.1cm of 5]{$G_2$};
\end{tikzpicture}}
\end{center}

A larger gray node labelled $\CT^{4d}_i $ denotes an isolated 4d SCFT while a smaller gray node labelled $G_i$ 
denotes a conformally gauged 4d vector multiplet with gauge group $G_i$. An isolated SCFT connected to a $G_i$ vector multiplet 
must have a flavor symmetry subgroup $G_i$. The number of vector multiplets matches the number of marginal couplings 
(or equivalently, the dimension of the conformal manifold) of the SCFT.  The above description of the SCFT is at times called 
a \textit{partially weakly-coupled description}. 

Given such an S-duality frame, one can again compactify the theory on a circle and reduce to 3d. The 
conformally gauged vector multiplet descends to a 3d vector multiplet of the same gauge group, while the 
3d Lagrangian theories corresponding to the isolated SCFTs can be derived as before. Putting the components 
together, one therefore has a 3d $\CN = 4$ quiver gauge theory of the following form: 

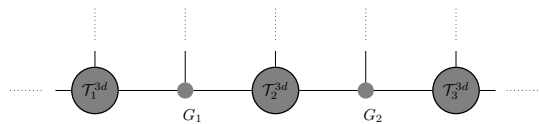
\begin{center}
\scalebox{0.6}{\begin{tikzpicture}
\node[] (0) at (0,0){};
\node[] (1) at (1,0){};
\node[rnode] (2) at (2,0){$\CT^{3d}_1$};
\node[unode, gray] (3) at (4,0){};
\node[rnode] (4) at (6,0){$\CT^{3d}_2$};
\node[unode, gray] (5) at (8,0){};
\node[rnode] (6) at (10,0){$\CT^{3d}_3$};
\node[] (7) at (11,0){};
\node[] (8) at (12,0){};
\node[] (9) at (2,1){};
\node[] (10) at (2,2){};
\node[] (11) at (4,1){};
\node[] (12) at (4,2){};
\node[] (13) at (6,1){};
\node[] (14) at (6,2){};
\node[] (15) at (8,1){};
\node[] (16) at (8,2){};
\node[] (17) at (10,1){};
\node[] (18) at (10,2){};
\draw[-, dotted] (0) -- (1);
\draw[-] (1) -- (2);
\draw[-] (2)-- (3);
\draw[-] (3) -- (4);
\draw[-] (4) -- (5);
\draw[-] (5) -- (6);
\draw[-] (6) -- (7);
\draw[-, dotted] (7) -- (8);
\draw[-] (2)-- (9);
\draw[-, dotted] (9)-- (10);
\draw[-] (3) -- (11);
\draw[-, dotted] (11)-- (12);
\draw[-] (4) -- (13);
\draw[-, dotted] (13)-- (14);
\draw[-] (5) -- (15);
\draw[-, dotted] (15)-- (16);
\draw[-] (6) -- (17);
\draw[-, dotted] (17)-- (18);
\node[text width= 0.1 cm](11) [below=0.1cm of 3]{$G_1$};
\node[text width= 0.1 cm](12) [below=0.1cm of 5]{$G_2$};
\end{tikzpicture}}
\end{center}

A larger gray node labelled $\CT^{3d}_i $ denotes a 3d quiver while a smaller gray node denotes a 3d $G_i$ 
vector multiplet. The 4d HB is then given by the HB of this 3d $\CN = 4$ quiver gauge 
theory and is therefore realized as a \hk quotient. \\

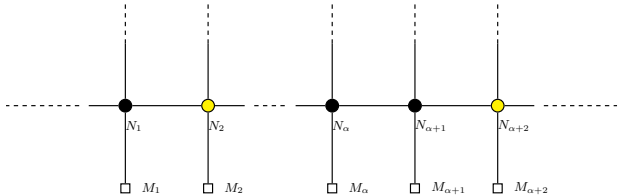
\begin{figure}[htbp]
\begin{center}
\scalebox{0.55}{\begin{tikzpicture}
\node[] (100) at (-3,0) {};
\node[] (1) at (-1,0) {};
\node[unode] (2) at (0,0) {};
\node[text width=.2cm](31) at (0.1,-0.5){$N_1$};
\node[sunode] (3) at (2,0) {};
\node[text width=.2cm](32) at (2.1,-0.5){$N_2$};
\node[] (4) at (3,0) {};
\node[] (5) at (4,0) {};
\node[unode] (6) at (5,0) {};
\node[text width=.2cm](33) at (5.1,-0.5){$N_{\alpha}$};
\node[unode] (7) at (7,0) {};
\node[text width=.2cm](34) at (7.1,-0.5){$N_{\alpha+1}$};
\node[sunode] (8) at (9,0) {};
\node[text width=.2cm] (40) at (9.1,-0.5) {$N_{\alpha+2}$};
\node[fnode] (9) at (9,-2) {};
\node[text width=.2cm] (40) at (9.5,-2) {$M_{\alpha+2}$};
\node[] (10) at (10,0) {};
\node[] (11) at (12,0) {};
\node[fnode] (20) at (0,-2) {};
\node[fnode] (21) at (2,-2) {};
\node[fnode] (22) at (5,-2) {};
\node[fnode] (27) at (7,-2) {};
\node[text width=.2cm](23) at (0.5,-2){$M_1$};
\node[text width=.2cm](24) at (2.5,-2){$M_2$};
\node[text width=.2cm](25) at (5.5,-2){$M_\alpha$};
\node[text width=.2cm](26) at (7.5,-2){$M_{\alpha +1}$};
\draw[thick] (1) -- (2);
\draw[thick] (2) -- (3);
\draw[thick] (3) -- (4);
\draw[thick,dashed] (4) -- (5);
\draw[thick] (5) -- (6);
\draw[thick] (6) -- (7);
\draw[thick] (7) -- (8);
\draw[thick] (7) -- (27);
\draw[thick] (8) -- (9);
\draw[thick] (8) -- (10);
\draw[thick,dashed] (10) -- (11);
\draw[thick,dashed] (1) -- (100);
\draw[thick] (2) -- (0,1.5);
\draw[thick, dashed] (0,1.5) -- (0,2.5);
\draw[thick] (2,1.5) -- (3);
\draw[thick, dashed] (2,1.5) -- (2,2.5);
\draw[thick] (5,1.5) -- (6);
\draw[thick, dashed] (5,1.5) -- (5,2.5);
\draw[thick] (7,1.5) -- (7);
\draw[thick, dashed] (7,1.5) -- (7, 2.5);
\draw[thick] (9,1.5) -- (8);
\draw[thick,dashed] (9,1.5) -- (9,2.5);
\draw[thick] (2) -- (20);
\draw[thick] (3) -- (21);
\draw[thick] (6) -- (22);
\end{tikzpicture}}
\end{center}
\caption{\footnotesize{A generic quiver with unitary/special unitary gauge nodes and (bi)fundamental matter. The yellow and the black nodes denote the special unitary vector multiplets and the unitary vector multiplets respectively, with the black lines connected to gauge nodes denote 
fundamental/bifundamental hypermultiplets.}}
\label{fig: USUgen}
\end{figure}

The construction presented in this paper will apply to the subclass of 4d SCFTs which admits a partially weakly-coupled description 
as described above, with the constraint that the gauge group for any conformally gauged vector multiplet has to be special unitary. 
In addition, we will restrict ourselves to the subclass of isolated SCFTs for which the associated 
3d quivers $\CT^{3d}_i$ are given by unitary quiver gauge theories with hypermultiplets in the fundamental and 
the bifundamental representations. The resultant 3d quiver has the generic form shown in \Figref{fig: USUgen}, 
which precisely belongs to the class that was studied in the context of IR N-ality in the papers \cite{Dey:2023xhq, Dey:2022abc}. 
To distinguish it from other 3d quivers that we will discuss later in the paper, 
we will refer to this quiver as the \textit{U-SU quiver} associated with the 4d SCFT. 

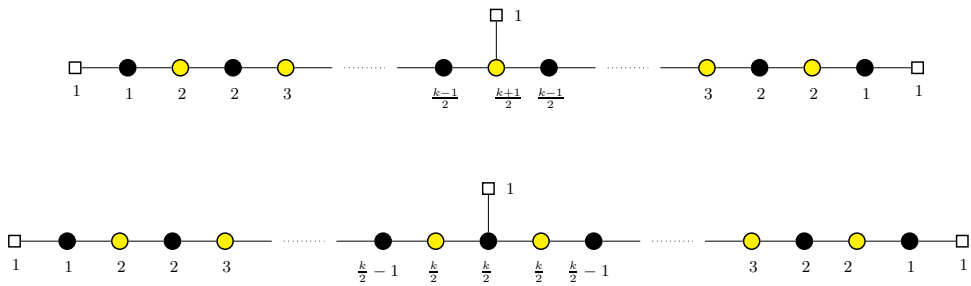
\begin{figure*}[htbp]
\begin{tabular}{c}
\scalebox{0.7}{\begin{tikzpicture}
\node[fnode] (0) at (0,0){};
\node[unode] (1) at (1,0){};
\node[sunode] (2) at (2,0){};
\node[unode] (3) at (3,0){};
\node[sunode] (4) at (4,0){};
\node[] (5) at (5,0){};
\node[] (6) at (6,0){};
\node[unode] (7) at (7,0){};
\node[sunode] (8) at (8,0){};
\node[fnode] (9) at (8, 1){};
\node[unode] (10) at (9,0){};
\node[] (11) at (10,0){};
\node[] (12) at (11,0){};
\node[sunode] (13) at (12,0){};
\node[unode] (14) at (13,0){};
\node[sunode] (15) at (14,0){};
\node[unode] (16) at (15,0){};
\node[fnode] (17) at (16,0){};
\draw[-] (0) -- (1);
\draw[-] (1) -- (2);
\draw[-] (2)-- (3);
\draw[-] (3) -- (4);
\draw[-] (4) -- (5);
\draw[-, dotted] (5) -- (6);
\draw[-] (6) -- (7);
\draw[-] (7) -- (8);
\draw[-] (8) -- (9);
\draw[-] (8) -- (10);
\draw[-] (10) -- (11);
\draw[-, dotted] (11) -- (12);
\draw[-] (12) -- (13);
\draw[-] (13) -- (14);
\draw[-] (14) -- (15);
\draw[-] (15) -- (16);
\draw[-] (16) -- (17);
\node[text width= 0.1 cm](41) [below=0.1cm of 0]{1};
\node[text width= 0.1 cm](42) [below=0.1cm of 1]{1};
\node[text width= 0.1 cm](43) [below=0.1cm of 2]{2};
\node[text width= 0.1 cm](44) [below=0.1cm of 3]{2};
\node[text width= 0.1 cm](45) [below=0.1cm of 4]{3};
\node[text width= 0.5 cm](46) [below=0.1cm of 7]{$\frac{k-1}{2}$};
\node[text width= 0.1 cm](47) [below=0.1cm of 8]{$\frac{k+1}{2}$};
\node[text width= 0.1 cm](48) [right=0.1cm of 9]{1};
\node[text width= 0.5 cm](49) [below=0.1cm of 10]{$\frac{k-1}{2}$};
\node[text width= 0.1 cm](50) [below=0.1cm of 13]{3};
\node[text width= 0.1 cm](51) [below=0.1cm of 14]{2};
\node[text width= 0.1 cm](52) [below=0.1cm of 15]{2};
\node[text width= 0.1 cm](53) [below=0.1cm of 16]{1};
\node[text width= 0.1 cm](54) [below=0.1cm of 17]{1};
\end{tikzpicture}}\\
\qquad \qquad  \\
\qquad \qquad  \\
\scalebox{0.7}{\begin{tikzpicture}
\node[fnode] (0) at (0,0){};
\node[unode] (1) at (1,0){};
\node[sunode] (2) at (2,0){};
\node[unode] (3) at (3,0){};
\node[sunode] (4) at (4,0){};
\node[] (5) at (5,0){};
\node[] (6) at (6,0){};
\node[unode] (7) at (7,0){};
\node[sunode] (8) at (8,0){};
\node[unode] (9) at (9,0){};
\node[sunode] (10) at (10,0){};
\node[unode] (11) at (11,0){};
\node[] (12) at (12,0){};
\node[] (13) at (13,0){};
\node[sunode] (14) at (14,0){};
\node[unode] (15) at (15,0){};
\node[sunode] (16) at (16,0){};
\node[unode] (17) at (17,0){};
\node[fnode] (18) at (18,0){};
\node[fnode] (19) at (9,1){};
\draw[-] (0) -- (1);
\draw[-] (1) -- (2);
\draw[-] (2)-- (3);
\draw[-] (3) -- (4);
\draw[-] (4) -- (5);
\draw[-, dotted] (5) -- (6);
\draw[-] (6) -- (7);
\draw[-] (7) -- (8);
\draw[-] (8) -- (9);
\draw[-] (9) -- (10);
\draw[-] (10) -- (11);
\draw[-] (11) -- (12);
\draw[-, dotted] (12) -- (13);
\draw[-] (13) -- (14);
\draw[-] (14) -- (15);
\draw[-] (15) -- (16);
\draw[-] (16) -- (17);
\draw[-] (17) -- (18);
\draw[-] (9) -- (19);
\node[text width= 0.1 cm](41) [below=0.1cm of 0]{1};
\node[text width= 0.1 cm](42) [below=0.1cm of 1]{1};
\node[text width= 0.1 cm](43) [below=0.1cm of 2]{2};
\node[text width= 0.1 cm](44) [below=0.1cm of 3]{2};
\node[text width= 0.1 cm](45) [below=0.1cm of 4]{3};
\node[text width= 1 cm](46) [below=0.1cm of 7]{$\frac{k}{2}-1$};
\node[text width= 0.3 cm](47) [below=0.1cm of 8]{$\frac{k}{2}$};
\node[text width= 0.3 cm](48) [below=0.1cm of 9]{$\frac{k}{2}$};
\node[text width= 0.3 cm](49) [below=0.1cm of 10]{$\frac{k}{2}$};
\node[text width= 1 cm](50) [below=0.1cm of 11]{$\frac{k}{2}-1$};
\node[text width= 0.1 cm](51) [below=0.1cm of 14]{3};
\node[text width= 0.1 cm](52) [below=0.1cm of 15]{2};
\node[text width= 0.5 cm](53) [below=0.1cm of 16]{2};
\node[text width= 0.1 cm](54) [below=0.1cm of 17]{1};
\node[text width= 0.1 cm](55) [below=0.1cm of 18]{1};
\node[text width= 0.1 cm](56) [right=0.1cm of 19]{1};
\end{tikzpicture}}
\caption{\footnotesize{The U-SU quivers for the $(A_k, A_k)$ SCFT for $k$ odd (top) and $k$ even (bottom) with $k > 2$.}}
\label{fig: USUquiv}
\end{tabular}
\end{figure*}

As an explicit example, we will write down the relevant S-duality frames of the $(A_k, A_k)$ SCFTs, which have been 
studied in the literature from the perspective of Hitchin systems \cite{Xie:2017vaf} as well as that of the Type IIB realization 
\cite{Buican:2014hfa, Giacomelli:2020ryy}. 
To do this, we will need the $D^b_p(SU(N))$ theories -- in particular, the subfamily of isolated SCFTs $D_2(SU(2n-1))$ 
for an integer $n > 1$. The relevant S-duality frame for an $(A_k, A_k)$ SCFT is shown in \Figref{fig: Sdualityframes}, for $k$ odd and $k$ even 
respectively. Note that there are precisely $k-2$ special unitary gauge nodes corresponding to the number of marginal 
couplings of the SCFT. The circle reduction of the $D_2(SU(2n-1))$ 
is discussed in \Appref{app: 3dLagDp} : the resultant 3d SCFT can be realized as the SCFT at the origin 
of the CB of a $U(n-1)$ SQCD with $N_f = 2n-1$. Combining this result with the S-duality frame in \Figref{fig: Sdualityframes}, 
one can write down the U-SU quivers for the $(A_k, A_k)$ SCFTs  -- given in \Figref{fig: USUquiv} for $k$ odd and $k$ even 
respectively. These quivers are linear quivers with unitary and special unitary gauge nodes and are therefore a special subclass of the 
quivers in \Figref{fig: USUgen}. In the next section, we will discuss how the \hk quotient N-ality arises from these U-SU quivers. 

\section{The Basic Construction : Duality Sequence and the Maximal Unitary Quiver} \label{sec: maxquiv}

Consider a U-SU quiver of the generic form given in \Figref{fig: USUgen}. We define a balance parameter $e_\alpha$
for a given $U(N_\alpha)$ or $SU(N_\alpha)$ gauge node as $e_\alpha = N^\alpha_{f/bf} - 2N_\alpha$, where 
$N^\alpha_{f/bf}$ denotes the total number of fundamental and bifundamental hypers associated with the gauge node. 
We refer to an $U(N_\alpha)$ node as balanced if $e_\alpha =0$, while an $SU(N_\alpha)$ node is balanced if $e_\alpha = -1$. 
Given this convention, we note that \textit{all} of the $k-2$ special unitary nodes in the linear U-SU quiver associated with the 
$(A_k, A_k)$ SCFT are balanced, while the unitary nodes are all overbalanced. One can additionally check 
that these quiver gauge theories are good theories in the Gaiotto-Witten sense. 
 
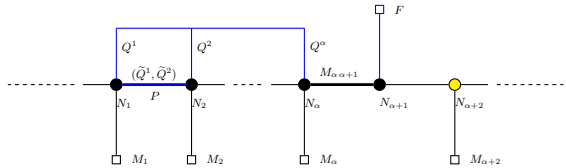
\begin{figure}[htbp]
\begin{center}
\scalebox{0.5}{\begin{tikzpicture}
\node[] (100) at (-3,0) {};
\node[] (1) at (-1,0) {};
\node[unode] (2) at (0,0) {};
\node[text width=.2cm](31) at (0.1,-0.5){$N_1$};
\node[unode] (3) at (2,0) {};
\node[text width=.2cm](32) at (2.1,-0.5){$N_2$};
\node[] (4) at (3,0) {};
\node[] (5) at (4,0) {};
\node[unode] (6) at (5,0) {};
\node[text width=.2cm](33) at (5.1,-0.5){$N_{\alpha}$};
\node[unode] (7) at (7,0) {};
\node[text width=.2cm](34) at (7.1,-0.5){$N_{\alpha+1}$};
\node[text width=.2cm](35) at (5.5,0.3){$M_{\alpha\,\alpha+1}$};
\node[sunode] (8) at (9,0) {};
\node[text width=.2cm] (40) at (9.1,-0.5) {$N_{\alpha+2}$};
\node[fnode] (9) at (9,-2) {};
\node[text width=.2cm] (40) at (9.5,-2) {$M_{\alpha+2}$};
\node[] (10) at (10,0) {};
\node[] (11) at (12,0) {};
\node[afnode] (12) at (7,2) {};
\node[text width=.2cm](36) at (7.5,2){$F$};
\node[fnode] (20) at (0,-2) {};
\node[fnode] (21) at (2,-2) {};
\node[fnode] (22) at (5,-2) {};
\node[text width=.2cm](23) at (0.5,-2){$M_1$};
\node[text width=.2cm](24) at (2.5,-2){$M_2$};
\node[text width=.2cm](25) at (5.5,-2){$M_\alpha$};
\draw[thick] (1) -- (2);
\draw[line width=0.75mm, blue] (2) -- (3);
\node[text width=.2cm](50) at (0.5,0.3){$(\wt{Q}^1,\wt{Q}^2)$};
\node[text width=.2cm](51) at (1,-0.3){$P$};
\draw[thick] (3) -- (4);
\draw[thick,dashed] (4) -- (5);
\draw[thick] (5) -- (6);
\draw[line width=0.75mm] (6) -- (7);
\draw[thick] (7) -- (8);
\draw[thick,blue] (7) -- (12);
\draw[thick] (8) -- (9);
\draw[thick] (8) -- (10);
\draw[thick,dashed] (10) -- (11);
\draw[thick,dashed] (1) -- (100);
\draw[thick, blue] (2) -- (0,1.5);
\draw[thick, blue] (0,1.5) -- (5,1.5);
\draw[thick, blue] (2,1.5) -- (3);
\draw[thick, blue] (5,1.5) -- (6);
\draw[thick] (2) -- (20);
\draw[thick] (3) -- (21);
\draw[thick] (6) -- (22);
\node[text width=.2cm](15) at (0.25,1){$Q^1$};
\node[text width=.2cm](16) at (2.25,1){$Q^2$};
\node[text width=.2cm](17) at (5.25,1){$Q^\alpha$};
\end{tikzpicture}}
\end{center}
\caption{\footnotesize{A generic quiver with unitary/special unitary gauge nodes, with (bi)fundamental and Abelian hypermultiplets.
A blue box with label $F$ denotes $F$ Abelian hypermultiplets in the determinant representation, while blue lines connecting 
multiple gauge nodes denote hypermultiplets charged under those gauge nodes with charges $\{Q^i \}$.}}
\label{fig: AbHMgen}
\end{figure}

It was shown in \cite{Dey:2023xhq, Dey:2022abc} that a U-SU quiver $\CT$ with at least one balanced SU gauge node has an emergent CB global 
symmetry: 
\be
\text{rk}(\frg^{\rm IR}_{\rm C}(\CT)) > \text{rk}(\frg^{\rm UV}_{\rm C}(\CT)), 
\ee
where $\frg^{\rm IR/UV}_{\rm C}(\CT)$ denotes the CB symmetry algebra for the theory $\CT$ in the IR/UV.
The UV symmetry algebra is simply given by the direct sum of the $\fru(1)$ topological symmetries i.e. 
$\frg^{\rm UV}_{\rm C}(\CT) = \oplus^L_{\alpha=1}\,\fru(1)_\alpha$, with $\alpha$ labelling the unitary gauge 
nodes in $\CT$. The existence of an emergent CB global symmetry 
implies that the 3d IR SCFT has mass deformations which are not visible in the quiver as FI parameters. 
This in turn implies that for the \hk quotient constructed using the U-SU quiver, one can turn on only a subset of 
resolution/deformation parameters of the HB geometry. It was also shown in \cite{Dey:2022abc} that a U-SU quiver with an 
emergent symmetry admits a sequence of IR dualities, generated by a set of four quiver mutations -- 
mutations $I$, $I'$, $II$ and $III$ -- which obey non-trivial closure relations. We review these mutations 
in \Appref{app: quivmut} and refer the reader to the papers \cite{Dey:2023xhq, Dey:2022abc} for further details. 

Each IR duality in this sequence maps the CB (HB) of one theory to the CB (HB) of the dual theory in the IR.
The duality sequence therefore generates a set of quiver gauge theories which flow to the same SCFT in the IR without 
an exchange of the CB and the HB -- this is dubbed as IR N-ality and the aforementioned set of theories is referred to 
as the N-al set. A generic theory in the N-al set, however, is not a U-SU quiver of the form given in \Figref{fig: USUgen} 
but it is a more general quiver gauge theory of the form given in \Figref{fig: AbHMgen}. In addition to unitary/special unitary gauge nodes and 
fundamental/bifundamental matter, the theory in \Figref{fig: AbHMgen} has hypermultiplets that transform in powers of the determinant 
and/or the anti-determinant representations of the unitary gauge nodes. We refer to these matter 
multiplets collectively as \textit{Abelian hypermultiplets}. Evidently, the class of quivers in \Figref{fig: USUgen} is a subclass 
of the quivers in \Figref{fig: AbHMgen}.

As reviewed in \Appref{app: quivmut}, the quiver mutation $I$ acts on a balanced $SU(N_\alpha)$ gauge node, 
while remaining mutations act on unitary gauge nodes connected to Abelian hypermultiplet(s). The mutations 
$I', II, III$ act on such a $U(N_\alpha)$ gauge node with balance parameters 
$e_\alpha=1,0,-1$ respectively. Given the quiver mutations, the recipe for writing down the 3d quivers associated with a 4d SCFT is 
straightforward. One first implements the circle reduction discussed in \Secref{sec: circlered} 
to determine the U-SU quiver gauge theory $\CT$ of the generic form given in \Figref{fig: USUgen}, which will generically have 
$l \geq 1$ balanced SU gauge nodes. Given the quiver $\CT$, one implements mutation $I$ at every balanced SU node giving rise 
to $l$ distinct quiver gauge theories. Each such gauge theory will have $l-1$ balanced SU nodes as well as unitary nodes which are
connected to an Abelian hypermultiplet and have different balance parameters $e_\alpha \geq -1$. In the next step, one implements all admissible quiver mutations on this gauge gauge theory to generate another set of quiver gauge theories. One continues this procedure until no new quiver gauge theory can be generated using the above mutations.\\

The quiver mutations $I$ and $III$ increase the number of $\fru(1)$ topological symmetries (equivalently the number of FI parameters)  
by 1. On the other hand, mutation $I'$ decreases this number by 1, and mutation ${II}$ keeps it invariant. Therefore, the 3d Lagrangians generated by the quiver mutations will generically differ among themselves in UV-manifest rank of the CB global symmetry. This implies that the associated \hk quotient realizations will differ in the number of resolution/deformation parameters that can be turned on. The U-SU quiver $\CT$ has the minimum number of FI parameters, since there does not exist a quiver mutation that can decrease the number of topological symmetries further 
on this quiver. At the other extreme, there exists a subset of quivers for which the number of FI parameters precisely matches 
the number of resolution/deformation parameters for the 4d HB. We will call a generic quiver in this subset a maximal unitary 
quiver and denote it as $\CT_{\rm maximal}$ for a given U-SU quiver $\CT$. In the duality sequence discussed above, a quiver can be 
identified as maximal unitary if none of its gauge nodes admit either a mutation $I$ or a mutation $III$. A distinct pair of maximal unitary 
quivers are related by a sequence of mutation $II$. 

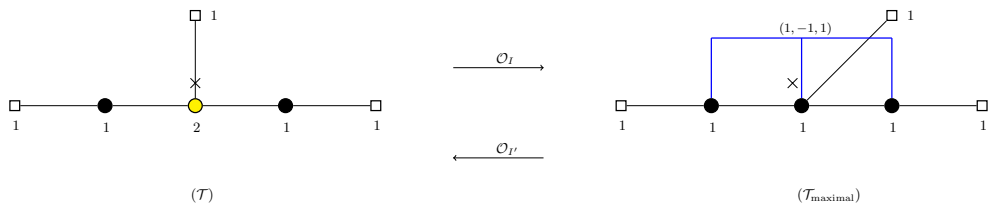
\begin{figure*}[htbp]
\begin{tabular}{ccccc}
\scalebox{0.6}{\begin{tikzpicture}
\node[fnode] (1) at (0,0){};
\node[unode] (2) at (2,0){};
\node[sunode] (3) at (4,0){};
\node[unode] (4) at (6,0){};
\node[fnode] (5) at (8,0){};
\node[fnode] (6) at (4,2){};
\node[cross, green] (40) at (4, 0.5){};
\draw[-] (1) -- (2);
\draw[-] (2)-- (3);
\draw[-] (3) -- (4);
\draw[-] (4) --(5);
\draw[-] (3) --(6);
\node[text width=.1cm](10) [below=0.1 cm of 1]{1};
\node[text width=.1cm](11) [below=0.1cm of 2]{1};
\node[text width=.1cm](12) [below=0.1cm of 3]{2};
\node[text width=.1cm](13) [below=0.1cm of 4]{1};
\node[text width=.1cm](14) [below=0.1cm of 5]{1};
\node[text width=.1cm](15) [right=0.1cm of 6]{1};
\node[text width=.2cm](20) at (4,-2){$(\CT)$};
\end{tikzpicture}}
& \qquad \qquad
& \scalebox{.6}{\begin{tikzpicture}
\draw[->] (0,0) -- (2, 0);
\draw[->] (2,-2) -- (0, -2);
\node[text width=0.1cm](29) at (1, 0.2) {$\CO_{I}$};
\node[text width=0.1cm](30) at (1, -1.8) {$\CO_{I'}$};
\node[](30) at (1, - 3) {};
\end{tikzpicture}} 
& \qquad \qquad
& \scalebox{0.6}{\begin{tikzpicture}
\node[fnode] (1) at (0,0){};
\node[unode] (2) at (2,0){};
\node[unode] (3) at (4,0){};
\node[unode] (4) at (6,0){};
\node[fnode] (5) at (8,0){};
\node[fnode] (6) at (6,2){};
\node[cross, green] (40) at (3.8, 0.5){};
\draw[-] (1) -- (2);
\draw[-] (2)-- (3);
\draw[-] (3) -- (4);
\draw[-] (4) --(5);
\draw[-] (3) --(6);
\draw[-, thick, blue] (2)--(2,1.5);
\draw[-, thick, blue] (3)--(4,1.5);
\draw[-, thick, blue] (4)--(6,1.5);
\draw[-, thick, blue] (2,1.5)--(6,1.5);
\node[text width=3 cm](10) at (5, 1.7){\footnotesize{$(1, -1, 1)$}};
\node[text width=.1cm](10) [below=0.1 cm of 1]{1};
\node[text width=.1cm](11) [below=0.1cm of 2]{1};
\node[text width=.1cm](12) [below=0.1cm of 3]{1};
\node[text width=.1cm](13) [below=0.1cm of 4]{1};
\node[text width=.1cm](14) [below=0.1cm of 5]{1};
\node[text width=.1cm](15) [right=0.1cm of 6]{1};
\node[text width=.2cm](20) at (4,-2){$(\CT_{\rm maximal})$};
\end{tikzpicture}}
\end{tabular}
\caption{\footnotesize{The duality sequence and the 3d quivers associated with the $(A_3, A_3)$ SCFT. For every quiver, the nodes which admit a mutation is marked by a cross.}}
\label{fig: maxquivEx1}
\end{figure*}

If all the SU nodes in $\CT$ are balanced, the theory $\CT_{\rm maximal}$ consists only of unitary 
gauge nodes with fundamental/bifundamental hypers as well as Abelian hypermultiplets. In addition, 
it has the property that every non-Abelian unitary node is either balanced or overbalanced. For the 
$(A_k, A_k)$ SCFTs, the theory $\CT_{\rm maximal}$ turns out to be a unique Abelian quiver gauge theory. 

The fact that the number of FI parameters in $\CT_{\rm maximal}$ matches the number of resolution/deformation parameters 
of the 4d HB can be verified from the 3d mirror. Under mirror symmetry, the HB of 
$\CT_{\rm maximal}$ is mapped to the CB of the 3d mirror in the deep IR, and in particular the FI parameters of the former 
are mapped to the mass parameters of the latter. Since turning on generic mass parameters completely resolves the CB of 
the 3d mirror as an algebraic variety, it follows from the above argument that the HB of $\CT_{\rm maximal}$ 
(and equivalently the 4d HB)  is also completely resolved/deformed by turning on generic FI parameters. 
Therefore, one simply needs to check that the number of FI parameters in $\CT_{\rm maximal}$ matches 
the number of independent mass parameters in the 3d mirror. We will perform this check explicitly 
for the $(A_k, A_k)$ SCFTs in the later sections.

\section{Hyperk\"ahler Quotient N-ality : $(A_k, A_k)$ SCFTs}\label{sec: IREx}

Let us now demonstrate how the general construction presented in \Secref{sec: maxquiv} works 
for the $(A_{k}, A_{k})$ SCFTs. We begin with the simplest SCFT with a non-trivial conformal 
manifold -- the $(A_{3}, A_{3})$ theory -- which has a conformal manifold of dimension one. 
From the top figure in \Figref{fig: Sdualityframes}, the S-duality frame can 
be seen to have the following form:

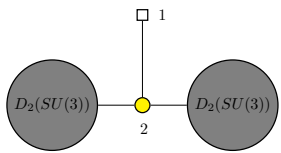
\begin{center}
\scalebox{0.6}{\begin{tikzpicture}
\node[rnode] (1) at (2,0){$D_2(SU(3))$};
\node[sunode] (2) at (4,0){};
\node[rnode] (3) at (6,0){$D_2(SU(3))$};
\node[fnode] (4) at (4, 2){};
\draw[-] (1) -- (2);
\draw[-] (2)-- (3);
\draw[-] (4) -- (2);
\node[text width= 0.1 cm](11) [below=0.1cm of 2]{2};
\node[text width= 0.1 cm](12) [right=0.1cm of 4]{1};
\end{tikzpicture}}
\end{center}

The U-SU quiver for the $(A_{3}, A_{3})$ SCFT can be read off from the S-duality frame as described in \Secref{sec: circlered}, 
and has the following form:

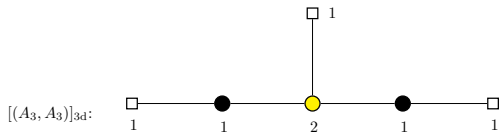
\begin{center}
\begin{tabular}{ccc}
\scalebox{0.6}{\begin{tikzpicture}
\node[text width = 2 cm] (0) at (-1,0){$[(A_3, A_3)]_{\rm 3d}$:};
\node[] at (-1,-0.4){};
\end{tikzpicture}}
& \quad
& \scalebox{0.6}{\begin{tikzpicture}
\node[fnode] (1) at (0,0){};
\node[unode] (2) at (2,0){};
\node[sunode] (3) at (4,0){};
\node[unode] (4) at (6,0){};
\node[fnode] (5) at (8,0){};
\node[fnode] (6) at (4,2){};
\draw[-] (1) -- (2);
\draw[-] (2)-- (3);
\draw[-] (3) -- (4);
\draw[-] (4) --(5);
\draw[-] (3) --(6);
\node[text width=.1cm](10) [below=0.1 cm of 1]{1};
\node[text width=.1cm](11) [below=0.1cm of 2]{1};
\node[text width=.1cm](12) [below=0.1cm of 3]{2};
\node[text width=.1cm](13) [below=0.1cm of 4]{1};
\node[text width=.1cm](14) [below=0.1cm of 5]{1};
\node[text width=.1cm](15) [right=0.1cm of 6]{1};
\end{tikzpicture}}
\end{tabular}
\end{center}

The central $SU(2)$ node is balanced and one can therefore implement mutation $I$ at this node 
as shown in \Figref{fig: maxquivEx1}. The resultant quiver is an Abelian gauge theory, and  
none of the gauge nodes admit either mutation $I$ or mutation $III$. Therefore, this quiver can be identified as the maximal unitary 
quiver. The central $U(1)$ node in $\CT_{\rm maximal}$ has balance parameter $e=1$ and therefore admits a mutation $I'$ which takes 
$\CT_{\rm maximal}$ back to the quiver $\CT$. In this case, the N-al set consists of two distinct Lagrangians. Therefore,
the \hk quotient N-ality associated with the $(A_3, A_3)$ SCFT is simply a duality. Note that 
$\text{rk}(\frg^{\rm UV}_{\rm C}(\CT_{\rm maximal}))=3$ corresponding to the three unitary gauge nodes, 
while $\text{rk}(\frg^{\rm UV}_{\rm C}(\CT)) =2$. One can therefore turn on 3 FI parameters in $\CT_{\rm maximal}$, 
while for $\CT$ one can turn on only 2. \\

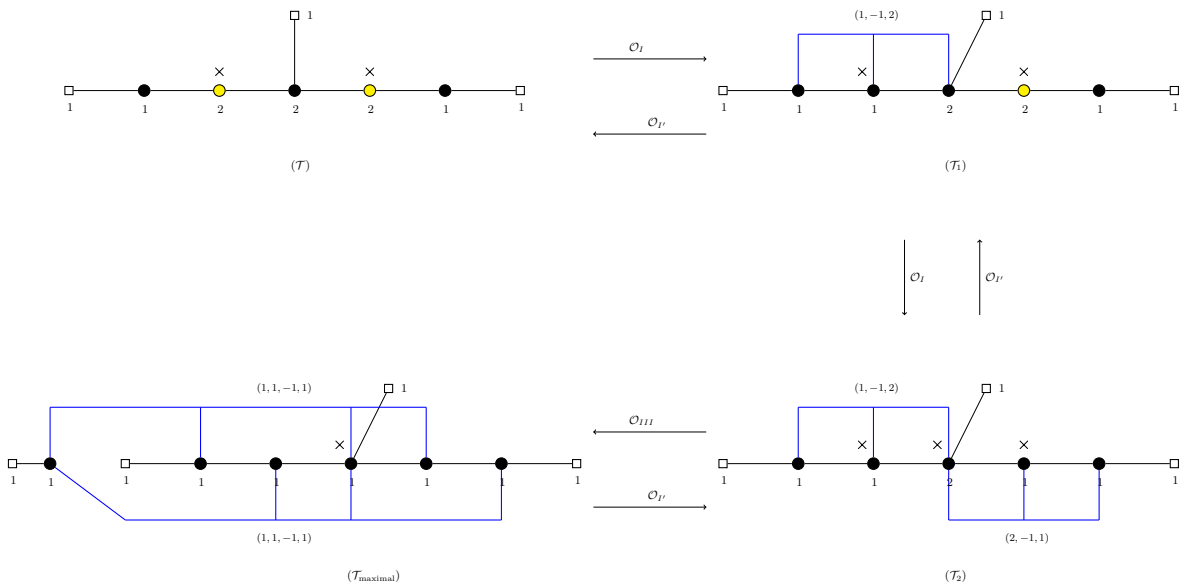
\begin{figure*}[htbp]
\begin{tabular}{ccc}
\scalebox{0.5}{\begin{tikzpicture}
\node[fnode] (1) at (0,0){};
\node[unode] (2) at (2,0){};
\node[sunode] (3) at (4,0){};
\node[unode] (4) at (6,0){};
\node[sunode] (5) at (8,0){};
\node[unode] (6) at (10,0){};
\node[fnode] (7) at (12, 0){};
\node[fnode] (8) at (6,2){};
\node[cross, red] (40) at (4, 0.5){};
\node[cross, red] (41) at (8, 0.5){};
\draw[-] (1) -- (2);
\draw[-] (2)-- (3);
\draw[-] (3) -- (4);
\draw[-] (4) --(5);
\draw[-] (5) --(6);
\draw[-] (6) --(7);
\draw[-] (4) --(8);
\node[text width=.1cm](10) [below=0.1 cm of 1]{1};
\node[text width=.1cm](11) [below=0.1cm of 2]{1};
\node[text width=.1cm](12) [below=0.1cm of 3]{2};
\node[text width=.1cm](13) [below=0.1cm of 4]{2};
\node[text width=.1cm](14) [below=0.1cm of 5]{2};
\node[text width=.1cm](15) [below=0.1cm of 6]{1};
\node[text width=.1cm](16) [below=0.1cm of 7]{1};
\node[text width=.1cm](17) [right=0.1cm of 8]{1};
\node[text width=.2cm](20) at (6,-2){$(\CT)$};
\end{tikzpicture}}
& \scalebox{.5}{\begin{tikzpicture}
\draw[->] (4,0) -- (7, 0);
\draw[->] (7, -2) -- (4, -2);
\node[text width=0.1cm](29) at (5, 0.3) {$\CO_{I}$};
\node[text width=0.1cm](29) at (5.5, -1.7) {$\CO_{I'}$};
\node[](30) at (5, -3) {};
\end{tikzpicture}}
& \scalebox{0.5}{\begin{tikzpicture}
\node[fnode] (1) at (0,0){};
\node[unode] (2) at (2,0){};
\node[unode] (3) at (4,0){};
\node[unode] (4) at (6,0){};
\node[sunode] (5) at (8,0){};
\node[unode] (6) at (10,0){};
\node[fnode] (7) at (12, 0){};
\node[fnode] (8) at (7,2){};
\node[cross, red] (40) at (8, 0.5){};
\node[cross, red] (41) at (3.7, 0.5){};
\draw[-] (1) -- (2);
\draw[-] (2)-- (3);
\draw[-] (3) -- (4);
\draw[-] (4) --(5);
\draw[-] (5) --(6);
\draw[-] (6) --(7);
\draw[-] (4) --(8);
\draw[-, thick, blue] (2)--(2,1.5);
\draw[-, thick, blue] (3)--(4,1.5);
\draw[-, thick, blue] (4)--(6,1.5);
\draw[-, thick, blue] (2,1.5)--(6,1.5);
\node[text width=3 cm](10) at (5, 2){\footnotesize{$(1, -1, 2)$}};
\node[text width=.1cm](10) [below=0.1 cm of 1]{1};
\node[text width=.1cm](11) [below=0.1cm of 2]{1};
\node[text width=.1cm](12) [below=0.1cm of 3]{1};
\node[text width=.1cm](13) [below=0.1cm of 4]{2};
\node[text width=.1cm](14) [below=0.1cm of 5]{2};
\node[text width=.1cm](15) [below=0.1cm of 6]{1};
\node[text width=.1cm](16) [below=0.1cm of 7]{1};
\node[text width=.1cm](17) [right=0.1cm of 8]{1};
\node[text width=.2cm](20) at (6,-2){$(\CT_1)$};
\end{tikzpicture}}\\
\qquad & \qquad & \qquad \\
\qquad & \qquad & \qquad \\
\qquad 
& \qquad
& \scalebox{.5}{\begin{tikzpicture}
\draw[->] (0,-1) -- (0, -3);
\draw[->] (2,-3) -- (2, -1);
\node[text width=0.1cm](29) at (0.2, -2) {$\CO_{I}$};
\node[text width=0.1cm](30) at (2.2, -2) {$\CO_{I'}$};
\end{tikzpicture}} \\
\qquad & \qquad & \qquad \\
\qquad & \qquad & \qquad \\
\scalebox{0.5}{\begin{tikzpicture}
\node[unode](100) at (-2,0){};
\node[fnode](101) at (-3,0){};
\node[fnode] (1) at (0,0){};
\node[unode] (2) at (2,0){};
\node[unode] (3) at (4,0){};
\node[unode] (4) at (6,0){};
\node[unode] (5) at (8,0){};
\node[unode] (6) at (10,0){};
\node[fnode] (7) at (12, 0){};
\node[fnode] (8) at (7,2){};
\node[cross, green] (40) at (5.7, 0.5){};
\draw[-] (1) -- (2);
\draw[-] (2)-- (3);
\draw[-] (3) -- (4);
\draw[-] (4) --(5);
\draw[-] (5) --(6);
\draw[-] (6) --(7);
\draw[-] (4) --(8);
\draw[-] (100) --(101);
\draw[-, thick, blue] (100) -- (-2, 1.5);
\draw[-, thick, blue] (2)--(2,1.5);
\draw[-, thick, blue] (4)--(6,1.5);
\draw[-, thick, blue] (5)--(8,1.5);
\draw[-, thick, blue] (-2,1.5)--(8,1.5);
\node[text width=3 cm](10) at (5, 2){\footnotesize{$(1, 1, -1, 1)$}};
\draw[-, thick, blue] (100) -- (0, -1.5);
\draw[-, thick, blue] (3)--(4, -1.5);
\draw[-, thick, blue] (4)--(6, -1.5);
\draw[-, thick, blue] (6)--(10, -1.5);
\draw[-, thick, blue] (0,-1.5)--(10, -1.5);
\node[text width=3 cm](10) at (5, -2){\footnotesize{$(1, 1, -1, 1)$}};
\node[text width=.1cm](10) [below=0.1 cm of 1]{1};
\node[text width=.1cm](11) [below=0.1cm of 2]{1};
\node[text width=.1cm](12) [below=0.1cm of 3]{1};
\node[text width=.1cm](13) [below=0.1cm of 4]{1};
\node[text width=.1cm](14) [below=0.1cm of 5]{1};
\node[text width=.1cm](15) [below=0.1cm of 6]{1};
\node[text width=.1cm](16) [below=0.1cm of 7]{1};
\node[text width=.1cm](17) [right=0.1cm of 8]{1};
\node[text width=.1cm](18) [below=0.1cm of 100]{1};
\node[text width=.1cm](19) [below=0.1cm of 101]{1};
\node[text width=.2cm](20) at (6,-3){$(\CT_{\rm maximal})$};
\end{tikzpicture}}
& \scalebox{.5}{\begin{tikzpicture}
\draw[->] (7,0) -- (4, 0);
\draw[->] (4, -2) -- (7, -2);
\node[text width=0.1cm](29) at (5, 0.3) {$\CO_{III}$};
\node[text width=0.1cm](29) at (5.5, -1.7) {$\CO_{I'}$};
\node[](30) at (5, -4) {};
\end{tikzpicture}}
& \scalebox{0.5}{\begin{tikzpicture}
\node[fnode] (1) at (0,0){};
\node[unode] (2) at (2,0){};
\node[unode] (3) at (4,0){};
\node[unode] (4) at (6,0){};
\node[unode] (5) at (8,0){};
\node[unode] (6) at (10,0){};
\node[fnode] (7) at (12, 0){};
\node[fnode] (8) at (7,2){};
\node[cross, red] (40) at (5.7, 0.5){};
\node[cross, red] (41) at (3.7, 0.5){};
\node[cross, red] (42) at (8, 0.5){};
\draw[-] (1) -- (2);
\draw[-] (2)-- (3);
\draw[-] (3) -- (4);
\draw[-] (4) --(5);
\draw[-] (5) --(6);
\draw[-] (6) --(7);
\draw[-] (4) --(8);
\draw[-, thick, blue] (2)--(2,1.5);
\draw[-, thick, blue] (3)--(4,1.5);
\draw[-, thick, blue] (4)--(6,1.5);
\draw[-, thick, blue] (2,1.5)--(6,1.5);
\node[text width=3 cm](10) at (5, 2){\footnotesize{$(1, -1, 2)$}};
\draw[-, thick, blue] (4)--(6, -1.5);
\draw[-, thick, blue] (5)--(8, -1.5);
\draw[-, thick, blue] (6)--(10, -1.5);
\draw[-, thick, blue] (6,-1.5)--(10, -1.5);
\node[text width=3 cm](10) at (9, -2){\footnotesize{$(2, -1, 1)$}};
\node[text width=.1cm](10) [below=0.1 cm of 1]{1};
\node[text width=.1cm](11) [below=0.1cm of 2]{1};
\node[text width=.1cm](12) [below=0.1cm of 3]{1};
\node[text width=.1cm](13) [below=0.1cm of 4]{2};
\node[text width=.1cm](14) [below=0.1cm of 5]{1};
\node[text width=.1cm](15) [below=0.1cm of 6]{1};
\node[text width=.1cm](16) [below=0.1cm of 7]{1};
\node[text width=.1cm](17) [right=0.1cm of 8]{1};
\node[text width=.2cm](20) at (6,-3){$(\CT_2)$};
\end{tikzpicture}}
\end{tabular}
\caption{\footnotesize{The duality sequence and the 3d quivers for the $(A_4, A_4)$ SCFT. For every quiver, the nodes which admit a mutation is marked by a cross.} }
\label{fig: maxquivEx2}
\end{figure*}

Let us consider the next non-trivial example -- the $(A_{4}, A_{4})$ SCFT, for which the 
conformal manifold has dimension 2.  From the bottom figure in \Figref{fig: Sdualityframes}, the S-duality frame can 
be seen to have the following form:

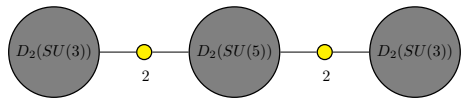
\begin{center}
\scalebox{0.6}{\begin{tikzpicture}
\node[rnode] (1) at (2,0){$D_2(SU(3))$};
\node[sunode] (2) at (4,0){};
\node[rnode] (3) at (6,0){$D_2(SU(5))$};
\node[sunode] (4) at (8,0){};
\node[rnode] (5) at (10,0){$D_2(SU(3))$};
\draw[-] (1) -- (2);
\draw[-] (2)-- (3);
\draw[-] (3) -- (4);
\draw[-] (4) -- (5);
\node[text width= 0.1 cm](11) [below=0.1cm of 2]{2};
\node[text width= 0.1 cm](12) [below=0.1cm of 4]{2};
\end{tikzpicture}}
\end{center}

The U-SU quiver for the $(A_{4}, A_{4})$ SCFT can be read off from the S-duality frame described in \Secref{sec: circlered}, 
and has the following form:
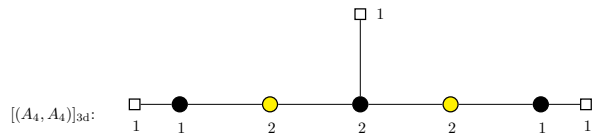
\begin{center}
\begin{tabular}{ccc}
\scalebox{0.6}{\begin{tikzpicture}
\node[text width = 2 cm] (0) at (-1,0){$[(A_4, A_4)]_{\rm 3d}$:};
\node[] at (-1,-0.4){};
\end{tikzpicture}}
& \quad
&\scalebox{0.6}{\begin{tikzpicture}
\node[fnode] (1) at (1,0){};
\node[unode] (2) at (2,0){};
\node[sunode] (3) at (4,0){};
\node[unode] (4) at (6,0){};
\node[sunode] (5) at (8,0){};
\node[unode] (6) at (10,0){};
\node[fnode] (7) at (11, 0){};
\node[fnode] (8) at (6,2){};
\draw[-] (1) -- (2);
\draw[-] (2)-- (3);
\draw[-] (3) -- (4);
\draw[-] (4) --(5);
\draw[-] (5) --(6);
\draw[-] (6) --(7);
\draw[-] (4) --(8);
\node[text width=.1cm](10) [below=0.1 cm of 1]{1};
\node[text width=.1cm](11) [below=0.1cm of 2]{1};
\node[text width=.1cm](12) [below=0.1cm of 3]{2};
\node[text width=.1cm](13) [below=0.1cm of 4]{2};
\node[text width=.1cm](14) [below=0.1cm of 5]{2};
\node[text width=.1cm](15) [below=0.1cm of 6]{1};
\node[text width=.1cm](16) [below=0.1cm of 7]{1};
\node[text width=.1cm](17) [right=0.1cm of 8]{1};
\end{tikzpicture}}
\end{tabular}
\end{center}

The theory has two balanced $SU(2)$ gauge nodes corresponding to the two marginal couplings in the SCFT, while the unitary nodes are overbalanced. Also note that the quiver is symmetric about the central $U(2)$ gauge node. Following the prescription in \Secref{sec: maxquiv}, the 
3d quivers constituting the N-al set can be found starting from the above quiver by the sequence of quiver mutations shown in \Figref{fig: maxquivEx2}:

\begin{enumerate}

\item We begin by implementing mutation $I$ at the $SU(2)$ node on the left which introduces an Abelian hypermultiplet. 
The resultant quiver $\CT_1$ has a single balanced $SU(2)$ node. Since the quiver $\CT$ is symmetric about the central 
$U(2)$ node, implementing mutation $I$ on the right $SU(2)$ node will produce the same quiver up to a discrete identification. 

\item The quiver $\CT_1$ admits mutation $I$ at the remaining balanced $SU(2)$ node which introduces a second Abelian 
hypermultiplet and gives the quiver $\CT_2$. The second $U(1)$ node from the left has balance parameter $e=1$ and 
therefore admits a mutation $I'$. This mutation takes the quiver $\CT_1$ back to the quiver $\CT$. It might appear that the 
central $U(2)$ node of $\CT_1$ admits a mutation $II$ but one can check that this operation amounts a change of variables 
of the Abelian vector multiplets and therefore does not produce a distinct Lagrangian. 

\item In the next step, we observe that the quiver $\CT_2$ has only unitary gauge nodes -- all of which are $U(1)$ nodes 
except for the central $U(2)$ gauge node which has balance parameter $-1$. The $U(2)$ node also connected to the two 
Abelian hypermultiplets and therefore admits a mutation $III$. Implementing this mutation gives an Abelian quiver gauge 
theory. This quiver does not admit mutation $I$ or $III$, and it can therefore be identified as the maximal unitary 
quiver. The second and the fourth gauge nodes from the left (both $U(1)$ nodes) have balance parameters $e=1$ and 
admit mutation $I'$. Implementing it at any of these nodes takes the quiver $\CT_2$ back to the quiver $\CT_1$. 

\item In the quiver $\CT_{\rm maximal}$, one of the $U(1)$ gauge nodes (marked with a cross) 
connected to the Abelian hypers has balance parameter $e=1$ and therefore admits a mutation $I'$. This mutation takes 
$\CT_{\rm maximal}$ back to the quiver $\CT_2$. 

\end{enumerate}

Therefore, the N-al set associated with the $(A_4, A_4)$ SCFT consists of four distinct quiver gauge theories. The \hk quotient N-ality  
in this case is a quadrality. Note that the UV-manifest rank of the CB global symmetry (or equivalently the number of FI parameters) 
in the quiver $\CT$ is 3, and it increases by 1 with every mutation $I$ and $III$ reaching the maximum of 6 for $\CT_{\rm maximal}$.\\

One can now see how the duality sequence works out for the case of a generic $k$. The U-SU quiver has $k-2$ balanced SU nodes 
and $k-1$ overbalanced unitary nodes. Mutation $I$ at the balanced nodes will bring a total of $k-2$ Abelian hypermultiplets, while 
applying Mutation $III$ sequentially leads to a maximal unitary quiver which is an Abelian gauge theory. The general structure of 
the quiver $\CT_{\rm maximal}$ is given as follows. One starts from the linear quiver with $2k-3$ $U(1)$ nodes:
\be \label{AkAk-LQ}
\scalebox{0.6}{\begin{tikzpicture}
\node[fnode] (1) at (0,0){};
\node[unode] (2) at (1,0){};
\node[unode] (3) at (2,0){};
\node[](4) at (2.5,0){};
\node[](5) at (3.5,0){};
\node[unode] (6) at (4,0){};
\node[fnode] (7) at (4,1){};
\node[](8) at (4.5,0){};
\node[](9) at (5.5,0){};
\node[unode](10) at (6,0){};
\node[unode](11) at (7,0){};
\node[fnode] (12) at (8,0){};
\draw[-] (1) -- (2);
\draw[-] (2)-- (3);
\draw[-] (3) -- (4);
\draw[-, dotted] (4) --(5);
\draw[-] (5) --(6);
\draw[-] (6) --(7);
\draw[-] (6) --(8);
\draw[-, dotted] (8) --(9);
\draw[-] (9) --(10);
\draw[-] (10) --(11);
\draw[-] (11) --(12);
\node[text width=.1cm](50) [below=0.1 cm of 1]{1};
\node[text width=.1cm](51) [below=0.1cm of 2]{1};
\node[text width=.1cm](52) [below=0.1cm of 3]{1};
\node[text width=.1cm](55) [below=0.1cm of 6]{1};
\node[text width=.1cm](56) [right=0.1cm of 7]{1};
\node[text width=.1cm](57) [below=0.1cm of 10]{1};
\node[text width=.1cm](58) [below=0.1cm of 11]{1};
\node[text width=.1cm](58) [below=0.1cm of 12]{1};
\end{tikzpicture}}
\ee
with the central $U(1)$ gauge node ($(k-1)$-th from the left or the right) having a $U(1)$ flavor node. The quiver $\CT_{\rm maximal}$ 
is then given by decorating the linear quiver with a total of $(k-2)$ Abelian hypermultiplets, which may in turn be attached 
to Abelian quiver tails. The precise charges of the Abelian hypers and the quiver tails may be worked out using the 
duality sequence for any given $k$.

\section{From the Maximal Unitary Quiver to the 3d Mirror}\label{sec: 3dmirr}

In this section, we write down the general recipe for constructing the 3d mirror of an AD-type SCFT, focusing on the 
subclass of theories for which the circle reduction of the SCFT gives a U-SU quiver $\CT$, which is a good theory in 
the Gaiotto-Witten sense. In addition, $\CT$ obeys the constraint that all the special unitary nodes are balanced. 
We will follow the treatment of \cite{Dey:2023vrt} where the problem of finding the 3d 
mirror was studied for a generic good 3d U-SU quiver gauge theory with at least one balanced special unitary node.

The starting point of the construction is the maximal unitary quiver $\CT_{\rm maximal}$ (or one of the maximal 
unitary quivers) which can be derived from $\CT$ using a sequence of the quiver mutations $I, III$ and $II$, as discussed above. 
Since mutation $I$ introduces an Abelian hyper and mutation $III$ introduces an Abelian hyper attached to a quiver tail, 
the quiver $\CT_{\rm maximal}$ always has the following general structure -- it is given by a standard quiver gauge theory $\CT_{\rm good}$
consisting of unitary gauge nodes and matter in the fundamental/bifundamental representation, decorated with Abelian hypermultiplets 
attached to Abelian quiver tails. For the $(A_k, A_k)$ SCFT, the quiver $\CT_{\rm good}$, given by \eref{AkAk-LQ}, is particularly simple -- 
a linear quiver with $U(1)$ gauge nodes which is a good theory. From the general rules of quiver mutations, 
one can check that the quiver $\CT_{\rm good}$ is at most an ugly theory, if $\CT$ is a good theory to begin with.

Given this general structure of $\CT_{\rm maximal}$, the 3d mirror can be found in the following fashion.
In the first step, the Abelian hypers and the Abelian quiver tails introduced by the quiver mutations 
are stripped off from $\CT_{\rm maximal}$ to obtain the quiver $\CT_{\rm good}$.
Next, we find the 3d mirror of $\CT_{\rm good}$, which will be denoted as $\wt{\CT}_{\rm good}$. 
If $\CT_{\rm good}$ is a linear quiver with unitary gauge nodes, $\wt{\CT}_{\rm good}$ will also belong to the same class. 
It will also be a good theory, plus a certain number of free hypermultiplets if $\CT_{\rm good}$ has ugly nodes. 

The theory $\CT_{\rm maximal}$ can be built up from of the theory $\CT_{\rm good}$ by a sequence of simple QFT operations, 
each of which can be thought of as a mild generalization of the $S$-operation \cite{Witten:2003ya}. Generically, if the theory $\CT$ has 
$l$ balanced special unitary nodes, the quiver $\CT_{\rm maximal}$ consists of $l$ Abelian hypermultiplets.
Starting from $\CT_{\rm good}$, we first add to the theory $\CT_{\rm good}$ a decoupled Abelian quiver gauge theory $\CT^{(1)}_{\rm decoupled}$ of the following form:
\begin{center}
\scalebox{0.6}{\begin{tikzpicture}
\node[text width=2 cm](0) at (-2,0){$\CT^{(1)}_{\rm decoupled}:$};
\node[fnode] (1) at (0,0){};
\node[unode] (2) at (1,0){};
\node[unode] (3) at (2,0){};
\node[] (4) at (3,0){};
\node[] (5) at (4,0){};
\node[unode] (6) at (5,0){};
\node[unode] (7) at (6,0){};
\draw[-] (1) -- (2);
\draw[-] (2)-- (3);
\draw[-] (3)-- (4);
\draw[-, dotted] (4)-- (5);
\draw[-] (5)-- (6);
\draw[-] (6)-- (7);
\node[text width=0.1cm](20) [above=0.1 cm of 1]{1};
\node[text width=0.1 cm](21) [above=0.1cm of 2]{1};
\node[text width=0.1 cm](22) [above=0.1cm of 3]{1};
\node[text width=0.1 cm](23) [above=0.1cm of 6]{1};
\node[text width=0.1 cm](24) [above=0.1cm of 7]{1};
\end{tikzpicture}}
\end{center}
where the number of $U(1)$ gauge nodes is $(s+1)$. Evidently, this quiver is simply another way of denoting a collection of $(s+1)$ 
twisted free hypermultiplets. Then, we gauge an $\fru(1)$ subalgebra of the topological symmetry algebra of the 
combined theory corresponding to the following generator :
\be \label{subalg-gauging}
J = J^{(1)}_{(s+1)} + \sum_{a}\, \frac{Q_a}{N_a}\,J_a, 
\ee
where $J^{(1)}_{(i)}$ is the generator for the topological symmetry associated with the $i$-th $U(1)$ gauge node 
of the $\CT^{(1)}_{\rm decoupled}$ ($i$ increasing from left to right), while $J_a$ is the topological symmetry 
generator associated with the $U(N_a)$ gauge node in $\CT_{\rm good}$. The sum on the RHS extends over 
all nodes of the quiver $\CT_{\rm good}$. The result of this operation is to attach an Abelian hyper to a collection 
of gauge nodes $\{ U(N_a) \}$ in $\CT_{\rm good}$ with respective charges $\{Q_a\}$, where the Abelian hyper 
is connected to a quiver tail of $s$ $U(1)$ gauge nodes. Also, the charge of the Abelian hyper under the $U(1)$ 
node of the quiver tail it is connected to is 1. These type of QFT operations (and certain generalized versions) were studied in 
 detail in \cite{Dey:2020hfe} and were used to construct new pairs of 3d mirrors from a given mirror pair. We refer the reader to 
 Section 4.1 of \cite{Dey:2020hfe} for further details.

 Let us call the quiver obtained by the above QFT operation $\CT^{(1)}_{\rm good}$, which is a unitary gauge theory 
 with a single Abelian hypermultiplet. 
 In the next step, one performs a similar operation now starting with  $\CT^{(1)}_{\rm good}$, adding a decoupled theory 
 $\CT^{(2)}_{\rm decoupled}$ of the same form as before and then gauging a $\fru(1)$ subalgebra of the topological 
 symmetry of the combined theory. This results in the quiver $\CT^{(2)}_{\rm good}$ which is a unitary gauge theory 
 with two Abelian hypers. Repeating the operation $l$ times, one obtains $\CT^{(l)}_{\rm good} = \CT_{\rm maximal}$, 
 which is quiver with $l$ Abelian hypermultiplets attached to Abelian quiver tails.

On the mirror side, one implements the dual sequence of QFT operations as follows. Starting from $\wt{\CT}_{\rm good}$, 
the first QFT operation involves adding a decoupled sector $\wt{\CT}^{(1)}_{\rm decoupled}$ of free hypermultiplets
-- the mirror dual of $\CT^{(1)}_{\rm decoupled}$ -- which we represent as : 
\begin{center}
\scalebox{0.6}{\begin{tikzpicture}
\node[text width=2 cm](0) at (-2,0){$\wt{\CT}^{(1)}_{\rm decoupled}:$};
\node[fnode] (1) at (0,0){};
\node[fnode] (2) at (2,0){};
\draw[-] (1) -- (2);
\node[text width=0.1cm](20) [above=0.1 cm of 1]{1};
\node[text width=1 cm](21) [above=0.1cm of 2]{$s+1$};
\end{tikzpicture}}.
\end{center}

Under mirror symmetry, the $\fru(1)$ subalgebra in \eref{subalg-gauging} maps to a $\fru(1)$ subalgebra of the HB global symmetry algebra 
for the combined theory $\wt{\CT}_{\rm good} \oplus \wt{\CT}^{(1)}_{\rm decoupled}$. The QFT operation then amounts 
to identifying a set of $U(1)$ flavor nodes in $\wt{\CT}_{\rm good}$ with the $U(1)$ flavor node in $\wt{\CT}^{(1)}_{\rm decoupled}$, 
and then gauging the identified flavor node. This operation is an example of an Abelian \textit{identification-flavoring-gauging} 
operation introduced in \cite{Dey:2020hfe}. The resultant quiver is the 3d mirror of $\CT^{(1)}_{\rm good}$ and we denote it by 
$\wt{\CT}^{(1)}_{\rm good}$. In the next step, we perform a similar operation starting with  $\wt{\CT}^{(1)}_{\rm good}$. After performing $l$ such operations, one ends up with the theory $\wt{\CT}^{(l)}_{\rm good}$, which may be identified with $\wt{\CT}_{\rm maximal}$ -- the 3d mirror of $\CT_{\rm maximal}$. Given IR N-ality, the quiver $\wt{\CT}_{\rm maximal}$ is the 3d mirror of all the theories in the N-al set, including the quiver $\CT$. We will explicitly show how this construction can be used to build the 3d mirrors of the $(A_k, A_k)$ SCFTs.

\section{3d mirrors for $(A_{k}, A_{k})$ SCFT : Complete graphs}\label{sec: mainexmirr}

In this section, we will apply the prescription described in \Secref{sec: 3dmirr} now to construct the 3d mirror of 
an $(A_{k}, A_{k})$ SCFT. Let us start with the case of $k=3$. The 3d mirror pair $(\CT_{\rm good}, \wt{\CT}_{\rm good})$ 
in this case is given as follows: 

\begin{center}
\begin{tabular}{ccc}
\scalebox{0.6}{\begin{tikzpicture}
\node[fnode] (1) at (0,0){};
\node[unode] (2) at (1,0){};
\node[unode] (3) at (2,0){};
\node[unode] (4) at (3,0){};
\node[fnode] (5) at (4,0){};
\node[fnode] (6) at (2,1){};
\draw[-] (1) -- (2);
\draw[-] (2)-- (3);
\draw[-] (3) -- (4);
\draw[-] (4) --(5);
\draw[-] (3) --(6);
\node[text width=.1cm](10) [below=0.1 cm of 1]{1};
\node[text width=.1cm](11) [below=0.1cm of 2]{1};
\node[text width=.1cm](12) [below=0.1cm of 3]{1};
\node[text width=.1cm](13) [below=0.1cm of 4]{1};
\node[text width=.1cm](14) [below=0.1cm of 5]{1};
\node[text width=.1cm](15) [right=0.1cm of 6]{1};
\node[text width=.2cm](20) at (2,-1.2){$(\CT_{\rm good})$};
\end{tikzpicture}}
& \qquad \qquad
& \scalebox{0.6}{\begin{tikzpicture}
\node[unode] (1) at (2,0){};
\node[unode] (2) at (4,0){};
\node[fnode] (3) at (2,-1){};
\node[fnode] (4) at (4,-1){};
\draw[-] (1) -- (2);
\draw[-] (1)-- (3);
\draw[-] (2) -- (4);
\node[text width=0.1cm](20) [above=0.1 cm of 1]{1};
\node[text width=0.1 cm](21) [above=0.1cm of 2]{1};
\node[text width=0.1 cm](22) [left=0.1cm of 3]{2};
\node[text width=0.1 cm](23) [right=0.1cm of 4]{2};
\node[text width=.2cm](31) at (3,-2){$(\wt{\CT}_{\rm good})$};
\end{tikzpicture}}
\end{tabular}
\end{center}

Next, we add a theory $\CT_{\rm decoupled}$ -- a $U(1)$ theory with a single hypermultiplet -- to the 
theory ${\CT}_{\rm good}$, while on the dual side we add the 3d mirror of $\CT_{\rm decoupled}$ -- 
a free hypermultiplet to the theory $\wt{\CT}_{\rm good}$. The QFT operation on the former side involves 
gauging a $\fru(1)$ subalgebra of topological symmetry of the combined theory, where the subalgebra 
corresponds to the generator:
\be
J = J_0 + J_1 - J_2 + J_3, \label{subalg-gaugingex1}
\ee
where $J_0$ is the generator of the topological symmetry for the $U(1)$ gauge node in $\CT_{\rm decoupled}$, 
while $J_i$ is the generator of the topological symmetry for the $i$-th $U(1)$ gauge node in ${\CT}_{\rm good}$ 
(with $i$ increasing from left to right). This gives the quiver $\CT_{\rm maximal}$ :

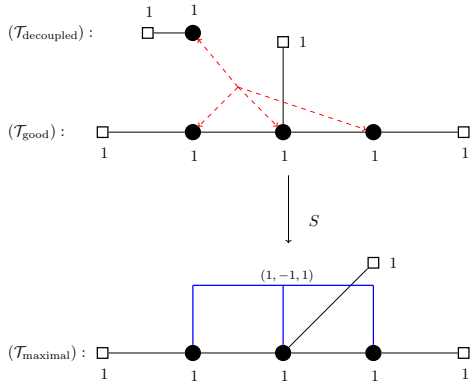
\begin{center}
\begin{tabular}{c}
\scalebox{0.6}{\begin{tikzpicture}
\node[unode] (100) at (2, 2.2){};
\node[fnode] (101) at (1, 2.2){};
\node[fnode] (1) at (0,0){};
\node[unode] (2) at (2,0){};
\node[unode] (3) at (4,0){};
\node[unode] (4) at (6,0){};
\node[fnode] (5) at (8,0){};
\node[fnode] (6) at (4,2){};
\draw[-] (1) -- (2);
\draw[-] (2)-- (3);
\draw[-] (3) -- (4);
\draw[-] (4) --(5);
\draw[-] (3) --(6);
\draw[-] (100) --(101);
\draw[->, red, dashed] (3,1) -- (100);
\draw[->, red, dashed] (3,1) -- (2);
\draw[->, red, dashed] (3,1) -- (3);
\draw[->, red, dashed] (3,1) -- (4);
\node[text width=.1cm](10) [below=0.1 cm of 1]{1};
\node[text width=.1cm](11) [below=0.1cm of 2]{1};
\node[text width=.1cm](12) [below=0.1cm of 3]{1};
\node[text width=.1cm](13) [below=0.1cm of 4]{1};
\node[text width=.1cm](14) [below=0.1cm of 5]{1};
\node[text width=.1cm](15) [right=0.1cm of 6]{1};
\node[text width=.1cm](18) [above=0.1cm of 100]{1};
\node[text width=.1cm](19) [above=0.1cm of 101]{1};
\node[text width=.2cm](20) at (-2, 0){$(\CT_{\rm good}):$};
\node[text width=.2cm](21) at (-2, 2.2){$(\CT_{\rm decoupled}):$};
\node[](21) at (11,0){};
\end{tikzpicture}}\\
\scalebox{.6}{\begin{tikzpicture}
\draw[->] (0,0) -- (0, -1.5);
\node[text width=0.1cm](29) at (0.5, -1) {$S$};
\end{tikzpicture}}\\
\scalebox{0.6}{\begin{tikzpicture}
\node[] (0) at (-1,0){};
\node[fnode] (1) at (0,0){};
\node[unode] (2) at (2,0){};
\node[unode] (3) at (4,0){};
\node[unode] (4) at (6,0){};
\node[fnode] (5) at (8,0){};
\node[fnode] (6) at (6,2){};
\draw[-] (1) -- (2);
\draw[-] (2)-- (3);
\draw[-] (3) -- (4);
\draw[-] (4) --(5);
\draw[-] (3) --(6);
\draw[-, thick, blue] (2)--(2,1.5);
\draw[-, thick, blue] (3)--(4,1.5);
\draw[-, thick, blue] (4)--(6,1.5);
\draw[-, thick, blue] (2,1.5)--(6,1.5);
\node[text width=3 cm](10) at (5, 1.7){\footnotesize{$(1, -1, 1)$}};
\node[text width=.1cm](10) [below=0.1 cm of 1]{1};
\node[text width=.1cm](11) [below=0.1cm of 2]{1};
\node[text width=.1cm](12) [below=0.1cm of 3]{1};
\node[text width=.1cm](13) [below=0.1cm of 4]{1};
\node[text width=.1cm](14) [below=0.1cm of 5]{1};
\node[text width=.1cm](15) [right=0.1cm of 6]{1};
\node[text width=.2cm](20) at (-2, 0){$(\CT_{\rm maximal}):$};
\node[](21) at (11,0){};
\end{tikzpicture}}
\end{tabular}
\end{center}

The red dashed lines mark the gauge nodes whose topological symmetry generators appear on the RHS of 
\eref{subalg-gaugingex1}. 
As mentioned in \Secref{sec: 3dmirr}, the operation on the mirror side therefore involves an identification of the 
$U(1)$ flavor nodes as shown below followed by a gauging operation of the identified node: 

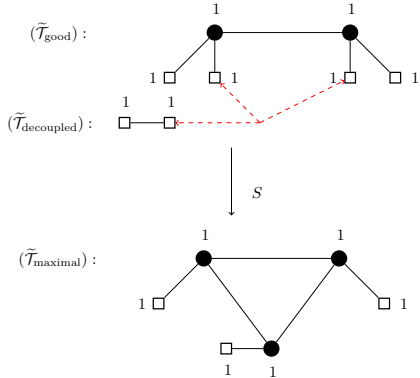
\begin{center}
\begin{tabular}{c}
\scalebox{0.6}{\begin{tikzpicture}
\node[unode] (1) at (2,0){};
\node[unode] (2) at (5,0){};
\node[fnode] (3) at (2,-1){};
\node[fnode] (4) at (1,-1){};
\node[fnode] (5) at (5,-1){};
\node[fnode] (6) at (6,-1){};
\node[fnode](101) at (1, -2){};
\node[fnode](102) at (0, -2){};
\draw[-] (1) -- (2);
\draw[-] (1)-- (3);
\draw[-] (1) -- (4);
\draw[-] (2) --(5);
\draw[-] (2) --(6);
\draw[-] (101) --(102);
\draw[->, red, dashed] (3,-2) -- (101);
\draw[->, red, dashed] (3,-2) -- (3);
\draw[->, red, dashed] (3,-2) -- (5);
\node[text width=0.1cm](20) [above=0.1 cm of 1]{1};
\node[text width=0.1 cm](21) [above=0.1cm of 2]{1};
\node[text width=0.1 cm](22) [right=0.1cm of 3]{1};
\node[text width=0.1 cm](23) [left=0.1cm of 4]{1};
\node[text width=0.1 cm](25) [left=0.1cm of 5]{1};
\node[text width=0.1 cm](26) [right=0.1cm of 6]{1};
\node[text width=.1cm](27) [above=0.1cm of 101]{1};
\node[text width=.1cm](28) [above=0.1cm of 102]{1};
\node[text width=.2cm](31) at (-2, 0){$(\wt{\CT}_{\rm good}):$};
\node[text width=.2cm](32) at (-2.5, -2){$(\wt{\CT}_{\rm decoupled}):$};
\node[](32) at (8,0){};
\end{tikzpicture}}\\
\scalebox{.6}{\begin{tikzpicture}
\draw[->] (0,0) -- (0, -1.5);
\node[text width=0.1cm](29) at (0.5, -1) {$S$};
\end{tikzpicture}}\\
\scalebox{0.6}{\begin{tikzpicture}
\node[unode] (1) at (2,0){};
\node[unode] (2) at (5,0){};
\node[unode] (3) at (3.5, -2){};
\node[fnode] (4) at (1,-1){};
\node[fnode] (5) at (6,-1){};
\node[fnode] (6) at (2.5,-2){};
\draw[-] (1) -- (2);
\draw[-] (1) -- (3);
\draw[-] (2) -- (3);
\draw[-] (1) -- (4);
\draw[-] (2) -- (5);
\draw[-] (3) --(6);
\node[text width=0.1cm](20) [above=0.1 cm of 1]{1};
\node[text width=0.1 cm](21) [above=0.1cm of 2]{1};
\node[text width=0.1 cm](22) [below=0.1cm of 3]{1};
\node[text width=0.1 cm](23) [left=0.1cm of 4]{1};
\node[text width=0.1 cm](24) [right=0.1cm of 5]{1};
\node[text width=0.1 cm](25) [below=0.1cm of 6]{1};
\node[text width=.2cm](31) at (-2, 0){$(\wt{\CT}_{\rm maximal}):$};
\node[](32) at (8,0){};
\end{tikzpicture}}
\end{tabular}
\end{center}

The resultant quiver $\wt{\CT}_{\rm maximal}$ is equivalent to the complete 
graph of with four vertices and edge multiplicity 1, after decoupling a $U(1)$ vector multiplet 
associated with one of the vertices: 

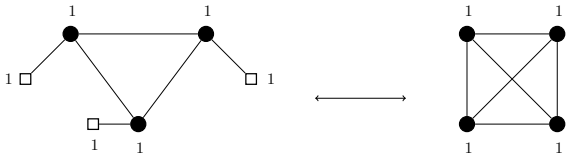
\begin{center}
\begin{tabular}{ccc}
\scalebox{0.6}{\begin{tikzpicture}
\node[unode] (1) at (2,0){};
\node[unode] (2) at (5,0){};
\node[unode] (3) at (3.5, -2){};
\node[fnode] (4) at (1,-1){};
\node[fnode] (5) at (6,-1){};
\node[fnode] (6) at (2.5,-2){};
\draw[-] (1) -- (2);
\draw[-] (1) -- (3);
\draw[-] (2) -- (3);
\draw[-] (1) -- (4);
\draw[-] (2) -- (5);
\draw[-] (3) --(6);
\node[text width=0.1cm](20) [above=0.1 cm of 1]{1};
\node[text width=0.1 cm](21) [above=0.1cm of 2]{1};
\node[text width=0.1 cm](22) [below=0.1cm of 3]{1};
\node[text width=0.1 cm](23) [left=0.1cm of 4]{1};
\node[text width=0.1 cm](24) [right=0.1cm of 5]{1};
\node[text width=0.1 cm](25) [below=0.1cm of 6]{1};
\end{tikzpicture}}
&\scalebox{.6}{\begin{tikzpicture}
\node[] at (3.5,0){};
\draw[<->] (4,0) -- (6, 0);
\node[](30) at (5, -1.2) {};
\end{tikzpicture}}
& \scalebox{0.6}{\begin{tikzpicture}
\node[] (1) at (1,0){};
\node[unode] (2) at (2,0){};
\node[unode] (3) at (4,0){};
\node[unode] (6) at (2,-2){};
\node[unode] (7) at (4,-2){};
\draw[-] (2)-- (3);
\draw[-] (2) --(6);
\draw[-] (2) --(7);
\draw[-] (3) --(6);
\draw[-] (3) --(7);
\draw[-] (6) --(7);
\node[text width=0.1 cm](21) [above=0.1cm of 2]{1};
\node[text width=0.1 cm](22) [above=0.1cm of 3]{1};
\node[text width=0.1 cm](25) [below=0.1cm of 6]{1};
\node[text width=0.1 cm](26) [below=0.1cm of 7]{1};
\end{tikzpicture}}
\end{tabular}
\end{center}

Given the 3d mirror, one can check that the rank of its HB global symmetry algebra is $\text{rk}(\frg_{\rm H}(\wt{\CT}_{\rm maximal}))= 3$. 
Therefore, the rank of the HB global symmetry algebra of the mirror (or equivalently the number of independent mass parameters) 
precisely matches the rank of the CB global symmetry algebra of $\CT_{\rm maximal}$ (or equivalently the number of FI parameters). 
We therefore have an explicit check that the maximal unitary quiver makes all the resolution/deformation parameters of the 4d HB manifest in this case.\\

\begin{figure*}[htbp]
\begin{tabular}{ccccc}
\scalebox{0.6}{\begin{tikzpicture}
\node[unode] (100) at (2, 2.2){};
\node[unode] (101) at (1, 2.2){};
\node[fnode] (102) at (0, 2.2){};
\node[fnode] (1) at (0,0){};
\node[unode] (2) at (1,0){};
\node[unode] (3) at (2,0){};
\node[unode] (4) at (3,0){};
\node[unode] (5) at (4,0){};
\node[unode] (6) at (5,0){};
\node[fnode] (7) at (6, 0){};
\node[fnode] (8) at (4,2){};
\draw[-] (1) -- (2);
\draw[-] (2)-- (3);
\draw[-] (3) -- (4);
\draw[-] (4) --(5);
\draw[-] (5) --(6);
\draw[-] (6) --(7);
\draw[-] (4) --(8);
\node[unode] (200) at (3, -2.2){};
\node[fnode] (201) at (4,-2.2){};
\draw[-] (200) --(201);
\draw[-] (100) --(101);
\draw[-] (101) --(102);
\draw[->, red, dashed] (2,1) -- (100);
\draw[->, red, dashed] (2,1) -- (2);
\draw[->, red, dashed] (2,1) -- (4);
\draw[->, red, dashed] (2,1) -- (5);
\node[text width=.1cm](10) [below=0.1 cm of 1]{1};
\node[text width=.1cm](11) [below=0.1cm of 2]{1};
\node[text width=.1cm](12) [below=0.1cm of 3]{1};
\node[text width=.1cm](13) [below=0.1cm of 4]{1};
\node[text width=.1cm](14) [below=0.1cm of 5]{1};
\node[text width=.1cm](15) [below=0.1cm of 6]{1};
\node[text width=.1cm](16) [below=0.1cm of 7]{1};
\node[text width=.1cm](17) [right=0.1cm of 8]{1};
\node[text width=.1cm](18) [above=0.1cm of 100]{1};
\node[text width=.1cm](19) [above=0.1cm of 101]{1};
\node[text width=.1cm](19) [above=0.1cm of 102]{1};
\node[text width= 5 cm](20) at (3,-3){$(\CT_{\rm good}\oplus \CT^{(1)}_{\rm decoupled} \oplus \CT^{(2)}_{\rm decoupled})$};
\end{tikzpicture}}
&\scalebox{.6}{\begin{tikzpicture}
\node[] at (3.5,0){};
\draw[->] (4,0) -- (6, 0);
\node[text width=0.1cm](29) at (5, 0.3) {$S_1$};
\node[](30) at (5, -3.2) {};
\end{tikzpicture}}
&\scalebox{0.6}{\begin{tikzpicture}
\node[unode](100) at (-1,0){};
\node[fnode](101) at (-2,0){};
\node[fnode] (1) at (0,0){};
\node[unode] (2) at (1,0){};
\node[unode] (3) at (2,0){};
\node[unode] (4) at (3,0){};
\node[unode] (5) at (4,0){};
\node[unode] (6) at (5,0){};
\node[fnode] (7) at (6, 0){};
\node[fnode] (8) at (4,2){};
\draw[-] (1) -- (2);
\draw[-] (2)-- (3);
\draw[-] (3) -- (4);
\draw[-] (4) --(5);
\draw[-] (5) --(6);
\draw[-] (6) --(7);
\draw[-] (4) --(8);
\draw[-] (100) --(101);
\draw[-, thick, blue] (100) -- (-1, 1.5);
\draw[-, thick, blue] (2)--(1,1.5);
\draw[-, thick, blue] (4)--(3,1.5);
\draw[-, thick, blue] (5)--(4,1.5);
\draw[-, thick, blue] (-1,1.5)--(4,1.5);
\node[text width=3 cm](10) at (3, 2){\footnotesize{$(1, 1, -1, 1)$}};
\node[unode] (200) at (3, -2.2){};
\node[fnode] (201) at (4,-2.2){};
\draw[-] (200) --(201);
\draw[->, red, dashed] (3,-1.5) -- (200);
\draw[->, red, dashed] (3,-1.5) -- (100);
\draw[->, red, dashed] (3,-1.5) -- (3);
\draw[->, red, dashed] (3,-1.5) -- (4);
\draw[->, red, dashed] (3,-1.5) -- (6);
\node[text width=.1cm](10) [below=0.1 cm of 1]{1};
\node[text width=.1cm](11) [below=0.1cm of 2]{1};
\node[text width=.1cm](12) [below=0.1cm of 3]{1};
\node[text width=.1cm](13) [below=0.1cm of 4]{1};
\node[text width=.1cm](14) [below=0.1cm of 5]{1};
\node[text width=.1cm](15) [below=0.1cm of 6]{1};
\node[text width=.1cm](16) [below=0.1cm of 7]{1};
\node[text width=.1cm](17) [right=0.1cm of 8]{1};
\node[text width=.1cm](18) [below=0.1cm of 100]{1};
\node[text width=.1cm](19) [below=0.1cm of 101]{1};
\node[text width= 3 cm](20) at (3,-3){$(\CT^{(1)}_{\rm good} \oplus \CT^{(2)}_{\rm decoupled})$};
\end{tikzpicture}}
&\scalebox{.6}{\begin{tikzpicture}
\node[] at (3.5,0){};
\draw[->] (4,0) -- (6, 0);
\node[text width=0.1cm](29) at (5, 0.3) {$S_2$};
\node[](30) at (5, -3.2) {};
\end{tikzpicture}}
& \scalebox{0.6}{\begin{tikzpicture}
\node[unode](100) at (-1,0){};
\node[fnode](101) at (-2,0){};
\node[fnode] (1) at (0,0){};
\node[unode] (2) at (1,0){};
\node[unode] (3) at (2,0){};
\node[unode] (4) at (3,0){};
\node[unode] (5) at (4,0){};
\node[unode] (6) at (5,0){};
\node[fnode] (7) at (6, 0){};
\node[fnode] (8) at (4,2){};
\draw[-] (1) -- (2);
\draw[-] (2)-- (3);
\draw[-] (3) -- (4);
\draw[-] (4) --(5);
\draw[-] (5) --(6);
\draw[-] (6) --(7);
\draw[-] (4) --(8);
\draw[-] (100) --(101);
\draw[-, thick, blue] (100) -- (-1, 1.5);
\draw[-, thick, blue] (2)--(1,1.5);
\draw[-, thick, blue] (4)--(3,1.5);
\draw[-, thick, blue] (5)--(4,1.5);
\draw[-, thick, blue] (-1,1.5)--(4,1.5);
\node[text width=3 cm](10) at (3, 2){\footnotesize{$(1, 1, -1, 1)$}};
\draw[-, thick, blue] (100) -- (0, -1.5);
\draw[-, thick, blue] (3)--(2, -1.5);
\draw[-, thick, blue] (4)--(3, -1.5);
\draw[-, thick, blue] (6)--(5, -1.5);
\draw[-, thick, blue] (0,-1.5)--(5, -1.5);
\node[text width=3 cm](10) at (3, -2){\footnotesize{$(1, 1, -1, 1)$}};
\node[text width=.1cm](10) [below=0.1 cm of 1]{1};
\node[text width=.1cm](11) [below=0.1cm of 2]{1};
\node[text width=.1cm](12) [below=0.1cm of 3]{1};
\node[text width=.1cm](13) [below=0.1cm of 4]{1};
\node[text width=.1cm](14) [below=0.1cm of 5]{1};
\node[text width=.1cm](15) [below=0.1cm of 6]{1};
\node[text width=.1cm](16) [below=0.1cm of 7]{1};
\node[text width=.1cm](17) [right=0.1cm of 8]{1};
\node[text width=.1cm](18) [below=0.1cm of 100]{1};
\node[text width=.1cm](19) [below=0.1cm of 101]{1};
\node[text width=.2cm](20) at (2,-3){$(\CT_{\rm maximal})$};
\end{tikzpicture}} \\
\qquad & \qquad & \qquad & \qquad & \qquad  \\
\qquad & \qquad & \qquad & \qquad & \qquad  \\
\scalebox{0.6}{\begin{tikzpicture}
\node[unode] (1) at (2,0){};
\node[unode] (2) at (5,0){};
\node[fnode] (3) at (2,-1){};
\node[fnode] (4) at (1,-1){};
\node[fnode] (5) at (5,-1){};
\node[fnode] (6) at (6,-1){};
\node[fnode](101) at (5, -2){};
\node[fnode](102) at (6, -2){};
\node[fnode](103) at (1, -2){};
\node[fnode](104) at (0, -2){};
\draw[-] (1) -- (2);
\draw[-] (1)-- (3);
\draw[-] (1) -- (4);
\draw[-] (2) --(5);
\draw[-] (2) --(6);
\draw[-] (101) --(102);
\draw[-] (103) --(104);
\draw[->, red, dashed] (3,-2) -- (103);
\draw[->, red, dashed] (3,-2) -- (3);
\draw[->, red, dashed] (3,-2) -- (5);
\node[text width=0.1cm](20) [above=0.1 cm of 1]{1};
\node[text width=0.1 cm](21) [above=0.1cm of 2]{1};
\node[text width=0.1 cm](22) [right=0.1cm of 3]{1};
\node[text width=0.1 cm](23) [left=0.1cm of 4]{2};
\node[text width=0.1 cm](25) [left=0.1cm of 5]{1};
\node[text width=0.1 cm](26) [right=0.1cm of 6]{2};
\node[text width=.1cm](27) [above=0.1cm of 101]{1};
\node[text width=.1cm](28) [above=0.1cm of 102]{1};
\node[text width=.1cm](29) [above=0.1cm of 103]{1};
\node[text width=.1cm](30) [above=0.1cm of 104]{2};
\node[text width= 5 cm](31) at (3,-3){$(\wt{\CT}_{\rm good}\oplus \wt{\CT}^{(1)}_{\rm decoupled}\oplus \wt{\CT}^{(2)}_{\rm decoupled})$};
\end{tikzpicture}}
&\scalebox{.6}{\begin{tikzpicture}
\node[] at (3.5,0){};
\draw[->] (4,0) -- (6, 0);
\node[text width=0.1cm](29) at (5, 0.3) {$S_1$};
\node[](30) at (5, -2) {};
\end{tikzpicture}}
&\scalebox{0.6}{\begin{tikzpicture}
\node[unode] (1) at (2,0){};
\node[unode] (2) at (5,0){};
\node[unode] (3) at (3.5, -2){};
\node[fnode] (4) at (2,-1){};
\node[fnode] (5) at (1,-1){};
\node[fnode] (6) at (5,-1){};
\node[fnode] (7) at (6,-1){};
\node[fnode] (8) at (4.5,-2){};
\node[fnode] (9) at (2.5,-2){};
\node[fnode](101) at (6.5, -2){};
\node[fnode](102) at (7.5, -2){};
\draw[-] (1) -- (2);
\draw[-] (1) -- (3);
\draw[-] (2) -- (3);
\draw[-] (1) -- (4);
\draw[-] (1) -- (5);
\draw[-] (2) --(6);
\draw[-] (2) --(7);
\draw[-] (3) --(8);
\draw[-] (3) --(9);
\draw[-] (101) --(102);
\draw[->, red, dashed] (5.5,-2) -- (101);
\draw[->, red, dashed] (5.5,-2) -- (4);
\draw[->, red, dashed] (5.5,-2) -- (6);
\draw[->, red, dashed] (5.5,-2) -- (8);
\node[text width=0.1cm](20) [above=0.1 cm of 1]{1};
\node[text width=0.1 cm](21) [above=0.1cm of 2]{1};
\node[text width=0.1 cm](22) [below=0.1cm of 3]{1};
\node[text width=0.1 cm](23) [left=0.1cm of 4]{1};
\node[text width=0.1 cm](24) [left=0.1cm of 5]{1};
\node[text width=0.1 cm](25) [right=0.1cm of 6]{1};
\node[text width=0.1 cm](26) [right=0.1cm of 7]{1};
\node[text width=0.1 cm](27) [below=0.1cm of 8]{1};
\node[text width=0.1 cm](28) [below=0.1cm of 9]{1};
\node[text width=.1cm](29) [above=0.1cm of 101]{1};
\node[text width=.1cm](30) [above=0.1cm of 102]{1};
\node[text width= 3 cm](31) at (4,-3){$(\wt{\CT}^{(1)}_{\rm good} \oplus \wt{\CT}^{(2)}_{\rm decoupled})$};
\end{tikzpicture}}
&\scalebox{.6}{\begin{tikzpicture}
\node[] at (3.5,0){};
\draw[->] (4,0) -- (6, 0);
\node[text width=0.1cm](29) at (5, 0.3) {$S_2$};
\node[](30) at (5, -2) {};
\end{tikzpicture}}
& \scalebox{0.6}{\begin{tikzpicture}
\node[fnode] (1) at (0,0){};
\node[unode] (2) at (2,0){};
\node[unode] (3) at (4,0){};
\node[fnode] (4) at (6,0){};
\node[fnode] (5) at (0,-2){};
\node[unode] (6) at (2,-2){};
\node[unode] (7) at (4,-2){};
\node[fnode] (8) at (6,-2){};
\draw[-] (1) -- (2);
\draw[-] (2)-- (3);
\draw[-] (3) -- (4);
\draw[-] (2) --(6);
\draw[-] (2) --(7);
\draw[-] (3) --(6);
\draw[-] (3) --(7);
\draw[-] (5) --(6);
\draw[-] (6) --(7);
\draw[-] (7) --(8);
\node[text width=0.1cm](20) [above=0.1 cm of 1]{1};
\node[text width=0.1 cm](21) [above=0.1cm of 2]{1};
\node[text width=0.1 cm](22) [above=0.1cm of 3]{1};
\node[text width=0.1 cm](23) [above=0.1cm of 4]{1};
\node[text width=0.1 cm](24) [below=0.1cm of 5]{1};
\node[text width=0.1 cm](25) [below=0.1cm of 6]{1};
\node[text width=0.1 cm](26) [below=0.1cm of 7]{1};
\node[text width=0.1 cm](27) [below=0.1cm of 8]{1};
\node[text width=.2cm](30) at (3,-3){$(\wt{\CT}_{\rm maximal})$};
\end{tikzpicture}}
\end{tabular}
\caption{\footnotesize{Construction of the 3d mirror of the $(A_4, A_4)$ SCFT.}}
\label{SOP-Ex2}
\end{figure*}
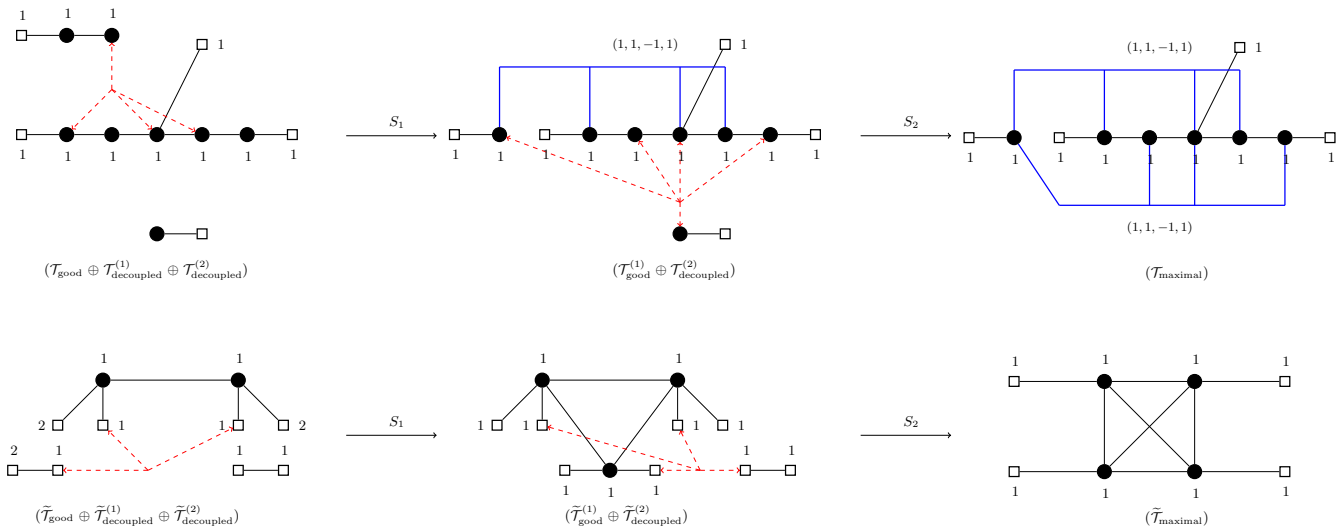

Next, consider the case of $(A_4, A_4)$ SCFT. In this case, the theory $\CT_{\rm good}$ and its 3d mirror 
$\wt{\CT}_{\rm good}$ are given as follows: 

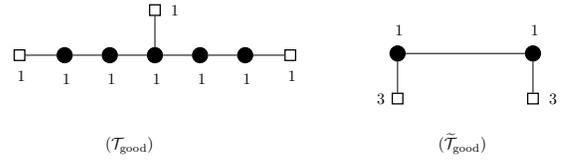
\begin{center}
\begin{tabular}{ccc}
\scalebox{0.6}{\begin{tikzpicture}
\node[fnode] (1) at (0,0){};
\node[unode] (2) at (1,0){};
\node[unode] (3) at (2,0){};
\node[unode] (4) at (3,0){};
\node[unode] (5) at (4,0){};
\node[unode] (6) at (5,0){};
\node[fnode] (7) at (6, 0){};
\node[fnode] (8) at (3,1){};
\draw[-] (1) -- (2);
\draw[-] (2)-- (3);
\draw[-] (3) -- (4);
\draw[-] (4) --(5);
\draw[-] (5) --(6);
\draw[-] (6) --(7);
\draw[-] (4) --(8);
\node[text width=.1cm](10) [below=0.1 cm of 1]{1};
\node[text width=.1cm](11) [below=0.1cm of 2]{1};
\node[text width=.1cm](12) [below=0.1cm of 3]{1};
\node[text width=.1cm](13) [below=0.1cm of 4]{1};
\node[text width=.1cm](14) [below=0.1cm of 5]{1};
\node[text width=.1cm](15) [below=0.1cm of 6]{1};
\node[text width=.1cm](16) [below=0.1cm of 7]{1};
\node[text width=.1cm](17) [right=0.1cm of 8]{1};
\node[text width=.2cm](20) at (2,-2){$(\CT_{\rm good})$};
\end{tikzpicture}}
& \qquad \qquad
& \scalebox{0.6}{\begin{tikzpicture}
\node[unode] (1) at (2,0){};
\node[unode] (2) at (5,0){};
\node[fnode] (3) at (2,-1){};
\node[fnode] (4) at (5,-1){};
\draw[-] (1) -- (2);
\draw[-] (1)-- (3);
\draw[-] (2) -- (4);
\node[text width=0.1cm](20) [above=0.1 cm of 1]{1};
\node[text width=0.1 cm](21) [above=0.1cm of 2]{1};
\node[text width=0.1 cm](22) [left=0.1cm of 3]{3};
\node[text width=0.1 cm](23) [right=0.1cm of 4]{3};
\node[text width=.2cm](31) at (3,-2){$(\wt{\CT}_{\rm good})$};
\end{tikzpicture}}
\end{tabular}
\end{center}

Starting from the mirror pair $({\CT}_{\rm good}, \wt{\CT}_{\rm good})$, the 3d mirror of 
$\CT_{\rm maximal}$ can be generated by implementing a sequence of two QFT operations 
which are shown in \Figref{SOP-Ex2}. To begin with, we add to the  theory ${\CT}_{\rm good}$ a 
decoupled theory $\CT^{(1)}_{\rm decoupled}$ where $\CT^{(1)}_{\rm decoupled}$ 
is an Abelian quiver tail of the following form:
\begin{center}
\scalebox{0.6}{\begin{tikzpicture}
\node[text width=2 cm](0) at (-2,0){$\CT^{(1)}_{\rm decoupled}:$};
\node[fnode] (1) at (0,0){};
\node[unode] (2) at (1,0){};
\node[unode] (3) at (2,0){};
\draw[-] (1) -- (2);
\draw[-] (2)-- (3);
\node[text width=0.1cm](20) [above=0.1 cm of 1]{1};
\node[text width=0.1 cm](21) [above=0.1cm of 2]{1};
\node[text width=0.1 cm](22) [above=0.1cm of 3]{1};
\end{tikzpicture}}. 
\end{center} 
On the dual side, we add to the theory $\wt{\CT}_{\rm good}$ the 3d mirror of $\CT^{(1)}_{\rm decoupled}$ -- 
a collection of 2 free hypermultiplets, which we represent as:
\begin{center}
\scalebox{0.6}{\begin{tikzpicture}
\node[text width=2 cm](0) at (-2,0){$\wt{\CT}^{(1)}_{\rm decoupled}:$};
\node[fnode] (1) at (0,0){};
\node[fnode] (2) at (1,0){};
\draw[-] (1) -- (2);
\node[text width=0.1cm](20) [above=0.1 cm of 1]{1};
\node[text width=0.1 cm](21) [above=0.1cm of 2]{2};
\end{tikzpicture}}.
\end{center}
The operation $S_1$ then acts on the combined theory $\CT_{\rm good} \oplus \CT^{(1)}_{\rm decoupled}$ by gauging a 
$\fru(1)$ subalgebra of the topological symmetry corresponding to the generator:
\be
J = {J}^{(1)}_2 + J_1 - J_3 + J_4,
\ee
where ${J}^{(1)}_k$ is the generator of the topological symmetry for the $U(1)_k$ gauge node ($k=1,2$ increasing from left to right) 
in $\CT^{(1)}_{\rm decoupled}$, while $J_i$ is the generator of the topological symmetry for the $i$-th $U(1)$ gauge node 
($i=1,\ldots,5$ increasing from left to right) in ${\CT}_{\rm good}$. This operation gives the quiver ${\CT}^{(1)}_{\rm good}$, 
as shown in the top row of \Figref{SOP-Ex2}. The operation $S_1$ on the dual side is shown in the bottom row and it leads to the quiver 
$\wt{\CT}^{(1)}_{\rm good}$.\\

In the next step, we introduce the second decoupled theory ${\CT}^{(2)}_{\rm decoupled}$ which is a $U(1)$ theory with a single 
hypermultiplet. On the mirror side, we add a free hypermultiplet which is the 3d mirror of ${\CT}^{(2)}_{\rm decoupled}$. 
The second operation $S_2$ then acts on the theory ${\CT}^{(1)}_{\rm good}\oplus \CT^{(2)}_{\rm decoupled}$ by gauging a 
$\fru(1)$ subalgebra of the topological symmetry corresponding to the generator:
\be
J' = {J}^{(2)} + {J}^{(1)}_1 + J_2 - J_3 + J_4,
\ee
where ${J}^{(2)}$ is the generator of the topological symmetry for the $U(1)$ gauge node in $\CT^{(2)}_{\rm decoupled}$. 
This operation gives the theory $\CT_{\rm maximal}$, as shown in the top row of \Figref{SOP-Ex2}. 
On the mirror side, the operation gives the quiver $\wt{\CT}_{\rm maximal}$ - the 3d mirror of $\CT_{\rm maximal}$. The quiver  
$\wt{\CT}_{\rm maximal}$ can be readily identified as the complete graph with five vertices and edge multiplicity 1, after decoupling a 
$U(1)$ vector multiplet associated with one of the vertices:

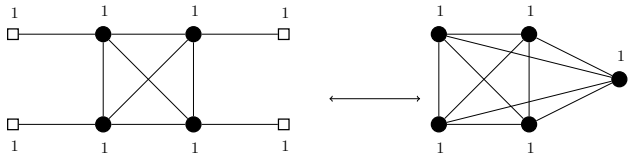
\begin{center}
\begin{tabular}{ccc}
\scalebox{0.6}{\begin{tikzpicture}
\node[fnode] (1) at (0,0){};
\node[unode] (2) at (2,0){};
\node[unode] (3) at (4,0){};
\node[fnode] (4) at (6,0){};
\node[fnode] (5) at (0,-2){};
\node[unode] (6) at (2,-2){};
\node[unode] (7) at (4,-2){};
\node[fnode] (8) at (6,-2){};
\draw[-] (1) -- (2);
\draw[-] (2)-- (3);
\draw[-] (3) -- (4);
\draw[-] (2) --(6);
\draw[-] (2) --(7);
\draw[-] (3) --(6);
\draw[-] (3) --(7);
\draw[-] (5) --(6);
\draw[-] (6) --(7);
\draw[-] (7) --(8);
\node[text width=0.1cm](20) [above=0.1 cm of 1]{1};
\node[text width=0.1 cm](21) [above=0.1cm of 2]{1};
\node[text width=0.1 cm](22) [above=0.1cm of 3]{1};
\node[text width=0.1 cm](23) [above=0.1cm of 4]{1};
\node[text width=0.1 cm](24) [below=0.1cm of 5]{1};
\node[text width=0.1 cm](25) [below=0.1cm of 6]{1};
\node[text width=0.1 cm](26) [below=0.1cm of 7]{1};
\node[text width=0.1 cm](27) [below=0.1cm of 8]{1};
\end{tikzpicture}}
& \scalebox{.6}{\begin{tikzpicture}
\node[] at (3.5,0){};
\draw[<->] (4,0) -- (6, 0);
\node[](30) at (5, -1.2) {};
\end{tikzpicture}}
& \scalebox{0.6}{\begin{tikzpicture}
\node[unode](1) at (6,-1){};
\node[unode] (2) at (2,0){};
\node[unode] (3) at (4,0){};
\node[unode] (6) at (2,-2){};
\node[unode] (7) at (4,-2){};
\draw[-] (1) -- (2);
\draw[-] (2)-- (3);
\draw[-] (3) -- (1);
\draw[-] (2) --(6);
\draw[-] (2) --(7);
\draw[-] (3) --(6);
\draw[-] (3) --(7);
\draw[-] (1) --(6);
\draw[-] (6) --(7);
\draw[-] (7) --(1);
\node[text width=0.1cm](20) [above=0.1 cm of 1]{1};
\node[text width=0.1 cm](21) [above=0.1cm of 2]{1};
\node[text width=0.1 cm](22) [above=0.1cm of 3]{1};
\node[text width=0.1 cm](25) [below=0.1cm of 6]{1};
\node[text width=0.1 cm](26) [below=0.1cm of 7]{1};
\end{tikzpicture}}
\end{tabular}
\end{center}

Given the 3d mirror, one can check that the rank of its HB global symmetry algebra is $\text{rk}(\frg_{\rm H}(\wt{\CT}_{\rm maximal}))= 6$. 
Therefore, the rank of the HB global symmetry algebra of the mirror (or equivalently the number of independent mass parameters) precisely 
matches the rank of the CB global symmetry algebra of $\CT_{\rm maximal}$ (or equivalently the number of FI parameters). 
This confirms that the maximal unitary quiver makes all the resolution/deformation parameters of the 4d HB manifest.\\

\section{Conclusions and Future Directions}\label{sec: concl}

We have shown that the 4d HB for a large class of  Argyres-Douglas theories generically admits multiple \hk quotient 
realizations as Higgs branches of 3d $\CN=4$ quiver gauge theories. The 3d quivers obey IR N-ality : they flow 
to the same 3d SCFT in the deep IR without any exchange of the CB and the HB, and may be related to each other 
by a sequence of four distinct types of quiver mutations. The 3d quivers differ amongst each other in the UV-manifest rank of the CB global symmetry, but 
there exists a special subset of quivers -- the maximal unitary quivers -- for which the ranks of the UV-manifest 
symmetry and the emergent IR symmetry match. For this subset of quivers, all the resolution/deformation parameters 
of the HB geometry are classically manifest as FI parameters. Given the maximal unitary quiver(s), one can deploy a 
sequence of Abelian QFT operations to engineer the 3d mirror of the SCFT. 

Starting from the Type IIB description of the 4d SCFT, we have presented a systematic construction to determine the associated 
N-al set of 3d quivers including the maximal unitary ones, leading to the 3d mirror. We should emphasize that 
our construction is limited to the subclass of of 4d SCFTs which after circle reduction, in a way described in \Secref{sec: circlered}, 
gives a 3d U-SU quiver (quiver with unitary/special unitary gauge nodes and matter in fundamental/bifundamental representation) 
with at least a single balanced special unitary node. This is the class of theories for which 
our tools of IR N-ality and quiver mutations, discussed in \Secref{sec: maxquiv}, 
can be applied to uncover 
the quotient N-ality. A generic quiver in this set of 3d N-al theories is a unitary/special unitary 
quiver gauge theory with Abelian hypermultiplets -- hypermultiplets transforming in powers of the 
determinant/anti-determinant representation of the unitary gauge groups. If all the special unitary 
nodes in the U-SU quiver are balanced, the maximal unitary quiver consists of unitary gauge nodes only. 

We demonstrate this construction for the $(A_k, A_k)$ SCFTs, focusing on the $k=3$ and the $k=4$ cases. For this family of 
SCFTs, all special unitary nodes in the U-SU quiver are balanced. For the case $k=3$, we find a duality of quotient realizations, 
while for the case $k=4$ we find a quadrality up to some discrete identification. The maximal unitary quiver 
turns out to be an Abelian quiver gauge theory with $\frac{k(k-1)}{2}$ $U(1)$ gauge nodes and with $(k-2)$ Abelian hypermultiplets, 
which we write down explicitly for $k=3$ and $k=4$. In addition, the 3d mirror is shown to be a complete graph of $k+1$ vertices 
with edge multiplicity 1. One can check that the rank of the UV-manifest CB symmetry (or equivalently the number of FI parameters) 
of the maximal unitary quiver precisely matches the rank of the UV-manifest HB symmetry (or equivalently the number of independent 
mass parameters) as expected. 

In \Appref{app: 3dLagDp} -- \Appref{app: DpEx}, we extend our construction to the $D^b_p(SU(N))$ SCFTs. The circle reduction 
of these SCFTs leads U-SU quivers which we review in \Appref{app: 3dLagDp}. In \Appref{app: Nalitycond}, we determine the 
subclass of SCFTs which admit \hk quotient N-ality. We show that this subclass coincides with the subclass of non-isolated SCFTs of 
the Argyres-Douglas type i.e. non-isolated SCFTs with fractional scaling dimensions for the CB operators. Every special unitary node 
in the U-SU quiver associated with any SCFT of this subclass is balanced. The maximal unitary quivers are therefore guaranteed 
to be unitary quiver gauge theories with Abelian hypermultiplets. In \Appref{app: DpEx}, we work out the maximal unitary quivers 
for the $D_6(SU(9))$ SCFT as an explicit example, and then construct the 3d mirror. This construction of the 3d mirror is a surprisingly 
simple alternative to the Xie-Yau construction \cite{Xie:2016uqq} that appears in the literature \cite{Giacomelli:2020ryy}.\\

It should be evident from the above discussion that our construction can be readily applied to trinion SCFTs 
built out of the $D^b_p(SU(N))$ theories, some of which can also be realized as $(A,G)$ theories. However, 
unlike the families of SCFTs studied in this paper, the U-SU quivers associated to trinion SCFTs are generically 
bad theories in the Gaiotto-Witten sense, and quotient N-ality for such theories needs a more careful analysis.
It is also tantalizing to contemplate whether generic $(G, G')$ theories admit quotient N-alities similar to the 
$(A,A')$ theories. This will require understanding IR N-alities in 3d SO-Sp quiver gauge theories with emergent 
CB symmetry. \\

\noindent \textbf{Acknowledgements} The author would like to thank Simone Giacomelli, Amihay Hanany, Zohar Komargodski, and Andrew Neitzke 
for correspondence and discussion on related issues. The author would like to thank the organizers of the program ``Hyperkahler quotients, singularities, 
and quivers" at the Simons Center for Geometry and Physics where some of the results presented in this work were discussed. The author gratefully 
acknowledges the hospitality of the Simons Summer Workshop 2023 during which part of the work was done.

\bibliographystyle{apsrev4-1}
\bibliography{cpn1-1}

\begin{thebibliography}{22}%
\makeatletter
\providecommand \@ifxundefined [1]{%
 \@ifx{#1\undefined}
}%
\providecommand \@ifnum [1]{%
 \ifnum #1\expandafter \@firstoftwo
 \else \expandafter \@secondoftwo
 \fi
}%
\providecommand \@ifx [1]{%
 \ifx #1\expandafter \@firstoftwo
 \else \expandafter \@secondoftwo
 \fi
}%
\providecommand \natexlab [1]{#1}%
\providecommand \enquote  [1]{``#1''}%
\providecommand \bibnamefont  [1]{#1}%
\providecommand \bibfnamefont [1]{#1}%
\providecommand \citenamefont [1]{#1}%
\providecommand \href@noop [0]{\@secondoftwo}%
\providecommand \href [0]{\begingroup \@sanitize@url \@href}%
\providecommand \@href[1]{\@@startlink{#1}\@@href}%
\providecommand \@@href[1]{\endgroup#1\@@endlink}%
\providecommand \@sanitize@url [0]{\catcode `\\12\catcode `\$12\catcode
  `\&12\catcode `\#12\catcode `\^12\catcode `\_12\catcode `\%12\relax}%
\providecommand \@@startlink[1]{}%
\providecommand \@@endlink[0]{}%
\providecommand \url  [0]{\begingroup\@sanitize@url \@url }%
\providecommand \@url [1]{\endgroup\@href {#1}{\urlprefix }}%
\providecommand \urlprefix  [0]{URL }%
\providecommand \Eprint [0]{\href }%
\providecommand \doibase [0]{http://dx.doi.org/}%
\providecommand \selectlanguage [0]{\@gobble}%
\providecommand \bibinfo  [0]{\@secondoftwo}%
\providecommand \bibfield  [0]{\@secondoftwo}%
\providecommand \translation [1]{[#1]}%
\providecommand \BibitemOpen [0]{}%
\providecommand \bibitemStop [0]{}%
\providecommand \bibitemNoStop [0]{.\EOS\space}%
\providecommand \EOS [0]{\spacefactor3000\relax}%
\providecommand \BibitemShut  [1]{\csname bibitem#1\endcsname}%
\let\auto@bib@innerbib\@empty
\bibitem [{\citenamefont {Argyres}\ \emph {et~al.}(1996)\citenamefont
  {Argyres}, \citenamefont {Ronen~Plesser}, \citenamefont {Seiberg},\ and\
  \citenamefont {Witten}}]{Argyres:1995xn}%
  \BibitemOpen
  \bibfield  {author} {\bibinfo {author} {\bibfnamefont {P.~C.}\ \bibnamefont
  {Argyres}}, \bibinfo {author} {\bibfnamefont {M.}~\bibnamefont
  {Ronen~Plesser}}, \bibinfo {author} {\bibfnamefont {N.}~\bibnamefont
  {Seiberg}}, \ and\ \bibinfo {author} {\bibfnamefont {E.}~\bibnamefont
  {Witten}},\ }\href {\doibase 10.1016/0550-3213(95)00671-0} {\bibfield
  {journal} {\bibinfo  {journal} {Nucl. Phys.}\ }\textbf {\bibinfo {volume}
  {B461}},\ \bibinfo {pages} {71} (\bibinfo {year} {1996})},\ \Eprint
  {http://arxiv.org/abs/hep-th/9511154} {arXiv:hep-th/9511154} \BibitemShut
  {NoStop}%
\bibitem [{\citenamefont {Argyres}\ and\ \citenamefont
  {Douglas}(1995)}]{Argyres:1995jj}%
  \BibitemOpen
  \bibfield  {author} {\bibinfo {author} {\bibfnamefont {P.~C.}\ \bibnamefont
  {Argyres}}\ and\ \bibinfo {author} {\bibfnamefont {M.~R.}\ \bibnamefont
  {Douglas}},\ }\href {\doibase 10.1016/0550-3213(95)00281-V} {\bibfield
  {journal} {\bibinfo  {journal} {Nucl. Phys.}\ }\textbf {\bibinfo {volume}
  {B448}},\ \bibinfo {pages} {93} (\bibinfo {year} {1995})},\ \Eprint
  {http://arxiv.org/abs/hep-th/9505062} {arXiv:hep-th/9505062} \BibitemShut
  {NoStop}%
\bibitem [{\citenamefont {Klemm}\ \emph {et~al.}(1996)\citenamefont {Klemm},
  \citenamefont {Lerche}, \citenamefont {Mayr}, \citenamefont {Vafa},\ and\
  \citenamefont {Warner}}]{Klemm:1996bj}%
  \BibitemOpen
  \bibfield  {author} {\bibinfo {author} {\bibfnamefont {A.}~\bibnamefont
  {Klemm}}, \bibinfo {author} {\bibfnamefont {W.}~\bibnamefont {Lerche}},
  \bibinfo {author} {\bibfnamefont {P.}~\bibnamefont {Mayr}}, \bibinfo {author}
  {\bibfnamefont {C.}~\bibnamefont {Vafa}}, \ and\ \bibinfo {author}
  {\bibfnamefont {N.~P.}\ \bibnamefont {Warner}},\ }\href {\doibase
  10.1016/0550-3213(96)00353-7} {\bibfield  {journal} {\bibinfo  {journal}
  {Nucl. Phys. B}\ }\textbf {\bibinfo {volume} {477}},\ \bibinfo {pages} {746}
  (\bibinfo {year} {1996})},\ \Eprint {http://arxiv.org/abs/hep-th/9604034}
  {arXiv:hep-th/9604034} \BibitemShut {NoStop}%
\bibitem [{\citenamefont {Katz}\ \emph {et~al.}(1997)\citenamefont {Katz},
  \citenamefont {Klemm},\ and\ \citenamefont {Vafa}}]{Katz:1996fh}%
  \BibitemOpen
  \bibfield  {author} {\bibinfo {author} {\bibfnamefont {S.~H.}\ \bibnamefont
  {Katz}}, \bibinfo {author} {\bibfnamefont {A.}~\bibnamefont {Klemm}}, \ and\
  \bibinfo {author} {\bibfnamefont {C.}~\bibnamefont {Vafa}},\ }\href {\doibase
  10.1016/S0550-3213(97)00282-4} {\bibfield  {journal} {\bibinfo  {journal}
  {Nucl. Phys. B}\ }\textbf {\bibinfo {volume} {497}},\ \bibinfo {pages} {173}
  (\bibinfo {year} {1997})},\ \Eprint {http://arxiv.org/abs/hep-th/9609239}
  {arXiv:hep-th/9609239} \BibitemShut {NoStop}%
\bibitem [{\citenamefont {Shapere}\ and\ \citenamefont
  {Vafa}(1999)}]{Shapere:1999xr}%
  \BibitemOpen
  \bibfield  {author} {\bibinfo {author} {\bibfnamefont {A.~D.}\ \bibnamefont
  {Shapere}}\ and\ \bibinfo {author} {\bibfnamefont {C.}~\bibnamefont {Vafa}},\
  }\href@noop {} {\  (\bibinfo {year} {1999})},\ \Eprint
  {http://arxiv.org/abs/hep-th/9910182} {arXiv:hep-th/9910182} \BibitemShut
  {NoStop}%
\bibitem [{\citenamefont {Argyres}\ \emph {et~al.}(2012)\citenamefont
  {Argyres}, \citenamefont {Maruyoshi},\ and\ \citenamefont
  {Tachikawa}}]{Argyres:2012fu}%
  \BibitemOpen
  \bibfield  {author} {\bibinfo {author} {\bibfnamefont {P.~C.}\ \bibnamefont
  {Argyres}}, \bibinfo {author} {\bibfnamefont {K.}~\bibnamefont {Maruyoshi}},
  \ and\ \bibinfo {author} {\bibfnamefont {Y.}~\bibnamefont {Tachikawa}},\
  }\href {\doibase 10.1007/JHEP10(2012)054} {\bibfield  {journal} {\bibinfo
  {journal} {JHEP}\ }\textbf {\bibinfo {volume} {10}},\ \bibinfo {pages} {054}
  (\bibinfo {year} {2012})},\ \Eprint {http://arxiv.org/abs/1206.4700}
  {arXiv:1206.4700 [hep-th]} \BibitemShut {NoStop}%
\bibitem [{\citenamefont {Gaiotto}\ \emph {et~al.}(2010)\citenamefont
  {Gaiotto}, \citenamefont {Neitzke},\ and\ \citenamefont
  {Tachikawa}}]{Gaiotto:2009jjh}%
  \BibitemOpen
  \bibfield  {author} {\bibinfo {author} {\bibfnamefont {D.}~\bibnamefont
  {Gaiotto}}, \bibinfo {author} {\bibfnamefont {A.}~\bibnamefont {Neitzke}}, \
  and\ \bibinfo {author} {\bibfnamefont {Y.}~\bibnamefont {Tachikawa}},\ }\href
  {\doibase 10.1007/s00220-009-0938-6} {\bibfield  {journal} {\bibinfo
  {journal} {Commun. Math. Phys.}\ }\textbf {\bibinfo {volume} {294}},\
  \bibinfo {pages} {389} (\bibinfo {year} {2010})},\ \Eprint
  {http://arxiv.org/abs/0810.4541} {arXiv:0810.4541 [hep-th]} \BibitemShut
  {NoStop}%
\bibitem [{\citenamefont {Cecotti}\ \emph {et~al.}(2010)\citenamefont
  {Cecotti}, \citenamefont {Neitzke},\ and\ \citenamefont
  {Vafa}}]{Cecotti:2010fi}%
  \BibitemOpen
  \bibfield  {author} {\bibinfo {author} {\bibfnamefont {S.}~\bibnamefont
  {Cecotti}}, \bibinfo {author} {\bibfnamefont {A.}~\bibnamefont {Neitzke}}, \
  and\ \bibinfo {author} {\bibfnamefont {C.}~\bibnamefont {Vafa}},\ }\href@noop
  {} {\  (\bibinfo {year} {2010})},\ \Eprint {http://arxiv.org/abs/1006.3435}
  {arXiv:1006.3435 [hep-th]} \BibitemShut {NoStop}%
\bibitem [{\citenamefont {Nanopoulos}\ and\ \citenamefont
  {Xie}(2011)}]{Nanopoulos:2010bv}%
  \BibitemOpen
  \bibfield  {author} {\bibinfo {author} {\bibfnamefont {D.}~\bibnamefont
  {Nanopoulos}}\ and\ \bibinfo {author} {\bibfnamefont {D.}~\bibnamefont
  {Xie}},\ }\href {\doibase 10.1007/JHEP05(2011)071} {\bibfield  {journal}
  {\bibinfo  {journal} {JHEP}\ }\textbf {\bibinfo {volume} {05}},\ \bibinfo
  {pages} {071} (\bibinfo {year} {2011})},\ \Eprint
  {http://arxiv.org/abs/1011.1911} {arXiv:1011.1911 [hep-th]} \BibitemShut
  {NoStop}%
\bibitem [{\citenamefont {Closset}\ \emph {et~al.}(2021)\citenamefont
  {Closset}, \citenamefont {Giacomelli}, \citenamefont {Schafer-Nameki},\ and\
  \citenamefont {Wang}}]{Closset:2020afy}%
  \BibitemOpen
  \bibfield  {author} {\bibinfo {author} {\bibfnamefont {C.}~\bibnamefont
  {Closset}}, \bibinfo {author} {\bibfnamefont {S.}~\bibnamefont {Giacomelli}},
  \bibinfo {author} {\bibfnamefont {S.}~\bibnamefont {Schafer-Nameki}}, \ and\
  \bibinfo {author} {\bibfnamefont {Y.-N.}\ \bibnamefont {Wang}},\ }\href
  {\doibase 10.1007/JHEP05(2021)274} {\bibfield  {journal} {\bibinfo  {journal}
  {JHEP}\ }\textbf {\bibinfo {volume} {05}},\ \bibinfo {pages} {274} (\bibinfo
  {year} {2021})},\ \Eprint {http://arxiv.org/abs/2012.12827} {arXiv:2012.12827
  [hep-th]} \BibitemShut {NoStop}%
\bibitem [{\citenamefont {Giacomelli}\ \emph {et~al.}(2021)\citenamefont
  {Giacomelli}, \citenamefont {Mekareeya},\ and\ \citenamefont
  {Sacchi}}]{Giacomelli:2020ryy}%
  \BibitemOpen
  \bibfield  {author} {\bibinfo {author} {\bibfnamefont {S.}~\bibnamefont
  {Giacomelli}}, \bibinfo {author} {\bibfnamefont {N.}~\bibnamefont
  {Mekareeya}}, \ and\ \bibinfo {author} {\bibfnamefont {M.}~\bibnamefont
  {Sacchi}},\ }\href {\doibase 10.1007/JHEP03(2021)242} {\bibfield  {journal}
  {\bibinfo  {journal} {JHEP}\ }\textbf {\bibinfo {volume} {03}},\ \bibinfo
  {pages} {242} (\bibinfo {year} {2021})},\ \Eprint
  {http://arxiv.org/abs/2012.12852} {arXiv:2012.12852 [hep-th]} \BibitemShut
  {NoStop}%
\bibitem [{\citenamefont {Dey}(2023{\natexlab{a}})}]{Dey:2023xhq}%
  \BibitemOpen
  \bibfield  {author} {\bibinfo {author} {\bibfnamefont {A.}~\bibnamefont
  {Dey}},\ }\href@noop {} {\  (\bibinfo {year} {2023}{\natexlab{a}})},\ \Eprint
  {http://arxiv.org/abs/2307.02525} {arXiv:2307.02525 [hep-th]} \BibitemShut
  {NoStop}%
\bibitem [{\citenamefont {Dey}(2022)}]{Dey:2022abc}%
  \BibitemOpen
  \bibfield  {author} {\bibinfo {author} {\bibfnamefont {A.}~\bibnamefont
  {Dey}},\ }\href@noop {} {\  (\bibinfo {year} {2022})},\ \Eprint
  {http://arxiv.org/abs/2210.09319} {arXiv:2210.09319 [hep-th]} \BibitemShut
  {NoStop}%
\bibitem [{\citenamefont {Cecotti}\ and\ \citenamefont
  {Del~Zotto}(2013)}]{Cecotti:2012jx}%
  \BibitemOpen
  \bibfield  {author} {\bibinfo {author} {\bibfnamefont {S.}~\bibnamefont
  {Cecotti}}\ and\ \bibinfo {author} {\bibfnamefont {M.}~\bibnamefont
  {Del~Zotto}},\ }\href {\doibase 10.1007/JHEP01(2013)191} {\bibfield
  {journal} {\bibinfo  {journal} {JHEP}\ }\textbf {\bibinfo {volume} {01}},\
  \bibinfo {pages} {191} (\bibinfo {year} {2013})},\ \Eprint
  {http://arxiv.org/abs/1210.2886} {arXiv:1210.2886 [hep-th]} \BibitemShut
  {NoStop}%
\bibitem [{\citenamefont {Cecotti}\ \emph {et~al.}(2013)\citenamefont
  {Cecotti}, \citenamefont {Del~Zotto},\ and\ \citenamefont
  {Giacomelli}}]{Cecotti:2013lda}%
  \BibitemOpen
  \bibfield  {author} {\bibinfo {author} {\bibfnamefont {S.}~\bibnamefont
  {Cecotti}}, \bibinfo {author} {\bibfnamefont {M.}~\bibnamefont {Del~Zotto}},
  \ and\ \bibinfo {author} {\bibfnamefont {S.}~\bibnamefont {Giacomelli}},\
  }\href {\doibase 10.1007/JHEP04(2013)153} {\bibfield  {journal} {\bibinfo
  {journal} {JHEP}\ }\textbf {\bibinfo {volume} {04}},\ \bibinfo {pages} {153}
  (\bibinfo {year} {2013})},\ \Eprint {http://arxiv.org/abs/1303.3149}
  {arXiv:1303.3149 [hep-th]} \BibitemShut {NoStop}%
\bibitem [{\citenamefont {Dey}()}]{Dey:2024gen}%
  \BibitemOpen
  \bibfield  {author} {\bibinfo {author} {\bibfnamefont {A.}~\bibnamefont
  {Dey}},\ }\href@noop {} {\ }\Eprint {http://arxiv.org/abs/In preparation} {In
  preparation} \BibitemShut {NoStop}%
\bibitem [{\citenamefont {Xie}\ and\ \citenamefont {Yau}(2017)}]{Xie:2017vaf}%
  \BibitemOpen
  \bibfield  {author} {\bibinfo {author} {\bibfnamefont {D.}~\bibnamefont
  {Xie}}\ and\ \bibinfo {author} {\bibfnamefont {S.-T.}\ \bibnamefont {Yau}},\
  }\href@noop {} {\  (\bibinfo {year} {2017})},\ \Eprint
  {http://arxiv.org/abs/1701.01123} {arXiv:1701.01123 [hep-th]} \BibitemShut
  {NoStop}%
\bibitem [{\citenamefont {Buican}\ \emph {et~al.}(2015)\citenamefont {Buican},
  \citenamefont {Giacomelli}, \citenamefont {Nishinaka},\ and\ \citenamefont
  {Papageorgakis}}]{Buican:2014hfa}%
  \BibitemOpen
  \bibfield  {author} {\bibinfo {author} {\bibfnamefont {M.}~\bibnamefont
  {Buican}}, \bibinfo {author} {\bibfnamefont {S.}~\bibnamefont {Giacomelli}},
  \bibinfo {author} {\bibfnamefont {T.}~\bibnamefont {Nishinaka}}, \ and\
  \bibinfo {author} {\bibfnamefont {C.}~\bibnamefont {Papageorgakis}},\ }\href
  {\doibase 10.1007/JHEP02(2015)185} {\bibfield  {journal} {\bibinfo  {journal}
  {JHEP}\ }\textbf {\bibinfo {volume} {02}},\ \bibinfo {pages} {185} (\bibinfo
  {year} {2015})},\ \Eprint {http://arxiv.org/abs/1411.6026} {arXiv:1411.6026
  [hep-th]} \BibitemShut {NoStop}%
\bibitem [{\citenamefont {Dey}(2023{\natexlab{b}})}]{Dey:2023vrt}%
  \BibitemOpen
  \bibfield  {author} {\bibinfo {author} {\bibfnamefont {A.}~\bibnamefont
  {Dey}},\ }\href@noop {} {\  (\bibinfo {year} {2023}{\natexlab{b}})},\ \Eprint
  {http://arxiv.org/abs/2308.12337} {arXiv:2308.12337 [hep-th]} \BibitemShut
  {NoStop}%
\bibitem [{\citenamefont {Witten}(2003)}]{Witten:2003ya}%
  \BibitemOpen
  \bibfield  {author} {\bibinfo {author} {\bibfnamefont {E.}~\bibnamefont
  {Witten}},\ }\href@noop {} {\  (\bibinfo {year} {2003})},\ \Eprint
  {http://arxiv.org/abs/hep-th/0307041} {arXiv:hep-th/0307041 [hep-th]}
  \BibitemShut {NoStop}%
\bibitem [{\citenamefont {Dey}(2021)}]{Dey:2020hfe}%
  \BibitemOpen
  \bibfield  {author} {\bibinfo {author} {\bibfnamefont {A.}~\bibnamefont
  {Dey}},\ }\href {\doibase 10.1007/JHEP07(2021)199} {\bibfield  {journal}
  {\bibinfo  {journal} {JHEP}\ }\textbf {\bibinfo {volume} {07}},\ \bibinfo
  {pages} {199} (\bibinfo {year} {2021})},\ \Eprint
  {http://arxiv.org/abs/2004.09738} {arXiv:2004.09738 [hep-th]} \BibitemShut
  {NoStop}%
\bibitem [{\citenamefont {Xie}\ and\ \citenamefont {Yau}(2016)}]{Xie:2016uqq}%
  \BibitemOpen
  \bibfield  {author} {\bibinfo {author} {\bibfnamefont {D.}~\bibnamefont
  {Xie}}\ and\ \bibinfo {author} {\bibfnamefont {S.-T.}\ \bibnamefont {Yau}},\
  }\href@noop {} {\  (\bibinfo {year} {2016})},\ \Eprint
  {http://arxiv.org/abs/1602.03529} {arXiv:1602.03529 [hep-th]} \BibitemShut
  {NoStop}%
\end{thebibliography}%

\clearpage

\onecolumngrid

\appendix

\section{The Quiver Mutations} \label{app: quivmut}

\begin{figure}[htbp]
\begin{tabular}{ccc}
\scalebox{0.6}{\begin{tikzpicture}
\node[] (1) at (1,0){};
\node[] (100) at (0,0){};
\node[unode] (2) at (2,0){};
\node[sunode] (3) at (4,0){};
\node[unode] (4) at (6,0){};
\node[unode] (51) at (2,2){};
\node[] (52) at (0,2){};
\node[] (53) at (1,2){};
\node[unode] (61) at (6,2){};
\node[] (62) at (7,2){};
\node[] (63) at (8,2){};
\node[] (5) at (7,0){};
\node[] (200) at (8,0){};
\node[cross, red] (6) at (4,0.5){};
\draw[-] (1) -- (2);
\draw[-] (2)-- (3);
\draw[-] (3) -- (4);
\draw[-] (4) --(5);
\draw[-] (3) --(51);
\draw[-] (3) --(61);
\draw[-, dotted] (1) -- (100);
\draw[-,dotted] (5) -- (200);
\draw[-, dotted] (52) -- (53);
\draw[-] (51) -- (53);
\draw[-, dotted] (62) -- (63);
\draw[-] (61) -- (62);
\node[text width=.2cm](11) [below=0.1cm of 2]{$N_{\alpha_2}$};
\node[text width=.2cm](12) at (4.1, -0.5){$N_\alpha$};
\node[text width=.2cm](13) [below=0.1cm of 4]{$N_{\alpha_3}$};
\node[text width=.2cm](15) [above=0.1cm of 51]{$N_{\alpha_1}$};
\node[text width=.2cm](16) [above=0.1cm of 61]{$N_{\alpha_4}$};
\node[text width=.2cm](20) at (4,-2){$(\CT)$};
\end{tikzpicture}}
& \scalebox{.6}{\begin{tikzpicture}
\draw[->] (0,0) -- (1.5, 0);
\node[text width=0.1cm](29) at (0.5, 0.3) {$\CO_{I}$};
\node[](30) at (0.5, - 2.0) {};
\end{tikzpicture}} 
& \scalebox{0.6}{\begin{tikzpicture}
\node[] (500) at (-2,0){};
\node[] (1) at (1,0){};
\node[] (100) at (0,0){};
\node[unode] (2) at (2,0){};
\node[unode] (3) at (4,0){};
\node[unode] (4) at (6,0){};
\node[unode] (51) at (2,2){};
\node[] (52) at (0,2){};
\node[] (53) at (1,2){};
\node[unode] (61) at (6,2){};
\node[] (62) at (7,2){};
\node[] (63) at (8,2){};
\node[] (5) at (7,0){};
\node[] (200) at (8,0){};
\draw[-] (1) -- (2);
\draw[-] (2)-- (3);
\draw[-] (3) -- (4);
\draw[-] (4) --(5);
\draw[-] (3) --(51);
\draw[-] (3) --(61);
\draw[-, dotted] (1) -- (100);
\draw[-,dotted] (5) -- (200);
\draw[-, dotted] (52) -- (53);
\draw[-] (51) -- (53);
\draw[-, dotted] (62) -- (63);
\draw[-] (61) -- (62);
\draw[-, thick, blue] (2)--(2,1.5);
\draw[-, thick, blue] (3)--(4,1.5);
\draw[-, thick, blue] (4)--(6,1.5);
\draw[-, thick, blue] (2,1.5)--(4,1.5);
\draw[-, thick, blue] (4,1.5)--(6,1.5);
\draw[-, thick, blue] (4,1.5)--(51);
\draw[-, thick, blue] (4,1.5)--(61);
\node[text width=.2cm](11) [below=0.1cm of 2]{$N_{\alpha_2}$};
\node[text width=1.5cm](12) at (4.1, -0.5){$N_\alpha -1$};
\node[text width=.2cm](13) [below=0.1cm of 4]{$N_{\alpha_3}$};
\node[text width=.2cm](15) [above=0.1cm of 51]{$N_{\alpha_1}$};
\node[text width=.2cm](16) [above=0.1cm of 61]{$N_{\alpha_4}$};
\node[text width=.2cm](20) at (4, -2){$(\CT^\vee)$};
\end{tikzpicture}}\\
\qquad & \qquad & \qquad \\
\qquad & \qquad & \qquad \\
\scalebox{0.6}{\begin{tikzpicture}
\node[] (1) at (1,0){};
\node[] (100) at (0,0){};
\node[unode] (2) at (2,0){};
\node[unode] (3) at (4,0){};
\node[unode] (4) at (6,0){};
\node[unode] (51) at (2,2){};
\node[] (52) at (0,2){};
\node[] (53) at (1,2){};
\node[unode] (61) at (6,2){};
\node[] (62) at (7,2){};
\node[] (63) at (8,2){};
\node[] (5) at (7,0){};
\node[] (200) at (8,0){};
\node[cross, red] (6) at (4,0.5){};
\draw[-] (1) -- (2);
\draw[-] (2)-- (3);
\draw[-] (3) -- (4);
\draw[-] (4) --(5);
\draw[-] (3) --(51);
\draw[-] (3) --(61);
\draw[-, dotted] (1) -- (100);
\draw[-,dotted] (5) -- (200);
\draw[-, dotted] (52) -- (53);
\draw[-] (51) -- (53);
\draw[-, dotted] (62) -- (63);
\draw[-] (61) -- (62);
\draw[dotted, thick, blue] (0,1.5)--(1,1.5);
\draw[-, thick, blue] (1,1.5)--(2,1.5);
\draw[-, thick, blue] (2)--(2,1.5);
\draw[-, thick, blue] (3)--(4,1.5);
\draw[-, thick, blue] (4)--(6,1.5);
\draw[-, thick, blue] (2,1.5)--(4,1.5);
\draw[-, thick, blue] (4,1.5)--(6,1.5);
\draw[-, thick, blue] (4,1.5)--(51);
\draw[-, thick, blue] (4,1.5)--(61);
\draw[-, thick, blue] (7,1.5)--(6,1.5);
\draw[dotted, thick, blue] (8,1.5)--(7,1.5);
\node[text width=.2cm](11) [below=0.1cm of 2]{$N_{\alpha_2}$};
\node[text width=.2cm](12) at (4.1, -0.5){$N_\alpha$};
\node[text width=.2cm](13) [below=0.1cm of 4]{$N_{\alpha_3}$};
\node[text width=.2cm](15) [above=0.1cm of 51]{$N_{\alpha_1}$};
\node[text width=.2cm](16) [above=0.1cm of 61]{$N_{\alpha_4}$};
\node[text width=.2cm](20) at (4, -2){$(\CT)$};
\end{tikzpicture}} 
&  \scalebox{.6}{\begin{tikzpicture}
\draw[->] (0,0) -- (1.5,0);
\node[text width=0.1cm](29) at (0.7, 0.5) {$\CO_{I'}$};
\node[](30) at (.5, -2) {};
\end{tikzpicture}}
& \scalebox{0.6}{\begin{tikzpicture}
\node[] (500) at (-2,0){};
\node[] (1) at (1,0){};
\node[] (100) at (0,0){};
\node[unode] (2) at (2,0){};
\node[unode] (3) at (4,0){};
\node[unode] (4) at (6,0){};
\node[unode] (51) at (2,2){};
\node[] (52) at (0,2){};
\node[] (53) at (1,2){};
\node[unode] (61) at (6,2){};
\node[] (62) at (7,2){};
\node[] (63) at (8,2){};
\node[] (5) at (7,0){};
\node[] (200) at (8,0){};
\draw[-] (1) -- (2);
\draw[-] (2)-- (3);
\draw[-] (3) -- (4);
\draw[-] (4) --(5);
\draw[-] (3) --(51);
\draw[-] (3) --(61);
\draw[-, dotted] (1) -- (100);
\draw[-,dotted] (5) -- (200);
\draw[-, dotted] (52) -- (53);
\draw[-] (51) -- (53);
\draw[-, dotted] (62) -- (63);
\draw[-] (61) -- (62);
\node[text width=.2cm](11) [below=0.1cm of 2]{$N_{\alpha_2}$};
\node[text width=1.5cm](12) at (4.1, -0.5){$N_\alpha +1$};
\node[text width=.2cm](13) [below=0.1cm of 4]{$N_{\alpha_3}$};
\node[text width=.2cm](15) [above=0.1cm of 51]{$N_{\alpha_1}$};
\node[text width=.2cm](16) [above=0.1cm of 61]{$N_{\alpha_4}$};
\node[text width=0.3 cm](10) at (8.5,0){$\Bigg/ U(1)$};
\node[text width=.2cm](20) at (4, -2){$(\CT^\vee)$};
\end{tikzpicture}}\\
\qquad & \qquad & \qquad \\
\qquad & \qquad & \qquad \\
\scalebox{0.6}{\begin{tikzpicture}
\node[] (1) at (1,0){};
\node[] (100) at (0,0){};
\node[unode] (2) at (2,0){};
\node[unode] (3) at (4,0){};
\node[unode] (4) at (6,0){};
\node[unode] (51) at (2,2){};
\node[] (52) at (0,2){};
\node[] (53) at (1,2){};
\node[unode] (61) at (6,2){};
\node[] (62) at (7,2){};
\node[] (63) at (8,2){};
\node[] (5) at (7,0){};
\node[] (200) at (8,0){};
\node[cross, red] (6) at (4,0.5){};
\draw[-] (1) -- (2);
\draw[-] (2)-- (3);
\draw[-] (3) -- (4);
\draw[-] (4) --(5);
\draw[-] (3) --(51);
\draw[-] (3) --(61);
\draw[-, dotted] (1) -- (100);
\draw[-,dotted] (5) -- (200);
\draw[-, dotted] (52) -- (53);
\draw[-] (51) -- (53);
\draw[-, dotted] (62) -- (63);
\draw[-] (61) -- (62);
\draw[dotted, thick, blue] (0,1.5)--(1,1.5);
\draw[-, thick, blue] (1,1.5)--(2,1.5);
\draw[-, thick, blue] (2)--(2,1.5);
\draw[-, thick, blue] (3)--(4,1.5);
\draw[-, thick, blue] (4)--(6,1.5);
\draw[-, thick, blue] (2,1.5)--(4,1.5);
\draw[-, thick, blue] (4,1.5)--(6,1.5);
\draw[-, thick, blue] (4,1.5)--(51);
\draw[-, thick, blue] (4,1.5)--(61);
\draw[-, thick, blue] (7,1.5)--(6,1.5);
\draw[dotted, thick, blue] (8,1.5)--(7,1.5);
\node[text width=.2cm](11) [below=0.1cm of 2]{$N_{\alpha_2}$};
\node[text width=.2cm](12) at (4.1, -0.5){$N_\alpha$};
\node[text width=.2cm](13) [below=0.1cm of 4]{$N_{\alpha_3}$};
\node[text width=.2cm](15) [above=0.1cm of 51]{$N_{\alpha_1}$};
\node[text width=.2cm](16) [above=0.1cm of 61]{$N_{\alpha_4}$};
\node[text width=0.1 cm](10) at (4, 2){\footnotesize{$\vec Q$}};
\node[text width=.2cm](20) at (4, -2){$(\CT)$};
\end{tikzpicture}}
& \scalebox{.6}{\begin{tikzpicture}
\draw[->] (0,0) -- (1.5,0);
\node[text width=0.1cm](29) at (0.7, 0.5) {$\CO_{II}$};
\node[](30) at (.5, -2) {};
\end{tikzpicture}}
& \scalebox{0.6}{\begin{tikzpicture}
\node[] (500) at (-2,0){};
\node[] (1) at (1,0){};
\node[] (100) at (0,0){};
\node[unode] (2) at (2,0){};
\node[unode] (3) at (4,0){};
\node[unode] (4) at (6,0){};
\node[unode] (51) at (2,2){};
\node[] (52) at (0,2){};
\node[] (53) at (1,2){};
\node[unode] (61) at (6,2){};
\node[] (62) at (7,2){};
\node[] (63) at (8,2){};
\node[] (5) at (7,0){};
\node[] (200) at (8,0){};
\draw[-] (1) -- (2);
\draw[-] (2)-- (3);
\draw[-] (3) -- (4);
\draw[-] (4) --(5);
\draw[-] (3) --(51);
\draw[-] (3) --(61);
\draw[-, dotted] (1) -- (100);
\draw[-,dotted] (5) -- (200);
\draw[-, dotted] (52) -- (53);
\draw[-] (51) -- (53);
\draw[-, dotted] (62) -- (63);
\draw[-] (61) -- (62);
\draw[dotted, thick, blue] (0,1.5)--(1,1.5);
\draw[-, thick, blue] (1,1.5)--(2,1.5);
\draw[-, thick, blue] (2)--(2,1.5);
\draw[-, thick, blue] (3)--(4,1.5);
\draw[-, thick, blue] (4)--(6,1.5);
\draw[-, thick, blue] (2,1.5)--(4,1.5);
\draw[-, thick, blue] (4,1.5)--(6,1.5);
\draw[-, thick, blue] (4,1.5)--(51);
\draw[-, thick, blue] (4,1.5)--(61);
\draw[-, thick, blue] (7,1.5)--(6,1.5);
\draw[dotted, thick, blue] (8,1.5)--(7,1.5);
\node[text width=.2cm](11) [below=0.1cm of 2]{$N_{\alpha_2}$};
\node[text width=.2cm](12) at (4.1, -0.5){$N_\alpha$};
\node[text width=.2cm](13) [below=0.1cm of 4]{$N_{\alpha_3}$};
\node[text width=.2cm](15) [above=0.1cm of 51]{$N_{\alpha_1}$};
\node[text width=.2cm](16) [above=0.1cm of 61]{$N_{\alpha_4}$};
\node[text width=0.1 cm](10) at (4, 2){\footnotesize{$\vec Q'$}};
\node[text width=.2cm](20) at (4, -2){$(\CT^\vee)$};
\end{tikzpicture}}\\
\qquad & \qquad & \qquad \\
\qquad & \qquad & \qquad \\
\scalebox{0.6}{\begin{tikzpicture}
\node[] (1) at (1,0){};
\node[] (100) at (0,0){};
\node[unode] (2) at (2,0){};
\node[unode] (3) at (4,0){};
\node[unode] (4) at (6,0){};
\node[unode] (51) at (2,2){};
\node[] (52) at (0,2){};
\node[] (53) at (1,2){};
\node[unode] (61) at (6,2){};
\node[] (62) at (7,2){};
\node[] (63) at (8,2){};
\node[] (5) at (7,0){};
\node[] (200) at (8,0){};
\node[cross, red] (6) at (4,0.5){};
\draw[-] (1) -- (2);
\draw[-] (2)-- (3);
\draw[-] (3) -- (4);
\draw[-] (4) --(5);
\draw[-] (3) --(51);
\draw[-] (3) --(61);
\draw[-, dotted] (1) -- (100);
\draw[-,dotted] (5) -- (200);
\draw[-, dotted] (52) -- (53);
\draw[-] (51) -- (53);
\draw[-, dotted] (62) -- (63);
\draw[-] (61) -- (62);
\draw[dotted, thick, blue] (0,1.5)--(1,1.5);
\draw[-, thick, blue] (1,1.5)--(2,1.5);
\draw[-, thick, blue] (2)--(2,1.5);
\draw[-, thick, blue] (3)--(4,1.5);
\draw[-, thick, blue] (4)--(6,1.5);
\draw[-, thick, blue] (2,1.5)--(4,1.5);
\draw[-, thick, blue] (4,1.5)--(6,1.5);
\draw[-, thick, blue] (4,1.5)--(51);
\draw[-, thick, blue] (4,1.5)--(61);
\draw[-, thick, blue] (7,1.5)--(6,1.5);
\draw[dotted, thick, blue] (8,1.5)--(7,1.5);
\node[text width=.2cm](11) [below=0.1cm of 2]{$N_{\alpha_2}$};
\node[text width=.2cm](12) at (4.1, -0.5){$N_\alpha$};
\node[text width=.2cm](13) [below=0.1cm of 4]{$N_{\alpha_3}$};
\node[text width=.2cm](15) [above=0.1cm of 51]{$N_{\alpha_1}$};
\node[text width=.2cm](16) [above=0.1cm of 61]{$N_{\alpha_4}$};
\node[text width=0.1 cm](10) at (4, 2){\footnotesize{$\vec Q$}};
\node[text width=.2cm](20) at (4, -2){$(\CT)$};
\end{tikzpicture}}
& \scalebox{.6}{\begin{tikzpicture}
\draw[->] (0,0) -- (1.5, 0);
\node[text width=0.1cm](29) at (0.5, 0.3) {$\CO_{III}$};
\node[](30) at (0.5, - 2.0) {};
\end{tikzpicture}}
&  \scalebox{0.6}{\begin{tikzpicture}
\node[] (500) at (-2,0){};
\node[] (1) at (1,0){};
\node[] (100) at (0,0){};
\node[unode] (2) at (2,0){};
\node[unode] (3) at (4,0){};
\node[unode] (4) at (6,0){};
\node[unode] (9) at (4,3.5){};
\node[fnode] (10) at (2,3.5){};
\node[unode] (51) at (2,2){};
\node[] (52) at (0,2){};
\node[] (53) at (1,2){};
\node[unode] (61) at (6,2){};
\node[] (62) at (7,2){};
\node[] (63) at (8,2){};
\node[] (5) at (7,0){};
\node[] (200) at (8,0){};
\draw[-] (1) -- (2);
\draw[-] (2)-- (3);
\draw[-] (3) -- (4);
\draw[-] (4) --(5);
\draw[-] (3) --(51);
\draw[-] (3) --(61);
\draw[-] (9) --(10);
\draw[-, dotted] (1) -- (100);
\draw[-,dotted] (5) -- (200);
\draw[-, dotted] (52) -- (53);
\draw[-] (51) -- (53);
\draw[-, dotted] (62) -- (63);
\draw[-] (61) -- (62);
\draw[dotted, thick, blue] (0,1.5)--(1,1.5);
\draw[-, thick, blue] (1,1.5)--(2,1.5);
\draw[-, thick, blue] (2)--(2,1.5);
\draw[-, thick, blue] (3)--(4,1.5);
\draw[-, thick, blue] (4)--(6,1.5);
\draw[-, thick, blue] (2,1.5)--(4,1.5);
\draw[-, thick, blue] (4,1.5)--(6,1.5);
\draw[-, thick, blue] (4,1.5)--(51);
\draw[-, thick, blue] (4,1.5)--(61);
\draw[-, thick, blue] (9)--(4,1.5);
\draw[-, thick, blue] (7,1.5)--(6,1.5);
\draw[dotted, thick, blue] (8,1.5)--(7,1.5);
\node[text width=.2cm](11) [below=0.1cm of 2]{$N_{\alpha_2}$};
\node[text width=1.5cm](12) at (4.1, -0.5){$N_\alpha -1$};
\node[text width=.2cm](13) [below=0.1cm of 4]{$N_{\alpha_3}$};
\node[text width=.2cm](24) [right=0.1cm of 9]{$1$};
\node[text width=.2cm](25) [left=0.1cm of 10]{$1$};
\node[text width=.2cm](15) [above=0.1cm of 51]{$N_{\alpha_1}$};
\node[text width=.2cm](16) [above=0.1cm of 61]{$N_{\alpha_4}$};
\node[text width=2 cm](30) at (5.5, 2){\footnotesize{$(1,\vec Q')$}};
\node[text width=.2cm](20) at (4, -2){$(\CT^\vee)$};
\end{tikzpicture}}
\end{tabular}
\caption{\footnotesize{The four quiver mutations.}}
\label{fig: Mutations}
\end{figure}
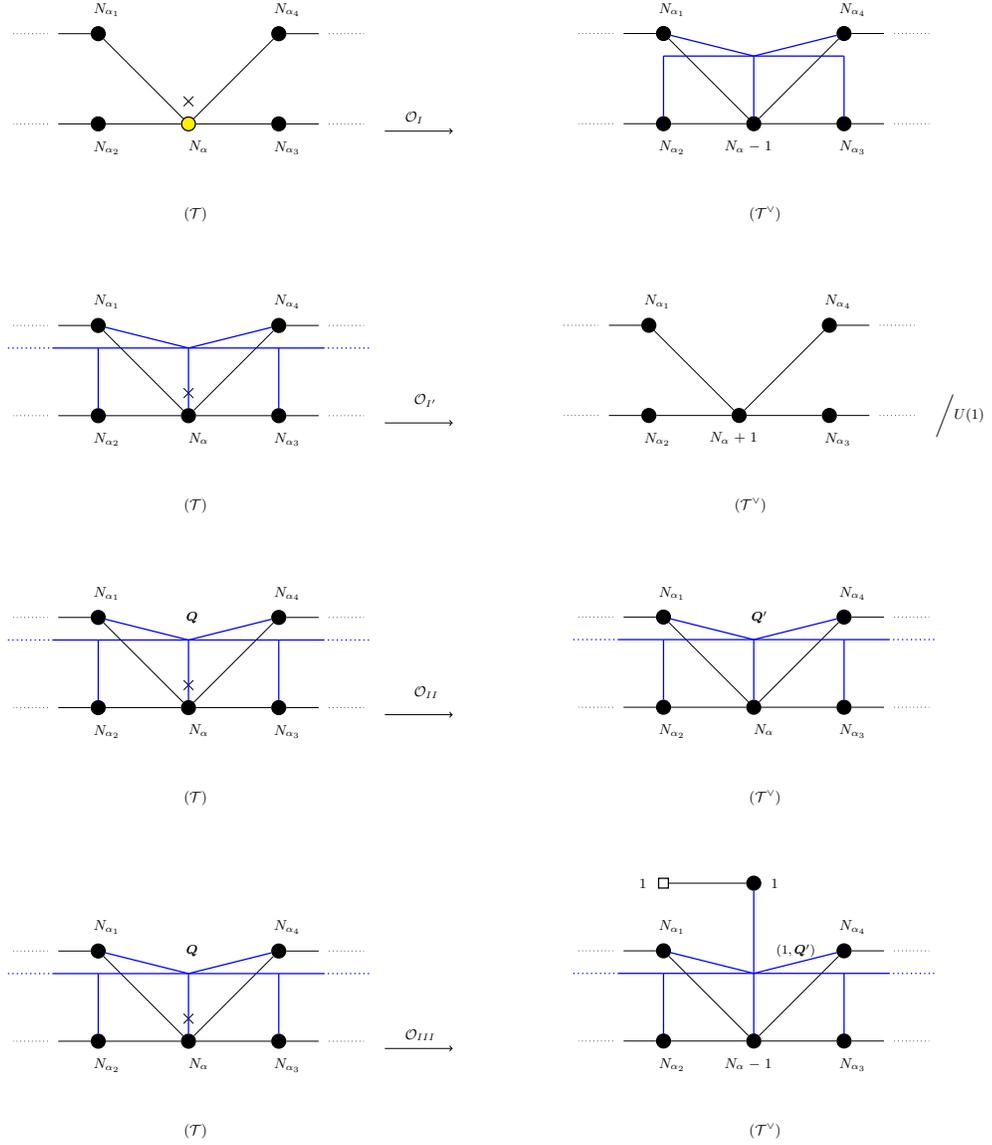

Consider a generic theory $\CT$ in the class of quivers given in \Figref{fig: AbHMgen}. We will require that at least 
one of the special unitary gauge nodes in $\CT$ is balanced, while the remaining ones are overbalanced. The unitary gauge nodes are either 
balanced or overbalanced, and connected to any number of Abelian hypers. One can then introduce four quiver mutations 
on $\CT$ -- shown in \Figref{fig: Mutations} -- as follows:

\begin{enumerate}

\item Mutation $I$ ($\CO_I$), shown in the first row of \Figref{fig: Mutations}, involves replacing a balanced 
$SU(N_\alpha)$ node by a $U(N_\alpha-1)$ node and a single Abelian hypermultiplet with the following charge 
vector $\vec Q$:
\begin{align} \label{charge-1}
\vec Q=\Big(0,\ldots, N_{\alpha _1}, N_{\alpha _2}, -(N_\alpha -1), N_{\alpha _3}, N_{\alpha _4}, \ldots,0 \Big), 
\end{align}
where $-(N_\alpha -1)$ is the charge under the $U(N_\alpha -1)$ gauge node, and $\{N_{\alpha_i}\}$ are the charges 
under those gauge nodes $\{U(N_{\alpha_i})\}$ which are connected to $U(N_\alpha -1)$ by bifundamental hypers. The 
hypermultiplet is not charged under any other gauge node in the quiver.

\item Mutation $I'$ ($\CO_{I'}$), shown in the second row of \Figref{fig: Mutations}, acts on a $U(N_\alpha)$ gauge node with balance parameter $e_\alpha =1$ and a single Abelian hypermultiplet. Without loss of generality, we choose the latter to have charge 
$N_\alpha$ under the $U(N_\alpha)$ node and charges $\{ Q_a\}$ under the other unitary gauge groups. The mutation deletes the Abelian hyper, replaces the $U(N_\alpha)$ node with a $U(N_\alpha +1)$ node, and ungauges a specific $U(1)$ symmetry of the quiver. The particular $U(1)$ symmetry to be ungauged corresponds to the following generator: 
\be
J_G \equiv \sum_{a \neq \alpha}\, \Big(\frac{Q_a}{N_a}\Big)\, J_a  + \sum_i\, J_{\alpha_i}  - J_\alpha, 
\ee
where $J_i$ denotes the generator for the $\fru(1)$ topological symmetry associated with the gauge group $U(N_i)$. On the RHS, the first sum 
extends over all the gauge nodes in $\CT$ under which the Abelian hypermultiplet is charged, except for the $U(N_\alpha)$ node. 
The second sum extends over all the gauge nodes which are connected to $U(N_\alpha)$ by bifundamental 
hypers. In the special case where the Abelian hyper is only charged under the latter set of gauge nodes with charges $\{-N_i\}$, the 
ungauging operation gives an $SU(N_\alpha +1)$. In this case, the $\CO_{I'}$ operation therefore reduces to the inverse of the operation $\CO_I$. For the generic case, the ungauging operation with respect to $\fru(1)_G$ is denoted by $``\Bigg/ U(1)"$ in the quiver.

\item Mutation $II$ ($\CO_{II}$), shown in the third row of \Figref{fig: Mutations}, acts on a $U(N_\alpha)$ gauge node with balance parameter $e_\alpha =0$ and $P \geq 1$ Abelian hypermultiplets. Without loss of generality, we choose the latter to have charge 
$N_\alpha$ under the $U(N_\alpha)$ node and charges $\{ Q^l_a\}_{l=1,\ldots,P}$ under the other unitary gauge groups, 
such that the charge vector has the form:
 \begin{align}
 \vec Q^l= (Q^l_1, \ldots, Q^l_{\alpha_1}, Q^l_{\alpha_2},  N_\alpha, Q^l_{\alpha_3}, Q^l_{\alpha_4}, \ldots, Q^l_L), \label{charge conv}
 \end{align}
 for  $l=1,\ldots,P$, $L$ being the total number of nodes in the quiver, and $\{N_{\alpha_i}\}$ being the charges 
under those gauge nodes $\{U(N_{\alpha_i})\}$ which are connected to $U(N_\alpha)$ by bifundamental hypers. 
 Under this mutation, the gauge and flavor nodes remain invariant, while the $P$ Abelian hypermultiplets in $\CT$
 are mapped to $P$ Abelian hypermultiplets with the following charge vectors: 
 \begin{align}
 \vec Q'^l=  (-Q^l_1, \ldots, -Q^l_{\alpha_1} - N_{\alpha_1}, -Q^l_{\alpha_2} - N_{\alpha_2}, N_\alpha, -Q^l_{\alpha_3} - N_{\alpha_3} , -Q^l_{\alpha_4} - N_{\alpha_4}, \ldots, -Q^l_L).
 \end{align}
 One can check that this operation squares to an identity operation.

\item Mutation $III$ ($\CO_{III}$), shown in the fourth row of \Figref{fig: Mutations}, acts on a $U(N_\alpha)$ gauge node with balance parameter $e_\alpha =-1$ and $P \geq 1$ Abelian hypermultiplets. The charges of the Abelian hypermultiplets may be chosen as in 
\eref{charge conv}. The mutation replaces a $U(N_\alpha)$ gauge node by a $U(N_\alpha-1)$ node and a $U(1)$ node where the latter node has a single fundamental hyper. In addition, the $P $ Abelian hypermultiplets get mapped by the mutation to $P $ Abelian hypermultiplets with the following charge vectors:
\begin{align} \label{charge-3}
\vec Q'^l  =\Big( 1,Q^l_1, \ldots, Q^l_{\alpha_1} + N_{\alpha_1}, Q^l_{\alpha_2} + N_{\alpha_2}, -(N_\alpha-1), Q^l_{\alpha_3} + N_{\alpha_3} , Q^l_{\alpha_4} + N_{\alpha_4}, \ldots, Q^l_L \Big),
\end{align}
where the first entry denotes the charge under the new $U(1)$ node and $-(N_\alpha-1)$ is the charge under the 
$U(N_\alpha -1)$ node. For the other nodes, only the charges associated with those connected to $U(N_\alpha-1)$ 
with bifundamental hypers get transformed under the mutation. Finally, one can check that the composition of 
$\CO_{I'}$ with $\CO_{III}$, with the $\CO_{I'}$ acting on the  $U(N_\alpha-1)$ node, gives the identity operation.

\end{enumerate}

\section{Circle reduction and U-SU quivers of $D^b_p(SU(N))$ SCFTs}\label{app: 3dLagDp}

In this section, we review the 3d U-SU quivers that arise from the circle-reduction of the $D^b_p(SU(N))$ 
theories, following the results in \cite{Closset:2020afy}. The quiver data can be 
summarized as follows:

\begin{center}
\begin{tabular}{ccc}
{\begin{tikzpicture}
\node[text width=3cm](20) at (0,0){$[D^b_p(SU(N))]_{3d}$\,:};
\end{tikzpicture}}
&\scalebox{0.75}{\begin{tikzpicture}
\node[fnode] (1) at (0,0){};
\node[unode, gray] (2) at (2,0){};
\node[unode, gray] (3) at (4,0){};
\node[] (4) at (5,0){};
\node[] (5) at (7,0){};
\node[unode, gray] (6) at (8,0){};
\node[unode, gray] (7) at (10,0){};
\draw[-] (1) -- (2);
\draw[-] (2)-- (3);
\draw[-] (3) -- (4);
\draw[dashed] (4) --(5);
\draw[-] (5) --(6);
\draw[-] (6) --(7);
\node[text width=.1cm](10) [below=0.1 cm of 1]{$N$};
\node[text width=.2cm](11) [below=0.1cm of 2]{$n_1$};
\node[text width=.1cm](12) [below=0.1cm of 3]{$n_2$};
\node[text width=.1cm](13) [below=0.1cm of 6]{$n_{l-1}$};
\node[text width=.1cm](14) [below=0.1cm of 7]{$n_l$};
\end{tikzpicture}}
&\quad {\begin{tikzpicture}
\node[text width=4cm](20) at (0,0){if $b=N$ and $\frac{p}{N} \not\in \BZ$};
\node[](21) at (0,-0.4){};
\end{tikzpicture}}\\
\qquad
& \qquad
& \qquad \\
\qquad
& \scalebox{0.75}{\begin{tikzpicture}
\node[fnode] (1) at (0,0){};
\node[unode, gray] (2) at (2,0){};
\node[unode, gray] (3) at (4,0){};
\node[] (4) at (5,0){};
\node[] (5) at (7,0){};
\node[unode, gray] (6) at (8,0){};
\node[unode, gray] (7) at (10,0){};
\node[fnode] (8) at (12,0){};
\draw[-] (1) -- (2);
\draw[-] (2)-- (3);
\draw[-] (3) -- (4);
\draw[dashed] (4) --(5);
\draw[-] (5) --(6);
\draw[-] (6) --(7);
\draw[-] (7) --(8);
\node[text width=.1cm](10) [below=0.1 cm of 1]{$N$};
\node[text width=.2cm](11) [below=0.1cm of 2]{$n_1$};
\node[text width=.1cm](12) [below=0.1cm of 3]{$n_2$};
\node[text width=.1cm](13) [below=0.1cm of 6]{$n_{l-1}$};
\node[text width=.1cm](14) [below=0.1cm of 7]{$n_l$};
\node[text width=.1cm](15) [below=0.1cm of 8]{1};
\end{tikzpicture}}
&\quad {\begin{tikzpicture}
\node[text width=2cm](20) at (0,0){otherwise};
\node[](21) at (0,-0.4){};
\end{tikzpicture}}
\end{tabular}
\end{center}
The integers $\{n_k\}$ which label the various gauge nodes of the above linear quiver are given as:
\be \label{LQ-rank}
n_k = \lfloor {N - \frac{k\,b}{p}} \rfloor, \qquad k \leq l,
\ee
where $l$ is the number of gauge nodes in the linear quiver. A gauge node labelled by an integer $n_k$ is  
a $U(n_k)$ node if $\frac{k\,b}{p} \not\in \BZ$ or an $SU(n_k)$ node $\frac{k\,b}{p} \in \BZ$ respectively. 
Finally, the length of the quiver $l$ is given by:
\begin{align} \label{formula-l}
l = \begin{cases}  & p-1-\frac{p}{N} \quad \text{if} \quad b=N\,\text{and}\,\frac{p}{N} \in \BZ, \\ 
& \lfloor p - \frac{p}{N} \rfloor  \quad \text{if} \quad b=N\,\text{and}\,\frac{p}{N} \not\in \BZ,\\
& p-1 \quad \text{if} \quad b=N-1. \end{cases}
\end{align}

Given the above rules, one can readily check that the quiver associated to the $D_2(SU(2n-1))$ SCFT is a 
$U(n-1)$ SQCD with $N_f = 2n-1$. A slightly more non-trivial example is that of the $D_6(SU(9))$ SCFT 
which we study in \Secref{app: DpEx}. In this case, the length of the quiver is $l= \lfloor p - \frac{p}{N} \rfloor = 5$, 
with the labels of the nodes being : $\{n_k =  \lfloor 9 - \frac{3k}{2} \rfloor \}_{k=1,\ldots,5}= \{7, 6, 4, 3, 1 \}$. Since 
$\frac{k\,b}{p} = \frac{3k}{2}$, the nodes corresponding to $k=2$ and $k=4$ are special unitary nodes while the rest 
are unitary nodes. This leads to the final form of the quiver:

\be \label{USU-ExDpN}
\scalebox{0.6}{\begin{tikzpicture}
\node[text width=3cm](30) at (-2,0){$[D_6(SU(9))]_{3d}$\,:};
\node[fnode] (1) at (0,0){};
\node[unode] (2) at (2,0){};
\node[sunode] (3) at (4,0){};
\node[unode] (4) at (6,0){};
\node[sunode] (5) at (8,0){};
\node[unode] (6) at (10,0){};
\draw[-] (1) -- (2);
\draw[-] (2)-- (3);
\draw[-] (3) -- (4);
\draw[-] (4) --(5);
\draw[-] (5) --(6);
\node[text width=.1cm](10) [below=0.1 cm of 1]{9};
\node[text width=.1cm](11) [below=0.1cm of 2]{7};
\node[text width=.1cm](12) [below=0.1cm of 3]{6};
\node[text width=.1cm](13) [below=0.1cm of 4]{4};
\node[text width=.1cm](14) [below=0.1cm of 5]{3};
\node[text width=.1cm](15) [below=0.1cm of 6]{1};
\end{tikzpicture}}
\ee

\section{Condition for Hyperk\"ahler Quotient N-ality: $D^b_p(SU(N))$ SCFTs}\label{app: Nalitycond}

In this section, we find the subclass of $D^b_p(SU(N))$ SCFTs which have the property of \hk quotient N-ality. 
The general form of the 3d U-SU quiver that arises from the circle-reduction of a $D^b_p(SU(N))$ SCFT is given 
in \Appref{app: 3dLagDp}. A generic U-SU quiver will have an emergent CB symmetry and admit an IR  duality 
sequence if at least one of the SU nodes is balanced. Therefore, our task is to find the condition involving the 
parameters $b,p$ and $N$ that ensures that the U-SU quiver has one or more balanced SU nodes. 

We begin by looking at the larger subclass of $D^b_p(SU(N))$ SCFTs which have SU nodes (balanced or overbalanced). 
Since the SU nodes are associated with the marginal couplings, this is in fact the subclass of non-isolated SCFTs. 
From \Appref{app: 3dLagDp}, we know that the $k$-th node in the quiver with label $n_k$ is special unitary, if the integer 
k satisfies the condition:
\be
\frac{k\,b}{p} \in \BZ, \quad \text{where} \quad k \leq l, \label{SU-node-cond}
\ee
with $l$ being the length of quiver given in \eref{formula-l}. The larger subclass therefore consists of SCFTs for 
which the associated U-SU quiver satisfies the above condition for at least one value of $k$. For such a $k$, 
one has a $SU(n_k)$ node with $n_k = \lfloor N - \frac{k\,b}{p} \rfloor = (N - \frac{k\,b}{p})$. This node will be 
balanced if $n_{k-1} + n_{k+1}= 2n_k -1$ which translates to the condition:
\be
\lfloor n_k + \frac{b}{p} \rfloor + \lfloor n_k - \frac{b}{p} \rfloor =2n_k -1. 
\ee
This condition is satisfied for a generic $k$ satisfying \eref{SU-node-cond} if the parameters $b$ and $p$ obey the constraint :
\be
\frac{b}{p} \not\in \BZ. \label{SU-bal-cond}
\ee

Note that this condition is independent of $k$. This implies that for a given theory in the subclass of non-isolated $D^b_p(SU(N))$ SCFTs, 
every SU node in the 3d U-SU quiver will be balanced provided the above condition is satisfied. Following the discussion in 
\Secref{sec: maxquiv}, any such quiver will admit a duality sequence leading to maximal unitary quivers which only 
consist of unitary gauge nodes (since all SU nodes are balanced and therefore admit mutation $I$) and a total of $GCD(b,p) -1$ 
Abelian hypermultiplets -- one each from the $GCD(b,p) -1$ balanced special unitary nodes, in addition to fundamental/bifundamental 
hypers. 

Let us investigate in a bit more detail what the condition \eref{SU-bal-cond} really means. 
In the Type IIB picture, the deformed singularity for the $D^b_p(SU(N))$ has the following form:
\be
\widehat{F}(\vec x,z) = F(\vec x, z) + \sum_{\substack{0 \leq i \leq p-1 \\ 0 \leq j \leq N-2}}\, u_{ij}\, x^{j}_3\, z^{i} =0,
\ee
where the polynomial $F(\vec x, z)$ has the form given in \eref{DpSUN-undef}, and the coefficients $\{ u_{ij}\}$ have the scaling dimensions:
\be
\Big[u_{ij} \Big] =\Big(N- j - i\,\frac{b}{p} \Big).
\ee
For $\frac{b}{p} \in \BZ$, none of the CB operators can have fractional scaling dimensions, while for $\frac{b}{p} \in \BZ$ the SCFT generically 
has CB operators with fractional scaling dimensions and are of the Argyres-Douglas type. Therefore, we conclude that the subclass of $D^b_p(SU(N))$ 
SCFTs which admit \hk quotient N-ality in the sense discussed in \Secref{sec: maxquiv} is given by the subclass of non-isolated SCFTs of the Argyres-Douglas 
type.

\section{Illustrative Example : Maximal Unitary Quivers and the 3d Mirror of $D_6(SU(9))$}\label{app: DpEx}

In this section, we will present a concrete example where we determine the maximal unitary quiver and the 3d mirror of a 
$D^b_p(SU(N))$ SCFT which admits \hk quotient N-ality. We choose the SCFT $D_6(SU(9))$ which has a conformal manifold 
of dimension 2 corresponding to two special unitary vector multiplets in the 3d U-SU quiver which is given in \eref{USU-ExDpN}. 

\begin{figure}[htbp]
\begin{tabular}{ccc}
\scalebox{0.6}{\begin{tikzpicture}
\node[fnode] (1) at (0,0){};
\node[unode] (2) at (2,0){};
\node[sunode] (3) at (4,0){};
\node[unode] (4) at (6,0){};
\node[sunode] (5) at (8,0){};
\node[unode] (6) at (10,0){};
\node[cross, red] (40) at (4, 0.5){};
\draw[-] (1) -- (2);
\draw[-] (2)-- (3);
\draw[-] (3) -- (4);
\draw[-] (4) --(5);
\draw[-] (5) --(6);
\node[text width=.1cm](10) [below=0.1 cm of 1]{9};
\node[text width=.1cm](11) [below=0.1cm of 2]{7};
\node[text width=.1cm](12) [below=0.1cm of 3]{6};
\node[text width=.1cm](13) [below=0.1cm of 4]{4};
\node[text width=.1cm](14) [below=0.1cm of 5]{3};
\node[text width=.1cm](15) [below=0.1cm of 6]{1};
\node[text width=.2cm](20) at (6,-2){$(\CT)$};
\end{tikzpicture}}
& \scalebox{.6}{\begin{tikzpicture}
\node[] at (3.5,0){};
\draw[->] (4,0) -- (7, 0);
\node[text width=0.1cm](29) at (5, 0.3) {$\CO_{I}$};
\node[](30) at (5, -2.2) {};
\end{tikzpicture}}
& \scalebox{0.6}{\begin{tikzpicture}
\node[fnode] (1) at (0,0){};
\node[unode] (2) at (2,0){};
\node[unode] (3) at (4,0){};
\node[unode] (4) at (6,0){};
\node[sunode] (5) at (8,0){};
\node[unode] (6) at (10,0){};
\node[cross, red] (40) at (8, 0.5){};
\draw[-] (1) -- (2);
\draw[-] (2)-- (3);
\draw[-] (3) -- (4);
\draw[-] (4) --(5);
\draw[-] (5) --(6);
\draw[-, thick, blue] (2)--(2,1.5);
\draw[-, thick, blue] (3)--(4,1.5);
\draw[-, thick, blue] (4)--(6,1.5);
\draw[-, thick, blue] (2,1.5)--(6,1.5);
\node[text width=3 cm](10) at (5, 2){\footnotesize{$(7, -5, 4)$}};
\node[text width=.1cm](10) [below=0.1 cm of 1]{9};
\node[text width=.1cm](11) [below=0.1cm of 2]{7};
\node[text width=.1cm](12) [below=0.1cm of 3]{5};
\node[text width=.1cm](13) [below=0.1cm of 4]{4};
\node[text width=.1cm](14) [below=0.1cm of 5]{2};
\node[text width=.1cm](15) [below=0.1cm of 6]{1};
\node[text width=.2cm](20) at (6,-2){$(\CT_1)$};
\end{tikzpicture}}\\
\qquad 
& \qquad
& \scalebox{.6}{\begin{tikzpicture}
\draw[->] (0,-1) -- (0, -3);
\node[text width=0.1cm](29) at (0.2, -2) {$\CO_{I}$};
\end{tikzpicture}} \\
\scalebox{0.6}{\begin{tikzpicture}
\node[unode](100) at (-1,0){};
\node[fnode](101) at (-2,0){};
\node[fnode] (1) at (0,0){};
\node[unode] (2) at (2,0){};
\node[unode] (3) at (4,0){};
\node[unode] (4) at (6,0){};
\node[unode] (5) at (8,0){};
\node[unode] (6) at (10,0){};
\draw[-] (1) -- (2);
\draw[-] (2)-- (3);
\draw[-] (3) -- (4);
\draw[-] (4) --(5);
\draw[-] (5) --(6);
\draw[-] (100) --(101);
\draw[-, thick, blue] (100) -- (-1, 1.5);
\draw[-, thick, blue] (2)--(2,1.5);
\draw[-, thick, blue] (4)--(6,1.5);
\draw[-, thick, blue] (5)--(8,1.5);
\draw[-, thick, blue] (-1,1.5)--(8,1.5);
\node[text width=3 cm](10) at (6, 2){\footnotesize{$(1, 7, -3, 2)$}};
\draw[-, thick, blue] (100) -- (0, -1.5);
\draw[-, thick, blue] (3)--(4, -1.5);
\draw[-, thick, blue] (4)--(6, -1.5);
\draw[-, thick, blue] (6)--(10, -1.5);
\draw[-, thick, blue] (0,-1.5)--(10, -1.5);
\node[text width=3 cm](10) at (6, -2){\footnotesize{$(1, 5, -3, 1)$}};
\node[text width=.1cm](10) [below=0.1 cm of 1]{9};
\node[text width=.1cm](11) [below=0.1cm of 2]{7};
\node[text width=.1cm](12) [below=0.1cm of 3]{5};
\node[text width=.1cm](13) [below=0.1cm of 4]{3};
\node[text width=.1cm](14) [below=0.1cm of 5]{2};
\node[text width=.1cm](15) [below=0.1cm of 6]{1};
\node[text width=.1cm](18) [below=0.1cm of 100]{1};
\node[text width=.1cm](19) [below=0.1cm of 101]{1};
\node[text width=.2cm](20) at (6,-3){$(\CT_{\rm maximal})$};
\end{tikzpicture}}
& \scalebox{.6}{\begin{tikzpicture}
\node[] at (3.5,0){};
\draw[->] (7,0) -- (4, 0);
\node[text width=0.1cm](29) at (5, 0.3) {$\CO_{III}$};
\node[](30) at (5, -3.2) {};
\end{tikzpicture}}
& \scalebox{0.6}{\begin{tikzpicture}
\node[fnode] (1) at (0,0){};
\node[unode] (2) at (2,0){};
\node[unode] (3) at (4,0){};
\node[unode] (4) at (6,0){};
\node[unode] (5) at (8,0){};
\node[unode] (6) at (10,0){};
\node[cross, red] (40) at (5.7, 0.5){};
\draw[-] (1) -- (2);
\draw[-] (2)-- (3);
\draw[-] (3) -- (4);
\draw[-] (4) --(5);
\draw[-] (5) --(6);
\draw[-, thick, blue] (2)--(2,1.5);
\draw[-, thick, blue] (3)--(4,1.5);
\draw[-, thick, blue] (4)--(6,1.5);
\draw[-, thick, blue] (2,1.5)--(6,1.5);
\node[text width=3 cm](10) at (5, 2){\footnotesize{$(7, -5, 4)$}};
\draw[-, thick, blue] (4)--(6, -1.5);
\draw[-, thick, blue] (5)--(8, -1.5);
\draw[-, thick, blue] (6)--(10, -1.5);
\draw[-, thick, blue] (6,-1.5)--(10, -1.5);
\node[text width=3 cm](10) at (9, -2){\footnotesize{$(4, -2, 1)$}};
\node[text width=.1cm](10) [below=0.1 cm of 1]{9};
\node[text width=.1cm](11) [below=0.1cm of 2]{7};
\node[text width=.1cm](12) [below=0.1cm of 3]{5};
\node[text width=.1cm](13) [below=0.1cm of 4]{4};
\node[text width=.1cm](14) [below=0.1cm of 5]{2};
\node[text width=.1cm](15) [below=0.1cm of 6]{1};
\node[text width=.2cm](20) at (6,-3){$(\CT_2)$};
\end{tikzpicture}}
\end{tabular}
\caption{\footnotesize{Derivation of $\CT_{\rm maximal}$ for the $D_6(SU(9))$ SCFT.} }
\label{IRdual-Ex1DN}
\end{figure}
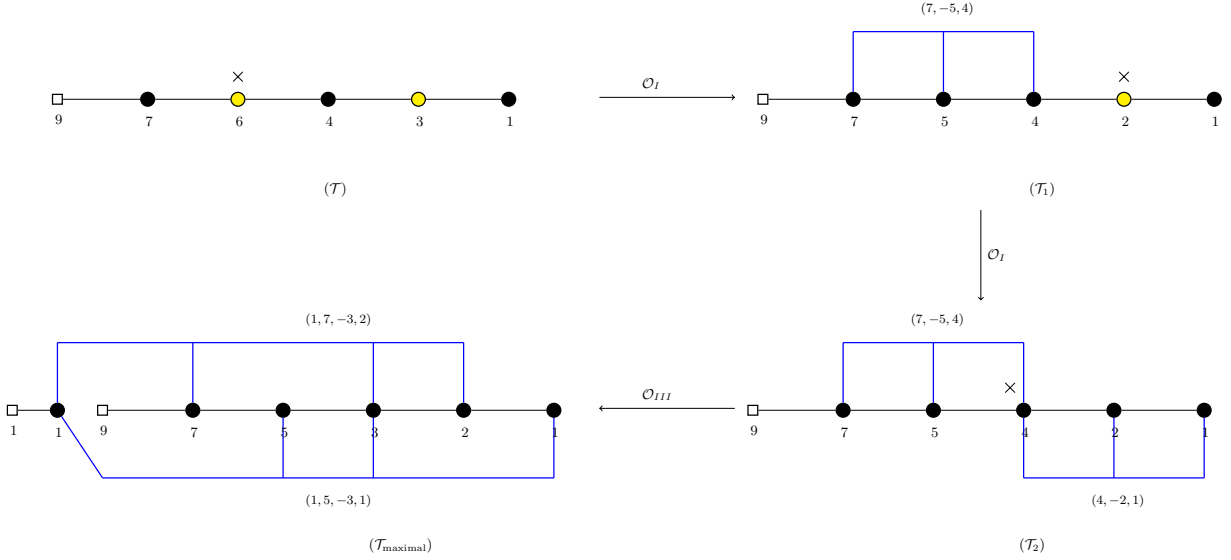

For the sake of brevity, we will not list the complete list of quivers in the N-al set, although they can be readily determined following 
the general prescription presented in \Secref{sec: maxquiv}. Instead, we will discuss a (sub)sequence of IR dualities that leads to the subset of 
maximal unitary quivers. The sequence of dualities is shown in \Figref{IRdual-Ex1DN}. We begin by implementing mutation $I$ at the
balanced $SU(6)$ node, followed by another mutation $I$ at the balanced $SU(4)$ node. Each mutation introduces an Abelian hypermultiplet 
with charges as shown, leading to the quiver $\CT_2$ which has only unitary nodes. The $U(4)$ node in $\CT_2$ has balance parameter $e=-1$ 
and therefore admits a mutation $III$. The resultant quiver (on the bottom left corner) has only unitary gauges nodes which are either balanced
or overbalanced and do not admit another mutation $III$. Therefore, one can identify this quiver as a maximal unitary quiver $\CT_{\rm maximal}$. 
The subset of maximal unitary quivers may be obtained from this quiver by enumerating the distinct Lagrangians that arise from implementing 
a sequence of mutation $II$ at the balanced nodes. 

Note that $\text{rk}(\frg^{\rm UV}_{\rm C}(\CT))=3$ which implies that one can only turn on 
two FI parameters for the quiver, and the rank increases along the duality sequence by 1 at each step. 
For the maximal unitary quiver, the number of FI parameters is maximized with $\text{rk}(\frg^{\rm UV}_{\rm C}(\CT_{\rm maximal}))=6$. \\

\begin{figure}[htbp]
\begin{tabular}{ccccc}
\scalebox{0.6}{\begin{tikzpicture}
\node[unode] (100) at (2, 2.2){};
\node[unode] (101) at (1, 2.2){};
\node[fnode] (102) at (0, 2.2){};
\node[fnode] (1) at (0,0){};
\node[unode] (2) at (1,0){};
\node[unode] (3) at (2,0){};
\node[unode] (4) at (3,0){};
\node[unode] (5) at (4,0){};
\node[unode] (6) at (5,0){};
\draw[-] (1) -- (2);
\draw[-] (2)-- (3);
\draw[-] (3) -- (4);
\draw[-] (4) --(5);
\draw[-] (5) --(6);
\node[unode] (200) at (3, -2.2){};
\node[fnode] (201) at (4,-2.2){};
\draw[-] (200) --(201);
\draw[-] (100) --(101);
\draw[-] (101) --(102);
\draw[->, red, dashed] (2,1) -- (100);
\draw[->, red, dashed] (2,1) -- (2);
\draw[->, red, dashed] (2,1) -- (4);
\draw[->, red, dashed] (2,1) -- (5);
\node[text width=.1cm](10) [below=0.1 cm of 1]{9};
\node[text width=.1cm](11) [below=0.1cm of 2]{7};
\node[text width=.1cm](12) [below=0.1cm of 3]{5};
\node[text width=.1cm](13) [below=0.1cm of 4]{3};
\node[text width=.1cm](14) [below=0.1cm of 5]{2};
\node[text width=.1cm](15) [below=0.1cm of 6]{1};
\node[text width=.1cm](18) [above=0.1cm of 100]{1};
\node[text width=.1cm](19) [above=0.1cm of 101]{1};
\node[text width=.1cm](20) [above=0.1cm of 102]{1};
\node[text width=.1cm](21) [below=0.1cm of 200]{1};
\node[text width=.1cm](22) [below=0.1cm of 201]{1};
\node[text width= 5 cm](23) at (2,-3.5){$(\CT_{\rm good}\oplus \CT^{(1)}_{\rm decoupled} \oplus \CT^{(2)}_{\rm decoupled})$};
\end{tikzpicture}}
&\scalebox{.6}{\begin{tikzpicture}
\node[] at (3.5,0){};
\draw[->] (4,0) -- (6, 0);
\node[text width=0.1cm](29) at (5, 0.3) {$S_1$};
\node[](30) at (5, -3.5) {};
\end{tikzpicture}}
&\scalebox{0.6}{\begin{tikzpicture}
\node[unode](100) at (-1,0){};
\node[fnode](101) at (-2,0){};
\node[fnode] (1) at (0,0){};
\node[unode] (2) at (1,0){};
\node[unode] (3) at (2,0){};
\node[unode] (4) at (3,0){};
\node[unode] (5) at (4,0){};
\node[unode] (6) at (5,0){};
\draw[-] (1) -- (2);
\draw[-] (2)-- (3);
\draw[-] (3) -- (4);
\draw[-] (4) --(5);
\draw[-] (5) --(6);
\draw[-] (100) --(101);
\draw[-, thick, blue] (100) -- (-1, 1.5);
\draw[-, thick, blue] (2)--(1,1.5);
\draw[-, thick, blue] (4)--(3,1.5);
\draw[-, thick, blue] (5)--(4,1.5);
\draw[-, thick, blue] (-1,1.5)--(4,1.5);
\node[text width=3 cm](10) at (3, 2){\footnotesize{$(1, 7, -3, 2)$}};
\node[unode] (200) at (3, -2.2){};
\node[fnode] (201) at (4,-2.2){};
\draw[-] (200) --(201);
\draw[->, red, dashed] (3,-1.5) -- (200);
\draw[->, red, dashed] (3,-1.5) -- (100);
\draw[->, red, dashed] (3,-1.5) -- (3);
\draw[->, red, dashed] (3,-1.5) -- (4);
\draw[->, red, dashed] (3,-1.5) -- (6);
\node[text width=.1cm](10) [below=0.1 cm of 1]{9};
\node[text width=.1cm](11) [below=0.1cm of 2]{7};
\node[text width=.1cm](12) [below=0.1cm of 3]{5};
\node[text width=.1cm](13) [below=0.1cm of 4]{3};
\node[text width=.1cm](14) [below=0.1cm of 5]{2};
\node[text width=.1cm](15) [below=0.1cm of 6]{1};
\node[text width=.1cm](18) [below=0.1cm of 100]{1};
\node[text width=.1cm](19) [below=0.1cm of 101]{1};
\node[text width=.1cm](20) [below=0.1cm of 200]{1};
\node[text width=.1cm](21) [below=0.1cm of 201]{1};
\node[text width= 3 cm](30) at (2,-3.5){$(\CT^{(1)}_{\rm good}\oplus \CT^{(2)}_{\rm decoupled})$};
\end{tikzpicture}}
&\scalebox{.6}{\begin{tikzpicture}
\node[] at (3.5,0){};
\draw[->] (4,0) -- (6, 0);
\node[text width=0.1cm](29) at (5, 0.3) {$S_2$};
\node[](30) at (5, -3.5) {};
\end{tikzpicture}}
& \scalebox{0.6}{\begin{tikzpicture}
\node[unode](100) at (-1,0){};
\node[fnode](101) at (-2,0){};
\node[fnode] (1) at (0,0){};
\node[unode] (2) at (1,0){};
\node[unode] (3) at (2,0){};
\node[unode] (4) at (3,0){};
\node[unode] (5) at (4,0){};
\node[unode] (6) at (5,0){};
\draw[-] (1) -- (2);
\draw[-] (2)-- (3);
\draw[-] (3) -- (4);
\draw[-] (4) --(5);
\draw[-] (5) --(6);
\draw[-] (100) --(101);
\draw[-, thick, blue] (100) -- (-1, 1.5);
\draw[-, thick, blue] (2)--(1,1.5);
\draw[-, thick, blue] (4)--(3,1.5);
\draw[-, thick, blue] (5)--(4,1.5);
\draw[-, thick, blue] (-1,1.5)--(4,1.5);
\node[text width=3 cm](10) at (3, 2){\footnotesize{$(1, 7, -3, 2)$}};
\draw[-, thick, blue] (100) -- (0, -1.5);
\draw[-, thick, blue] (3)--(2, -1.5);
\draw[-, thick, blue] (4)--(3, -1.5);
\draw[-, thick, blue] (6)--(5, -1.5);
\draw[-, thick, blue] (0,-1.5)--(5, -1.5);
\node[text width=3 cm](10) at (3, -2){\footnotesize{$(1, 5, -3, 1)$}};
\node[text width=.1cm](10) [below=0.1 cm of 1]{9};
\node[text width=.1cm](11) [below=0.1cm of 2]{7};
\node[text width=.1cm](12) [below=0.1cm of 3]{5};
\node[text width=.1cm](13) [below=0.1cm of 4]{3};
\node[text width=.1cm](14) [below=0.1cm of 5]{2};
\node[text width=.1cm](15) [below=0.1cm of 6]{1};
\node[text width=.1cm](18) [below=0.1cm of 100]{1};
\node[text width=.1cm](19) [below=0.1cm of 101]{1};
\node[text width=.2cm](20) at (2,-3.5){$(\CT_{\rm maximal})$};
\end{tikzpicture}} \\
\qquad & \qquad & \qquad & \qquad & \qquad  \\
\qquad & \qquad & \qquad & \qquad & \qquad  \\
\scalebox{0.6}{\begin{tikzpicture}
\node[unode] (1) at (2,0){};
\node[unode] (2) at (4,0){};
\node[fnode] (3) at (2,-1){};
\node[fnode] (4) at (1,-1){};
\node[fnode] (5) at (4,-1){};
\node[fnode] (6) at (5,-1){};
\node[rnode] (7) at (6,0){$T(U(6))$};
\node[fnode](101) at (4, -2){};
\node[fnode](102) at (5, -2){};
\node[fnode](103) at (1, -2){};
\node[fnode](104) at (0, -2){};
\draw[-] (1) -- (2);
\draw[-] (1)-- (3);
\draw[-] (1) -- (4);
\draw[-] (2) --(5);
\draw[-] (2) --(6);
\draw[-] (2) --(7);
\draw[-] (101) --(102);
\draw[-] (103) --(104);
\draw[->, red, dashed] (3,-2) -- (103);
\draw[->, red, dashed] (3,-2) -- (3);
\draw[->, red, dashed] (3,-2) -- (5);
\node[text width=0.1cm](20) [above=0.1 cm of 1]{5};
\node[text width=0.1 cm](21) [above=0.1cm of 2]{7};
\node[text width=0.1 cm](22) [right=0.1cm of 3]{1};
\node[text width=0.1 cm](23) [left=0.1cm of 4]{2};
\node[text width=0.1 cm](25) [left=0.1cm of 5]{1};
\node[text width=0.1 cm](26) [right=0.1cm of 6]{2};
\node[text width=.1cm](27) [above=0.1cm of 101]{1};
\node[text width=.1cm](28) [above=0.1cm of 102]{1};
\node[text width=.1cm](29) [above=0.1cm of 103]{1};
\node[text width=.1cm](30) [above=0.1cm of 104]{2};
\node[text width= 5 cm](31) at (3,-3){$(\wt{\CT}_{\rm good}\oplus \wt{\CT}^{(1)}_{\rm decoupled}\oplus \wt{\CT}^{(2)}_{\rm decoupled})$};
\end{tikzpicture}}
&\scalebox{.6}{\begin{tikzpicture}
\node[] at (3.5,0){};
\draw[->] (4,0) -- (6, 0);
\node[text width=0.1cm](29) at (5, 0.3) {$S_1$};
\node[](30) at (5, -2) {};
\end{tikzpicture}}
&\scalebox{0.6}{\begin{tikzpicture}
\node[unode] (1) at (2,0){};
\node[unode] (2) at (5,0){};
\node[unode] (3) at (3.5, -2){};
\node[fnode] (4) at (2,-1){};
\node[fnode] (5) at (1,-1){};
\node[fnode] (6) at (5,-1){};
\node[fnode] (7) at (6,-1){};
\node[fnode] (8) at (4.5,-2){};
\node[fnode] (9) at (2.5,-2){};
\node[rnode] (10) at (7, 0){$T(U(6))$};
\node[fnode](101) at (6.5, -2){};
\node[fnode](102) at (7.5, -2){};
\draw[-] (1) -- (2);
\draw[-] (1) -- (3);
\draw[-] (2) -- (3);
\draw[-] (1) -- (4);
\draw[-] (1) -- (5);
\draw[-] (2) --(6);
\draw[-] (2) --(7);
\draw[-] (2) --(10);
\draw[-] (3) --(8);
\draw[-] (3) --(9);
\draw[-] (101) --(102);
\draw[->, red, dashed] (5.5,-2) -- (101);
\draw[->, red, dashed] (5.5,-2) -- (4);
\draw[->, red, dashed] (5.5,-2) -- (6);
\draw[->, red, dashed] (5.5,-2) -- (8);
\node[text width=0.1cm](20) [above=0.1 cm of 1]{5};
\node[text width=0.1 cm](21) [above=0.1cm of 2]{7};
\node[text width=0.1 cm](22) [below=0.1cm of 3]{1};
\node[text width=0.1 cm](23) [left=0.1cm of 4]{1};
\node[text width=0.1 cm](24) [left=0.1cm of 5]{1};
\node[text width=0.1 cm](25) [right=0.1cm of 6]{1};
\node[text width=0.1 cm](26) [right=0.1cm of 7]{1};
\node[text width=0.1 cm](27) [below=0.1cm of 8]{1};
\node[text width=0.1 cm](28) [below=0.1cm of 9]{1};
\node[text width=.1cm](29) [above=0.1cm of 101]{1};
\node[text width=.1cm](30) [above=0.1cm of 102]{1};
\node[text width=3 cm](31) at (4,-3){$(\wt{\CT}^{(1)}_{\rm good}\oplus \wt{\CT}^{(2)}_{\rm decoupled})$};
\end{tikzpicture}}
&\scalebox{.6}{\begin{tikzpicture}
\node[] at (3.5,0){};
\draw[->] (4,0) -- (6, 0);
\node[text width=0.1cm](29) at (5, 0.3) {$S_2$};
\node[](30) at (5, -2) {};
\end{tikzpicture}}
& \scalebox{0.6}{\begin{tikzpicture}
\node[fnode] (1) at (1, 1){};
\node[unode] (2) at (2,0){};
\node[unode] (3) at (4,0){};
\node[fnode] (4) at (5,1){};
\node[fnode] (5) at (1,-2){};
\node[unode] (6) at (2,-2){};
\node[unode] (7) at (4,-2){};
\node[fnode] (8) at (5,-2){};
\node[rnode] (9) at (6,0){$T(U(6))$};
\draw[-] (1) -- (2);
\draw[-] (2)-- (3);
\draw[-] (3) -- (4);
\draw[-] (2) --(6);
\draw[-] (2) --(7);
\draw[-] (3) --(6);
\draw[-] (3) --(7);
\draw[-] (5) --(6);
\draw[-] (6) --(7);
\draw[-] (7) --(8);
\draw[-] (3) --(9);
\node[text width=0.1cm](20) [above=0.1 cm of 1]{1};
\node[text width=0.1 cm](21) [above=0.1cm of 2]{5};
\node[text width=0.1 cm](22) [above=0.1cm of 3]{7};
\node[text width=0.1 cm](23) [above=0.1cm of 4]{1};
\node[text width=0.1 cm](24) [below=0.1cm of 5]{1};
\node[text width=0.1 cm](25) [below=0.1cm of 6]{1};
\node[text width=0.1 cm](26) [below=0.1cm of 7]{1};
\node[text width=0.1 cm](27) [below=0.1cm of 8]{1};
\node[text width=.2cm](30) at (3,-3){$(\wt{\CT}_{\rm maximal})$};
\end{tikzpicture}}
\end{tabular}
\caption{\footnotesize{Construction of the 3d mirror for $D_6(SU(9))$. The gray node labelled denotes a $T(U(6))$ quiver tail.}}
\label{SOP-Ex1DN}
\end{figure}
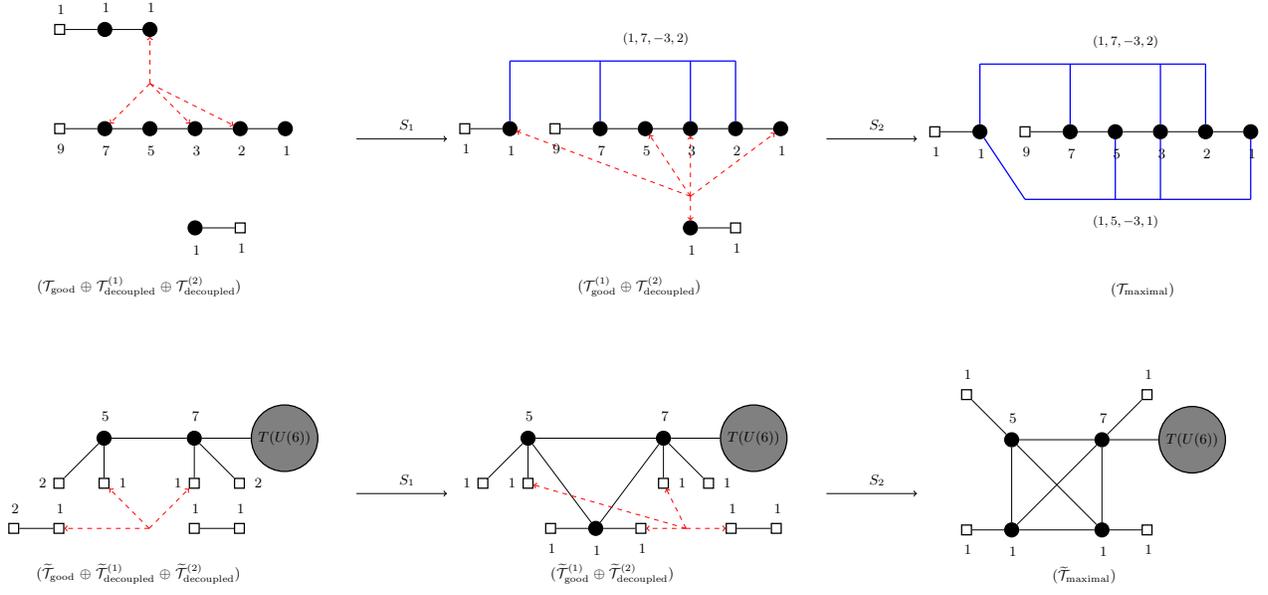

Next, let us construct the 3d mirror for $D_6(SU(9))$. The maximal unitary quiver $\CT_{\rm maximal}$ can again be thought of 
as a linear quiver gauge theory decorated with Abelian hypermultiplets attached in turn to Abelian quiver tails. The linear quiver 
$\CT_{\rm good}$, obtained by stripping off the Abelian hypermultiplets plus the quiver tails, and its 3d mirror $\wt{\CT}_{\rm good}$ 
are given as:

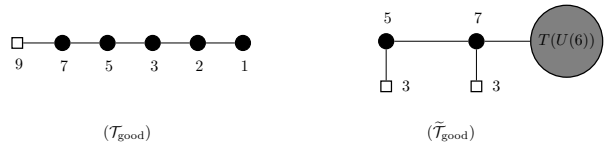
\begin{center}
\begin{tabular}{ccc}
\scalebox{0.6}{\begin{tikzpicture}
\node[] (100) at (3,1){};
\node[fnode] (1) at (0,0){};
\node[unode] (2) at (1,0){};
\node[unode] (3) at (2,0){};
\node[unode] (4) at (3,0){};
\node[unode] (5) at (4,0){};
\node[unode] (6) at (5,0){};
\draw[-] (1) -- (2);
\draw[-] (2)-- (3);
\draw[-] (3) -- (4);
\draw[-] (4) --(5);
\draw[-] (5) --(6);
\node[text width=.1cm](10) [below=0.1 cm of 1]{9};
\node[text width=.1cm](11) [below=0.1cm of 2]{7};
\node[text width=.1cm](12) [below=0.1cm of 3]{5};
\node[text width=.1cm](13) [below=0.1cm of 4]{3};
\node[text width=.1cm](14) [below=0.1cm of 5]{2};
\node[text width=.1cm](15) [below=0.1cm of 6]{1};
\node[text width=.2cm](23) at (2,-2){$(\CT_{\rm good})$};
\end{tikzpicture}}
& \qquad \qquad \qquad 
& \scalebox{0.6}{\begin{tikzpicture}
\node[unode] (1) at (2,0){};
\node[unode] (2) at (4,0){};
\node[fnode] (3) at (2,-1){};
\node[fnode] (5) at (4,-1){};
\node[rnode] (7) at (6,0){$T(U(6))$};
\draw[-] (1) -- (2);
\draw[-] (1)-- (3);
\draw[-] (2) --(5);
\draw[-] (2) --(7);
\node[text width=0.1cm](20) [above=0.1 cm of 1]{5};
\node[text width=0.1 cm](21) [above=0.1cm of 2]{7};
\node[text width=0.1 cm](22) [right=0.1cm of 3]{3};
\node[text width=0.1 cm](25) [right=0.1cm of 5]{3};
\node[text width=.2cm](31) at (3,-2){$(\wt{\CT}_{\rm good})$};
\end{tikzpicture}}
\end{tabular}
\end{center}

The gray node denotes a $T(U(6))$ tail : \scalebox{0.6}{\begin{tikzpicture}
\node[unode] (1) at (0,0){};
\node[unode] (2) at (1,0){};
\node[unode] (3) at (2,0){};
\node[unode] (4) at (3,0){};
\node[unode] (5) at (4,0){};
\node[unode] (6) at (5,0){};
\draw[-] (1) -- (2);
\draw[-] (2)-- (3);
\draw[-] (3) -- (4);
\draw[-] (4) --(5);
\draw[-] (5) --(6);
\node[text width=.1cm](10) [below=0.1 cm of 1]{6};
\node[text width=.1cm](11) [below=0.1cm of 2]{5};
\node[text width=.1cm](12) [below=0.1cm of 3]{4};
\node[text width=.1cm](13) [below=0.1cm of 4]{3};
\node[text width=.1cm](14) [below=0.1cm of 5]{2};
\node[text width=.1cm](15) [below=0.1cm of 6]{1};
\end{tikzpicture}}. Following the prescription in \Secref{sec: 3dmirr}, one can then proceed with implementing $S$-operations 
on the quiver $\CT_{\rm good}$, as shown in the top row of \Figref{SOP-Ex1DN}, to construct the quiver $\CT_{\rm maximal}$. 
The operations on the mirror side are shown in the bottom row and lead to the 3d mirror of the $D_6(SU(9))$ SCFT:
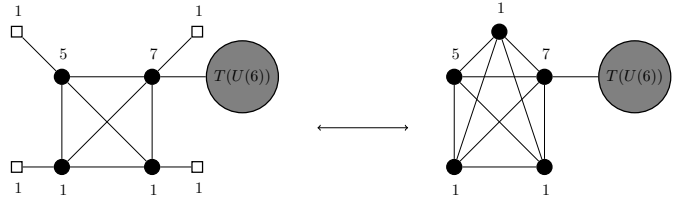
\begin{center}
\begin{tabular}{ccc}
\scalebox{0.6}{\begin{tikzpicture}
\node[fnode] (1) at (1, 1){};
\node[unode] (2) at (2,0){};
\node[unode] (3) at (4,0){};
\node[fnode] (4) at (5,1){};
\node[fnode] (5) at (1,-2){};
\node[unode] (6) at (2,-2){};
\node[unode] (7) at (4,-2){};
\node[fnode] (8) at (5,-2){};
\node[rnode] (9) at (6,0){$T(U(6))$};
\draw[-] (1) -- (2);
\draw[-] (2)-- (3);
\draw[-] (3) -- (4);
\draw[-] (2) --(6);
\draw[-] (2) --(7);
\draw[-] (3) --(6);
\draw[-] (3) --(7);
\draw[-] (5) --(6);
\draw[-] (6) --(7);
\draw[-] (7) --(8);
\draw[-] (3) --(9);
\node[text width=0.1cm](20) [above=0.1 cm of 1]{1};
\node[text width=0.1 cm](21) [above=0.1cm of 2]{5};
\node[text width=0.1 cm](22) [above=0.1cm of 3]{7};
\node[text width=0.1 cm](23) [above=0.1cm of 4]{1};
\node[text width=0.1 cm](24) [below=0.1cm of 5]{1};
\node[text width=0.1 cm](25) [below=0.1cm of 6]{1};
\node[text width=0.1 cm](26) [below=0.1cm of 7]{1};
\node[text width=0.1 cm](27) [below=0.1cm of 8]{1};
\end{tikzpicture}}
& \scalebox{.6}{\begin{tikzpicture}
\node[] at (3.5,0){};
\node[] at (6.5,0){};
\draw[<->] (4,0) -- (6, 0);
\node[](30) at (5, -1.5) {};
\end{tikzpicture}}
& \scalebox{0.6}{\begin{tikzpicture}
\node[unode] (1) at (3, 1){};
\node[unode] (2) at (2,0){};
\node[unode] (3) at (4,0){};
\node[unode] (6) at (2,-2){};
\node[unode] (7) at (4,-2){};
\node[rnode] (9) at (6,0){$T(U(6))$};
\draw[-] (1) -- (2);
\draw[-] (1) -- (3);
\draw[-] (1) -- (6);
\draw[-] (1) -- (7);
\draw[-] (2)-- (3);
\draw[-] (2) --(6);
\draw[-] (2) --(7);
\draw[-] (3) --(6);
\draw[-] (3) --(7);
\draw[-] (6) --(7);
\draw[-] (3) --(9);
\node[text width=0.1cm](20) [above=0.1 cm of 1]{1};
\node[text width=0.1 cm](21) [above=0.1cm of 2]{5};
\node[text width=0.1 cm](22) [above=0.1cm of 3]{7};
\node[text width=0.1 cm](25) [below=0.1cm of 6]{1};
\node[text width=0.1 cm](26) [below=0.1cm of 7]{1};
\end{tikzpicture}}
\end{tabular}
\end{center}

Given the 3d mirror quiver, one can readily check that the number of independent mass parameters is given by 
$\text{rk}(\frg^{\rm UV}_{\rm H}(\wt{\CT}_{\rm maximal}))=6$, which exactly matches the number of FI parameters 
in the quiver $\CT_{\rm maximal}$.

\appendix

\end{document}